\documentclass[12pt]{article}
\usepackage{authblk}   
\usepackage{setspace}  

\usepackage{amsfonts,amsthm,amsmath,amssymb,upgreek,bm}

\usepackage{hyperref}    
\hypersetup{
    colorlinks=true,     
    linkcolor=blue,      
    urlcolor=blue,       
    citecolor=blue       
}

\usepackage[paper=letterpaper,margin=.85in]{geometry}
\usepackage{graphicx}
\usepackage{units}
\usepackage{float}
\usepackage[export]{adjustbox}
\usepackage[dvipsnames]{xcolor}
\usepackage[shortlabels]{enumitem}
\usepackage[mathscr]{euscript}
\usepackage[normalem]{ulem}
\usepackage{bm}
\usepackage{tikz}
\usepackage{tikzsymbols}
\usepackage{pgfplots}
\usepackage{ytableau}
\usepackage{slashed}
\usepackage{mathtools}
\usepackage{bbold}
\usetikzlibrary{decorations.pathmorphing}
\usetikzlibrary{arrows.meta}
\usetikzlibrary{calc}
\usepgfplotslibrary{fillbetween}
\usepackage[nameinlink]{cleveref}
\pgfplotsset{compat=1.18}
\usepackage{framed}

\usepackage{tikz}
\usetikzlibrary{positioning,arrows.meta,calc}

\usepackage[normalem]{ulem}

\renewcommand{\tilde}{\widetilde}

\usepackage{inconsolata}
\newcommand{\lr}[1]{\left\langle #1 \right\rangle}
\usepackage[dvipsnames]{xcolor}
\usepackage{physics}

\numberwithin{equation}{section}

\def\Tr{\text{Tr}}

\newcommand{\syk}{\text{syk}}
\newcommand{\fake}{\text{fake}}
\newcommand{\naive}{\text{naive}}

\newcommand{\mZ}{\mathcal{Z}}

\newcommand{\mJ}{\mathcal{J}}
\newcommand{\mR}{\mathcal{R}}
\newcommand{\mS}{\mathcal{S}}
\newcommand{\mC}{\mathcal{C}}
\newcommand{\mK}{\mathcal{K}}
\renewcommand{\i}{\mathrm{i}}

\renewcommand{\t}{\texttt{t}}
\newcommand{\z}{\texttt{z}}
\newcommand{\qnm}{\text{qnm}}
\newcommand{\mh}{\mathfrak{h}}
\newcommand{\mA}{\mathcal{A}}

\renewcommand{\hat}{\widehat}

\newcommand{\ccline}{\rule[0.6ex]{0.9cm}{0.5pt}}

\usepackage{tocloft}
\begin{document}
\thispagestyle{empty}

\vspace*{2cm}
\begin{center}

{
{
\fontsize{25pt}{40pt}\selectfont 
A Two-Point Hologram for Everything}}

\begin{center}

\vspace{0.5cm}

{\fontsize{15.5pt}{1pt}\selectfont Tamra Nebabu${}^{1}$, Xiao-Liang Qi${}^2$, Haifeng Tang${}^{*2}$ and Huaijin Wang${}^{2}$}\\
\bigskip \rm

\bigskip 

{\fontsize{13pt}{1pt}\selectfont \textit{${}^1$ Department of Physics, Cornell University, Ithaca, New York} }
\,\vspace{0.5cm}\\
{\fontsize{13pt}{1pt}\selectfont\textit{${}^2$ Leinweber Institute for Theoretical Physics, Stanford University, Stanford, CA 94305, USA}}

\rm
  \end{center}

\vspace{1.5cm}
{\bf Abstract}
\end{center}
\begin{quotation}
\noindent

Known holographic dictionaries, especially AdS/CFT, rely on symmetry matching between the bulk and the boundary. We take a step toward a holographic dictionary with no symmetry requirement and without assuming the geometry being asymptotically AdS. Starting from \textit{any} interacting Majorana generalized free field on a $(0+1)$d boundary and its two-point function data, we derive a concise analytic formula for the dual $(1+1)$d bulk geometry, borrowing techniques from unitary matrix integral and inverse scattering. Using this formula, we compute the near-horizon curvature, give conditions for positive versus negative curvature, and identify simple boundary models with de Sitter or anti-de Sitter near-horizon duals. We also study the large-$q$ SYK model, finding an unusual temperature dependence of the near-horizon curvature, related to the discrepancy between physical temperature and the ``fake disk'' temperature. We also construct, directly from boundary operators, approximate algebras generated by null translations and boost that become exact at the bifurcate horizon.

\end{quotation}

\vspace*{-0cm}
\begin{figure}[H]
    
    \hspace*{6cm}
    \includegraphics[width=0.3\linewidth]{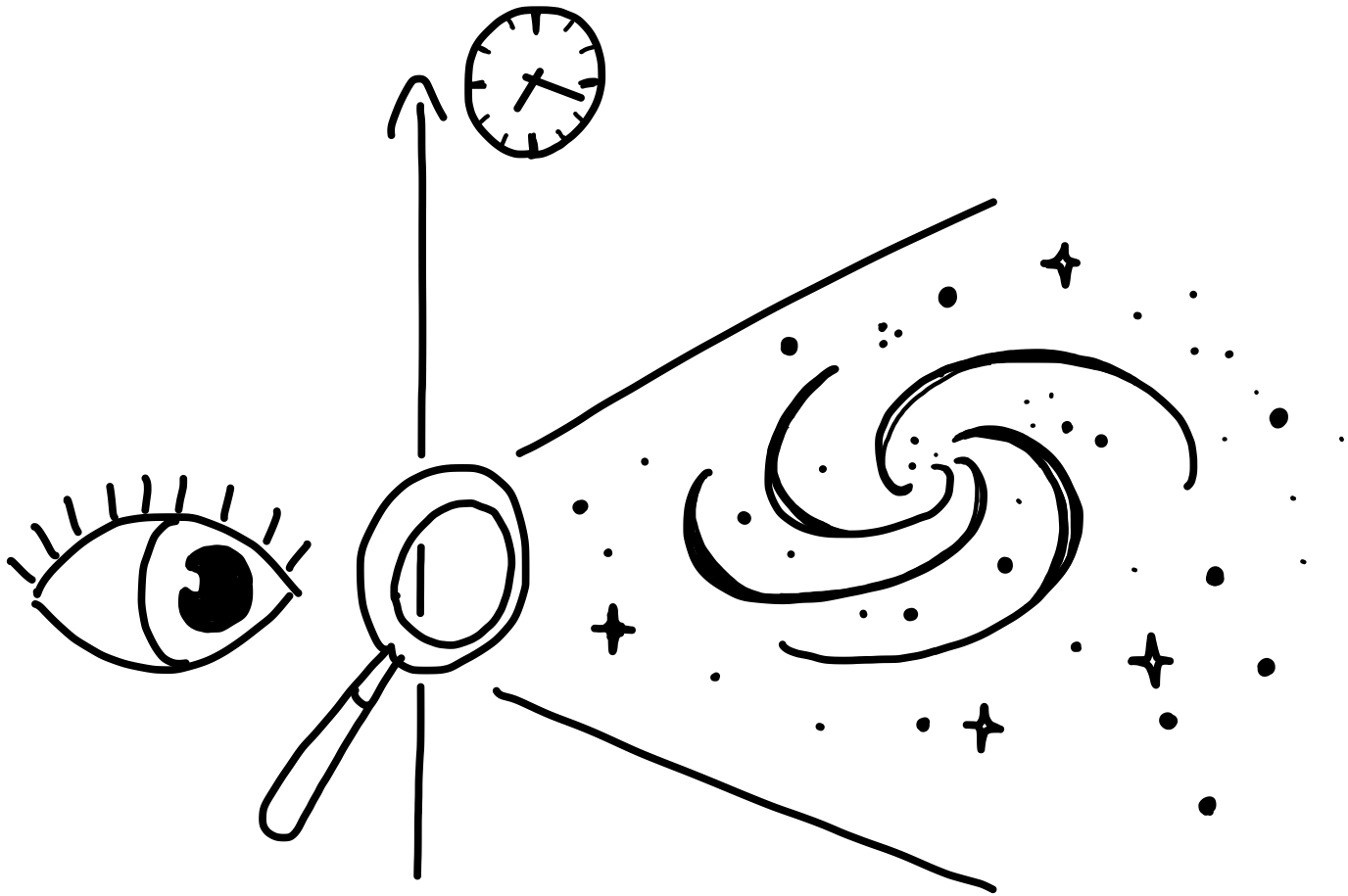}
\end{figure}



\vspace{-3pt}\noindent\rule{.4\columnwidth}{.4pt}

{\footnotesize${}^*$\texttt{hftang@stanford.edu}}

\setcounter{page}{0}
\setcounter{tocdepth}{2}
\setcounter{footnote}{0}

\setcounter{page}{0}
\setcounter{tocdepth}{2}
\setcounter{footnote}{0}

\newpage
\setcounter{page}{2}

\begingroup
\hypersetup{linkcolor=black} 
\tableofcontents
\endgroup

\newpage

\section{Introduction}
\label{sec: introduction}
Is there a gravitational dual for a \textit{generic} quantum many-body system? This question is motivated by the success of holographic principle~\cite{Susskind:1994vu}, which relates a bulk quantum gravity theory to a non-gravitational boundary theory in one lower dimension, applied to the \textit{special} setting of AdS (Anti-de Sitter)/CFT (conformal field theory) duality~\cite{Maldacena:1997re,Witten:1998qj,Gubser:1998bc,Aharony:1999ti}. A canonical example is the duality between supergravity on AdS$_D$ and $\mathcal N=4$ super Yang-Mills theory on Mink$_{D-1}$~\cite{Maldacena:1997re}.

In AdS/CFT, the dictionary is tightly constrained by symmetry matching: the isometry group of AdS${}_D$ equals the conformal group of Mink$_{D-1}$. It is therefore natural to ask whether a more general holographic principle~\cite{Susskind:1994vu} exists--- one with weaker, or even no, reliance on symmetry---particularly given the renewed interest in quantum gravity in de Sitter spacetime~\cite{Strominger:2001pn,Chandrasekaran:2022cip,Susskind:2021esx,Narovlansky:2023lfz,Coleman:2021nor,Maldacena:2024spf,Ivo:2025yek,Chen:2025jqm,Yang:2025lme,Collier:2025lux}.

Constructing such a dictionary at the fully \textit{non-perturbative} level is difficult: microscopic boundary models with quantum gravity bulk duals typically require large numbers of degrees of freedom and strong coupling~\cite{Banks:1996vh}. Here we instead pursue a \textit{phenomenological} dictionary: given only two-point correlation data on the boundary, what can be inferred about the semiclassical bulk geometry and about bulk operators expressed in terms of boundary ones?

A concrete step in this direction was taken in~\cite{Nebabu:2023iox}. Building on the Hamilton--Kabat--Lifschytz--Lowe (HKLL) construction~\cite{Hamilton:2005ju,Hamilton:2006az,Hamilton:2006fh}, 
they proposed a reconstruction scheme that uses only boundary data and assumes neither prior knowledge of the bulk geometry nor (approximate) conformal symmetry.

One limitation of~\cite{Nebabu:2023iox} is that the construction is numerical, relying on two-point functions on some discrete time points; the reconstructed bulk theory is specified as a circuit. In this work, we take the continuum limit and obtain analytic results. A main outcome is an explicit formula for the semiclassical bulk geometry reconstructed from the spectral function of a boundary generalized free field (GFF). Concretely, consider a time-translation-invariant Majorana GFF $\chi(t)$ in $(0+1)$d, and suppose we are given its anti-commutator two-point function $A(t,t'):=\lr{\{\chi(t),\chi(t')\}}$. We show that the dual $(1+1)$d bulk theory is a minimally coupled Majorana field of mass $m$ on a geometry with metric
\begin{equation}
ds^2=\Omega(z)^2(-dt^2+dz^2)
\label{eq: conformal coordinate}
\end{equation}
where the conformal factor $\Omega(z)$ is determined as follows. Denote the Fourier transformation of $A(t,t')$ as $\mathcal{A}(\omega)$. Assume\footnote{A more general formula without this assumption appears in \eqref{eq: main result3, contour integral of S},\eqref{eq: main result4, contour integral of S}.} that $\mA(\omega)$ has simple poles $\{\i J_n\}$ and zeros $\{\i\lambda_i\}$ in the upper half-plane, and associate to each $\lambda_i$ the coefficient
\begin{equation}
\beta_i:=2\lambda_i\times\prod_{k\neq i}\frac{\lambda_k+\lambda_i}{\lambda_k-\lambda_i}\times \prod_{n}\frac{J_n-\lambda_i}{J_n+\lambda_i}
\label{eq: main result1}
\end{equation}
Physically, $-\i\beta_i$ is the residue of the S-matrix $\mS(\omega)$ at the pole $\i\lambda_i$. Then
\begin{equation}
m^2\Omega(z)^2=-\partial_z^2\log\det[\textbf{1}-B(z)^2],\ \text{ with } B(z)_{ij}:=\frac{\beta_je^{-\lambda_j 2z}}{\lambda_i+\lambda_j}
\label{eq: main result2}
\end{equation}

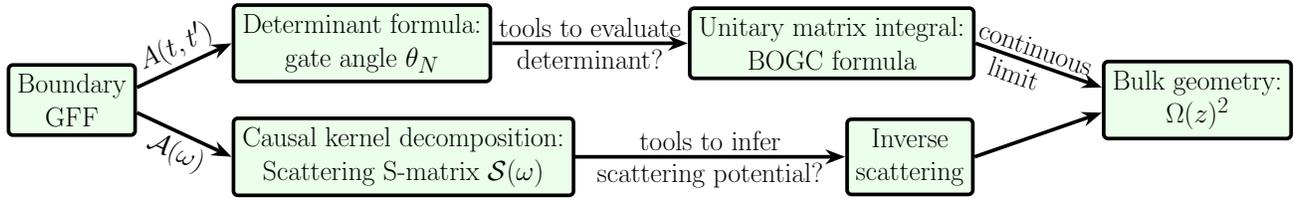
\begin{figure}[!t]
  \centering
  \resizebox{0.99\linewidth}{!}{%
    \begin{tikzpicture}[
        font=\Huge,
        >=Stealth,
        node distance=18mm and 22mm,
        box/.style={draw, rounded corners=4pt, line width=3.3pt, align=center, inner sep=8pt,
        fill=green!8},
        line/.style={->, line width=3.3pt}
    ]

    \node[box] (bnd) {Boundary\\GFF};

    \node[box, right=30mm of bnd, yshift=18mm] (det)
        {Determinant formula:\\ gate angle $\theta_N$};

    \node[box, right=62mm of det] (umi)
        {Unitary matrix integral:\\BOGC formula};

    \node[box, right=40mm of umi, yshift=-18mm] (bulk)
        {Bulk geometry:\\$\Omega(z)^2$};

    \node[box, right=30mm of bnd, yshift=-18mm] (causal)
        {Causal kernel decomposition:\\Scattering S-matrix $\mS(\omega)$};

    \node[box, right=85.5mm of causal] (inv)
        {Inverse\\scattering};

    \draw[line] ($(bnd.east)+(0,4mm)$) -- (det.west)
        node[midway, above, sloped] {$A(t,t')$};

    \draw[line] ($(bnd.east)+(0,-4mm)$) -- (causal.west)
        node[midway, below, sloped] {$\mA(\omega)$};

    \draw[line] (det.east) -- (umi.west) node[midway, above] {tools to evaluate }node[midway, below] {determinant?};
    \draw[line] (umi.east) -- ($(bulk.west)+(0,4mm)$)node[midway, above,sloped] { continuous   \ \     }node[midway, below,sloped] { limit\ \ \ \ \ \ };

    \draw[line] (causal.east) -- (inv.west) node[midway, above] {tools to infer }node[midway, below] {scattering potential?};
    \draw[line] (inv.east) -- ($(bulk.west)+(0,-4mm)$);
    
    \end{tikzpicture}%
  }
  \caption{Roadmap from the boundary GFF two-point function to the dual bulk geometry, via two complementary approaches.}
  \label{fig: roadmap}
\end{figure}

Equation~(\ref{eq: main result2}) is the main result of this work. Based on this result, we investigate several examples of two-point functions and various aspects of the bulk geometry. In particular, we derive a concise formula for the scalar curvature near the black hole horizon. 

The remainder of the paper is organized as follows. In \cref{sec: Derive bulk geometry from boundary correlator} we first overview the key results in Ref.~\cite{Nebabu:2023iox}, and then provide derivations of the main results \eqref{eq: main result1} and \eqref{eq: main result2} using the following two methods.
\begin{itemize}
\item Method I (unitary matrix integral): We start from the discrete Gaussian-circuit construction of~\cite{Nebabu:2023iox} (reviewed in \cref{sec: Review of Nebabu-Qi's discrete circuit construction}), and in \cref{sec: Method I: via unitary matrix integral} derive a determinant formula expressing the gate angles in terms of determinants of sub-matrices of $A(t,t')$, and evaluate the resulting Toeplitz determinants using the Borodin--Okounkov--Geronimo--Case (BOGC) formula~\cite{Borodin2000,Geronimo:1979iy}, familiar from giant graviton expansion in the context of  unitary matrix integral~\cite{Aharony:2003sx,Arai:2019aou,Arai:2019wgv,Arai:2019xmp,Arai:2020uwd,Gaiotto:2021xce,Imamura:2021ytr,Murthy:2022ien,Eniceicu:2023uvd,Chen:2024cvf}.

\item Method II (inverse scattering):  In \cref{sec: Method II: via inverse scattering}, we work directly in the continuum, interpret $\Omega(z)$ as the scattering potential, extract the scattering amplitude $\mS(\omega)$ from $\mA(\omega)$ via causal kernel decomposition, and reconstruct $\Omega(z)$ using inverse scattering~\cite{Newton1982Scattering}.
\end{itemize}
These methods are sketched in \cref{fig: roadmap}. We then study how the horizon curvature $R_\text{h}$ depends on the boundary correlator. 

In \cref{sec: Curvature near horizon and AdS universality}, we obtain the generic asymptotic behavior of metric and curvature near the horizon. We show that for generic $A(t,t')$ the near-horizon region is universally AdS, with $R_\text{h}/m^2=-2$, independent of microscopic details. Only if the first and second zero of $\mathcal{A}(\omega)$ has exactly ratio $3$, the curvature can take other values including dS, flat space or AdS with different value of $R_\text{h}/m^2$. 

In \cref{sec: Examples: finite QNMs}, we analyze correlators with finitely many quasi-normal modes (QNMs)~\cite{Horowitz:1999jd,Berti:2009kk}. For $N_\qnm=2$ the geometry is always constant-curvature AdS$_2$. For $N_\qnm=3$ we exhibit a family with $R_\text{h}/m^2>0$, realizing de Sitter near horizon. A generic feature of finite-QNM examples is that $\Omega(z)$ remains finite at the boundary (i.e. the boundary is not at the asymptotic AdS boundary), naturally realizing finite-cutoff holography~\cite{Batra:2024kjl,Coleman:2021nor,Anninos:2024xhc}. We also show how such spectra with finitely QNMs arise from Gaussian Lindbladian dynamics  as a boundary model, and in particular we record an explicit realization of such $N_\qnm=3$ model with de Sitter near horizon.

In \cref{sec: Examples: large-q SYK}, we turn to the large-$q$ SYK model~\cite{Sachdev:1992fk,Kitaev:2015talks,Maldacena:2016hyu,Polchinski:2016xgd}, which has infinitely many QNMs. In \cref{sec:syk_setup} we obtain the general formula for SYK model two-point function and curvature. In \cref{sec: zero temperature syk}, we study the zero-temperature limit and recover constant-curvature AdS$_2$ with the correct $R_\text h/m^2$. In \cref{sec: finite temperature syk}, we study finite temperature and find that $R_\text{h}/m^2$ has an unusual dependence on $T_\syk$, changing sign and diverging in an almost periodic fashion, due to collisions between poles and zeros of $\mS(\omega)$ induced by the mismatch between $T_\syk$ and the ``fake disk'' temperature $T_\fake$~\cite{Maldacena:2016hyu,Lin:2023trc}.

In \cref{sec: algebra near horizon}, we discuss the operator algebra in the near horizon region. For AdS$_2$ or dS$_2$ geometry there is an sl(2,$\mathbb R$) algebra generated by the null translations $P^\pm$ and boost $B$, while for flat space we expect to find a Poincar\'e algebra. We show that for a generic boundary theory  whose dual bulk geometry doesn't exhibit maximal symmetry, these three operators can also be defined, which form an approximate algebra; when acting on fermion modes at the bifurcate horizon, the algebra becomes exact. 

Finally, in \cref{sec: discussion} we present the conclusion and discuss potential future directions. 

\section{Derive bulk geometry from boundary correlator}
\label{sec: Derive bulk geometry from boundary correlator}
\subsection{Review of Nebabu-Qi's discrete circuit construction}
\label{sec: Review of Nebabu-Qi's discrete circuit construction}
In this section we briefly review the aspects of the discrete-circuit reconstruction of Nebabu--Qi~\cite{Nebabu:2023iox} needed for our continuum analysis.

We start from a $(0+1)$d boundary theory containing an interacting Majorana GFF $\chi$, with anti-commutator two-point function $A(t,t'):=\lr{\{\chi(t),\chi(t')\}}$. The goal is to map this data to a $(1+1)$d free massive Majorana field $\Psi_{L},\Psi_R$ living on the geometry~\eqref{eq: conformal coordinate}, with the boundary at $z=0$.

\begin{figure}[t]
    \centering
    \includegraphics[width=0.99\linewidth]{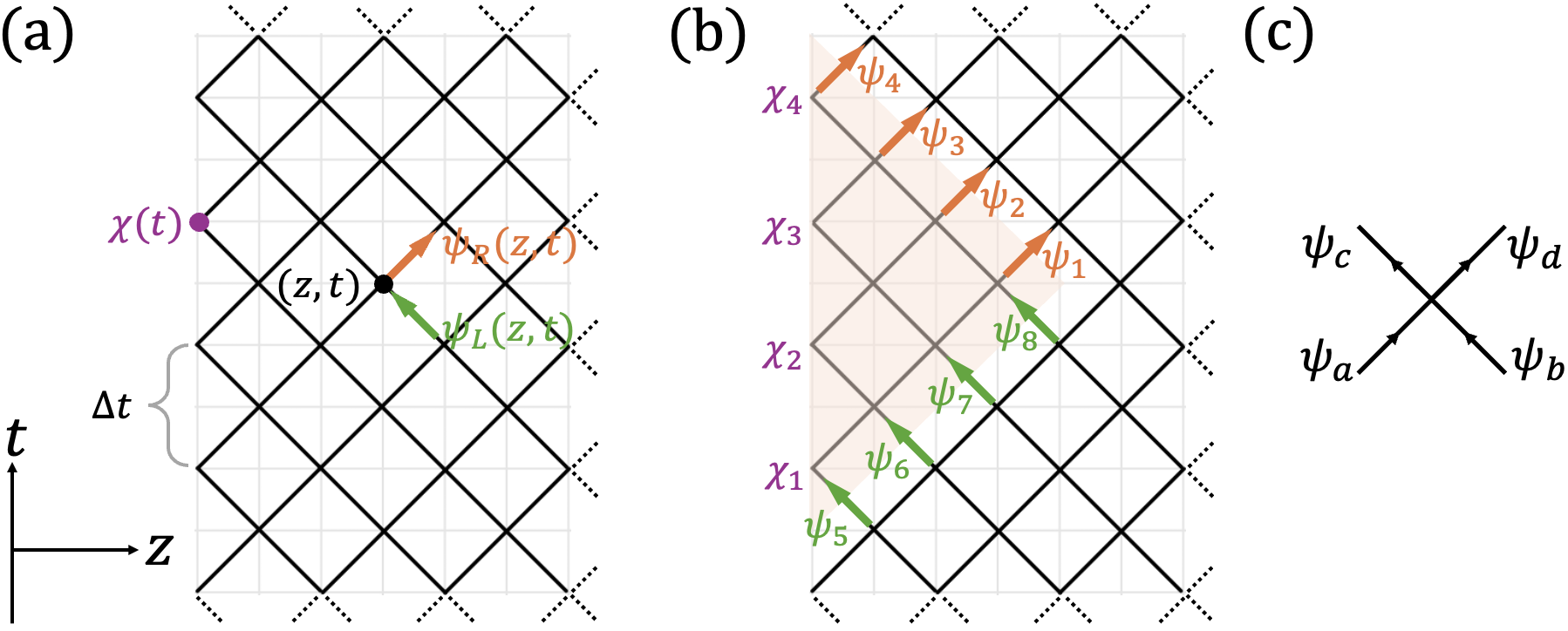}
    \caption{\textbf{(a)} The Nebabu--Qi circuit~\cite{Nebabu:2023iox}. \textbf{(b)} Causal-wedge reconstruction (cf.\ \eqref{eq: K is triangular}). \textbf{(c)} Each ``cross'' is an SO(2) gate, with the convention in \eqref{eq: definition of gate matrix}.}
    \label{fig: illustration of Nebabu-Qi circuit}
\end{figure}

Ref.~\cite{Nebabu:2023iox} focused on the two-point function at discretized time points $t:=\texttt t\Delta t$ with $\texttt{t}\in\mathbb Z$. The corresponding bulk coordinate is given by
\begin{align}
    u&=\texttt u\Delta t, v=\texttt v\Delta t\nonumber\\
    z&=\frac{v-u}2,~t=\frac{v+u}2
\end{align}
with $\texttt u,\texttt v\in\mathbb{Z}$, ${\texttt v}-{\texttt u}\geq 0$. Each value of $({\texttt u},{\texttt v})$ labels a gate in a semi-infinite ``brickwall" circuit, as is illustrated in \cref{fig: illustration of Nebabu-Qi circuit}(a). 
The circuit fermions satisfy the canonical anti-commutation relations
\begin{equation}
\{\psi_{a}(\z,\t),\psi_{b}(\z',\t)\}=\delta_{ab}\delta_{\z\z'},\ a,b\in\{L,R\}
\label{eq: canonical commutation relation of bulk fermion}
\end{equation}
where we distinguish $\psi_a(\z,\t)$ (circuit fermions) from $\Psi_a(z,t)$ (continuum QFT fields).

Since the bulk evolution is unitary, each gate in~\cref{fig: illustration of Nebabu-Qi circuit}(a) is an SO(2) rotation $U$ parametrized by an angle $\theta(\z,\t)$. For example, the gate in~\cref{fig: illustration of Nebabu-Qi circuit}(c) acts as
\begin{equation}
\begin{bmatrix}
\psi_c\\\psi_d
\end{bmatrix}=U\begin{bmatrix}
\psi_a\\\psi_b
\end{bmatrix}=
\begin{bmatrix}
\cos\theta &-\sin\theta\\
\sin\theta & \cos\theta
\end{bmatrix}
\begin{bmatrix}
\psi_a\\\psi_b
\end{bmatrix}
\label{eq: definition of gate matrix}
\end{equation}
More generally, there could be multiple flavors at each location. When there are $n$ fermions at each time labeled by $\chi_a(t),a=1,2,..,n$, there is the same number of fermions at each link of the bulk circuit, and the gate is an $SO(2n)$ matrix. 

The bulk fermions are linear combinations of boundary fermions via future/past kernels,
\begin{equation}
\psi_L(\z,\t)=\sum_{\texttt t'}K^p(\z,\t|\t')\chi(\t'),\ \psi_R(\z,\t)=\sum_{\texttt \t'}K^f(\z,\t|\t')\chi(\t')
\label{eq: kernel, circuit fermion}
\end{equation}
Following the HKLL construction, $K^{p,f}$: $K^{p,f}(\z,\t|\t')$ are required to be nonzero only when the bulk point $(\z,\t)$ is spacelike separated from the boundary point $(0,\t')$, as is shown in Fig.~\ref{fig: illustration of Nebabu-Qi circuit} (b). 

As shown in~\cite{Nebabu:2023iox}, the kernels can be constructed wedge-by-wedge. For the illustrative wedge in~\cref{fig: illustration of Nebabu-Qi circuit}(b), one reconstructs $\psi_{1\sim4},\psi_{5\sim8}$ from $\chi_{1\sim4}$, where $\psi_{1\sim4}$ and $\psi_{5\sim8}$ form two mutually orthonormal sets. The causal wedge condition enforces triangularity of $K^{p,f}$:
\begin{equation}
\begin{bmatrix}
\psi_5\\\psi_6\\\psi_7\\\psi_8
\end{bmatrix}=\begin{bmatrix}
1 & 0 & 0 & 0\\
K^p_{61} & K^p_{62} & 0 & 0\\
K^p_{71} & K^p_{72} & K^p_{73} & 0\\
K^p_{81} & K^p_{82} & K^p_{83} & K^p_{84}\\
\end{bmatrix}
\begin{bmatrix}
\chi_1\\\chi_2\\\chi_3\\\chi_4
\end{bmatrix},\ \ \ \
\begin{bmatrix}
\psi_1\\\psi_2\\\psi_3\\\psi_4
\end{bmatrix}=\begin{bmatrix}
K^f_{11} & K^f_{12} & K^f_{13} & K^f_{14}\\
0 & K^f_{22} & K^f_{23} & K^f_{24}\\
0 &0 & K^f_{33} & K^f_{34}\\
0 & 0& 0 & K^f_{44}\\
\end{bmatrix}
\begin{bmatrix}
\chi_1\\\chi_2\\\chi_3\\\chi_4
\end{bmatrix}
\label{eq: K is triangular}
\end{equation}

\noindent The canonical anti-commutation relations~\eqref{eq: canonical commutation relation of bulk fermion} then require
\begin{equation}
K^fA(K^f)^\intercal= \textbf{1},\ \ K^pA(K^p)^\intercal=\textbf{1}
\label{eq: KAK=1}
\end{equation}
Here $A_{ij}:=\lr{\{\chi_i,\chi_j\}}$ is the $4\times4$ submatrix of the boundary anti-commutator matrix $A(t,t')$.

Equations \eqref{eq: K is triangular} and \eqref{eq: KAK=1} implement a step-by-step orthonormalization procedure. For example, $\psi_4=\chi_4$; $\psi_3$ is the component of $\chi_3$ orthogonal to $\chi_4$; $\psi_2$ is the component of $\chi_2$ orthogonal to both $\chi_3$ and $\chi_4$; and so on. In linear algebra this is related to the Cholesky decomposition\footnote{\label{foot: positivity of A}For any positive semidefinite matrix, a Cholesky decomposition exists. The matrix $A(t,t')$ is positive semidefinite since $\sum_{t,t'}f(t)f(t')^*A(t,t')\geq0$ for any test function $f(t)$, and any submatrix inherits this property.}~\cite{wiki:CholeskyDecomposition}, which determines $K^{p,f}$: writing $A=R^f(R^f)^\intercal=R^p(R^p)^\intercal$ with $R^p$ lower triangular and $R^f$ upper triangular, one has\footnote{The inverse (when it exists) of an upper/lower triangular matrix is upper/lower triangular.} $K^p=(R^p)^{-1}$ and $K^f=(R^f)^{-1}$.

Once the kernels are fixed, the gate angles are fixed as well. For the gate in~\cref{fig: illustration of Nebabu-Qi circuit}(c), from \eqref{eq: definition of gate matrix} the orthogonal matrix $U$ (and hence $\theta$) is determined by
\begin{equation}
\begin{bmatrix}
\ccline \,K_c\,\ccline\\
\ \\
\ccline\,K_d\,\ccline
\end{bmatrix}
\begin{bmatrix}
\vert\\
\chi\\
\vert
\end{bmatrix}
=
U
\begin{bmatrix}
\ccline \,K_a\,\ccline\\
\ \\
\ccline\,K_b\,\ccline
\end{bmatrix}
\begin{bmatrix}
\vert\\
\chi\\
\vert
\end{bmatrix}\longrightarrow
U=\begin{bmatrix}
K_c^{\intercal} AK_a & K_c^{\intercal} AK_b\\
K_d^{\intercal} AK_a & K_d^{\intercal} AK_b
\end{bmatrix}
\label{eq: gate angle from kernel}
\end{equation}

The main goal of the current paper is to obtain more analytic results about the continuum limit of the circuit construction above. We would like to consider the following limit:
\begin{equation}
\Delta t\rightarrow0 \text{ and } \z,\t\rightarrow\infty,\text{ while } (\z\Delta t),(\t\Delta t)\rightarrow\text{const}.
\label{eq: definition of continuous limit}
\end{equation}
Since the bulk quantum circuit is Gaussian, and it has a sharp light cone, in the continuum limit we expect it to describe a free Majorana fermion in $(1+1)$d which is relativistic. To determine the continuum limit theory, we can compare the quantum circuit with a massive $(1+1)$d Majorana QFT minimally coupled to the metric \eqref{eq: conformal coordinate}, which has the action
\begin{equation}
S=\frac{1}{4\pi}\int dtdz\cdot \i\Psi_L\partial_t\Psi_L+\i\Psi_R\partial_t\Psi_R-\i\Psi_L\partial_z\Psi_L+\i\Psi_R\partial_z\Psi_R+2\i m\Omega(z)\Psi_L \Psi_R
\label{eq: action of QFT}
\end{equation}
The circuit and continuum fields are related by
\begin{equation}
\Psi_a(z,t)=\frac{1}{\sqrt{\Delta t}}\psi_a(\z,\t)
\label{eq: relation between circuit fermion and QFT fermion}
\end{equation}
such that $\Psi_a(z,t)$ satisfies the Dirac $\delta$-function canonical anti-commutation relation
\begin{align}
    \left\{\Psi_a(z,t),\Psi_b(z',t)\right\}=\delta_{ab}(2\pi)\delta(z-z')
\end{align}

Matching equations of motion in $\Delta t\rightarrow 0$ limit~\cite{Nebabu:2023iox} relates the geometry to the gate angle:
\begin{equation}
m\Omega(z,t)=\lim_{\Delta t\rightarrow0}\frac{\pi-2\theta(\z,\t)}{\Delta t}
\label{eq: relate gate angle to Omega}
\end{equation}
Physically, $m\Omega(z,t)\Delta t$ determines the small probability that a right-moving fermion becomes a left-moving fermion in $\Delta t$ time, which depends on the scale factor $\Omega(z,t)$ in the geometry. In the circuit language, this probability is given by the deviation of gate angle $\theta$ from $\pi/2$. $\theta=\pi/2$ corresonds to SWAP gate, which means the left-mover and right-mover completely decouple. The reflection probability amplitude is determined by $\cos\theta\simeq \frac{\pi}2-\theta$, and there is a factor of $2$ because in $\Delta t\times \Delta t$ area in the bulk there are two gates. Eq.~(\ref{eq: relate gate angle to Omega}) is the key relation that allows us to determine the curved geometry (up to a length unit $m$) from the bulk fermion dynamics. 

However, we would like to note that $\pi-2\theta(\texttt z,\texttt t)$ is not always guaranteed to be order  $O(\Delta t)$ in continuous limit. We are going to discuss different examples later in the draft. For the examples in \cref{sec: Examples: finite QNMs} and \cref{sec: Examples: large-q SYK}, this scaling holds strictly, ensuring $m\Omega(z,t)$ is well-defined. In \cref{app: Several examples of two-point function that doesn't have a continuous limit} we give explicit correlators for which $\theta$ does \emph{not} scale as $\pi/2-O(\Delta t)$ and hence admit no such continuum QFT limit.


While the reconstruction algorithm does not require time translation invariance, the remainder of this work focuses on the time translation invariant case, when $\theta(\z,\t)\equiv\theta(\z), \Omega(z,t)\equiv\Omega(z)$ only depends on $z$ or $\z$.

\subsection{Method I: via unitary matrix integral}
\label{sec: Method I: via unitary matrix integral}
In this section we derive \eqref{eq: main result1} and \eqref{eq: main result2} by taking the continuum limit of the discrete circuit. In \cref{sec: A determinant formula for gate angle}, we first obtain a determinant formula expressing the gate angle $\theta$ directly in terms of determinants of submatrices of the boundary correlator $A(\t,\t')$ (viewing $\t,\t'$ as matrix indices), bypassing the intermediate construction of $K^{p,f}$ in \eqref{eq: gate angle from kernel}. For a time-translation-invariant boundary theory, $A(\t,\t')$ and its  submatrices are Toeplitz\footnote{A Toeplitz matrix $M_{ij}$ depends only on $i-j$.}. In \cref{sec: Evaluate Toeplitz matrix determinant}, we evaluate these determinants using tools from unitary matrix integral~\cite{Aharony:2003sx,Murthy:2022ien,Eniceicu:2023uvd}, in particular the Borodin--Okounkov--Geronimo--Case (BOGC) formula~\cite{Borodin2000,Geronimo:1979iy}.
\begin{figure}[t]
    \centering
    \includegraphics[width=0.8\linewidth]{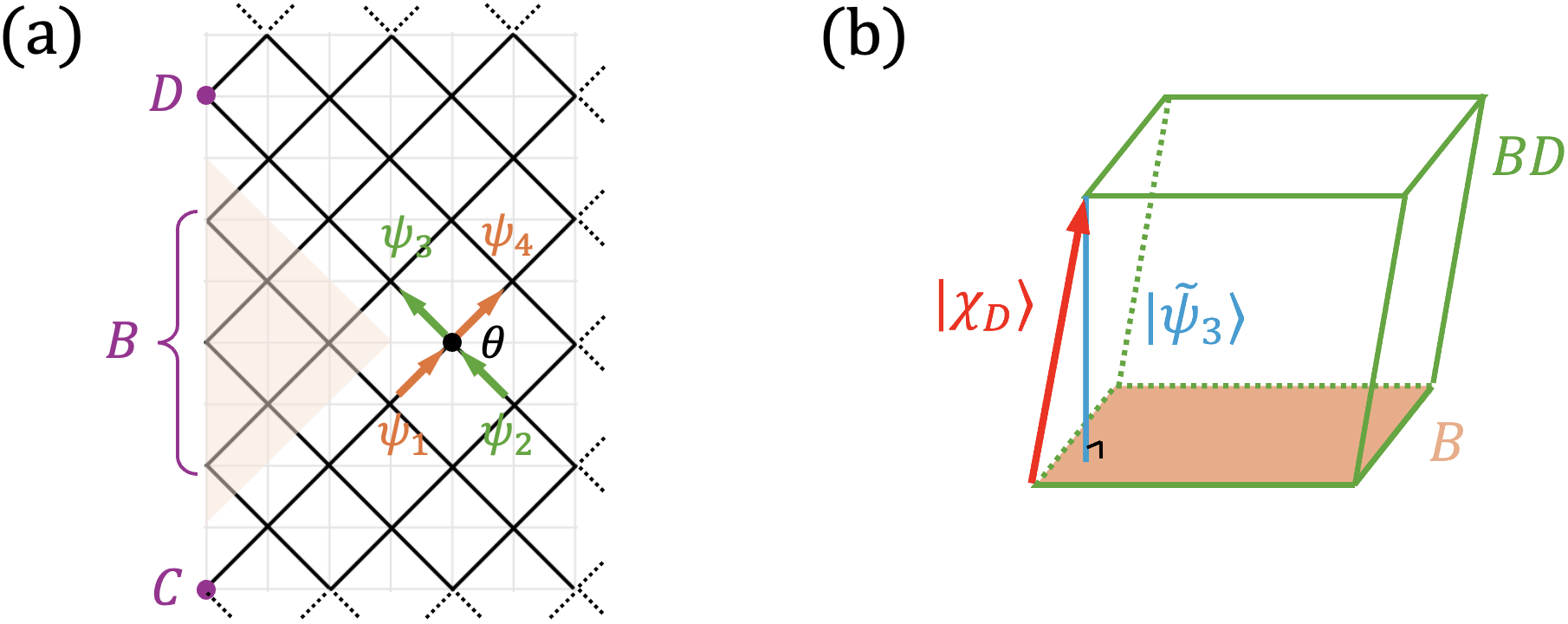}
    
    \caption{\textbf{(a)} An illustration of the determinant formula \eqref{eq: general determinant formula from region}. \textbf{(b)} The parallelepiped used in its derivation.}
    \label{fig:  determinant formula}
\end{figure}
\subsubsection{A determinant formula for gate angle}
\label{sec: A determinant formula for gate angle}
It is convenient to view the collection of boundary operators at different time points $\{\chi(\t),\t\in\mathbb Z\}$ as spanning an infinite dimensional linear space. We will sometimes use vector notation $|\chi\rangle$. This space carries the natural inner product $\lr{\chi(\t_1)|\chi(\t_2)}:=A(\t_1,\t_2)$ (see \cref{foot: positivity of A}). The bulk circuit fermions $\psi$ provide another basis, chosen so that spacelike separated $\psi$'s are orthonormal.

For the gate angle $\theta$ in \cref{fig: determinant formula}(a), causal wedge reconstruction implies that it depends only on data in the region $CBD$: $D$ and $C$ each contain a single time point, while $B$ contains the intervening time points. As in \cref{fig:  determinant formula}(a), we denote the input fermions of the gate as $\psi_{1,2}$ and the output fermions as $\psi_{3,4}$. By definition of gate angle \eqref{eq: definition of gate matrix},
\begin{equation}
\sin^2\theta=\lr{\psi_3|\psi_2}^2.
\end{equation}
As explained in \cref{sec: Review of Nebabu-Qi's discrete circuit construction}, the circuit implements step-by-step orthonormalization. Let $\tilde\psi_{2,3}$ denote the unnormalized vectors
\begin{equation}
|\tilde\psi_2\rangle=(1-P_{CB})|\chi_D\rangle,\qquad |\tilde \psi_3\rangle=(1-P_B)|\chi_D\rangle,
\end{equation}
where $P_{CB}$ projects $|\chi_D\rangle$ onto the span of $|\chi\rangle$ in region $CB$, and similarly for $P_B$. Using $(1-P_{CB})(1-P_B)=(1-P_{CB})^2$, one finds $\langle\tilde\psi_3|\tilde\psi_2\rangle=\langle\tilde\psi_2|\tilde\psi_2\rangle$, and hence
\begin{equation}
\sin^2\theta=\frac{\langle\tilde\psi_3|\tilde\psi_2\rangle^2}{\langle\tilde\psi_3|\tilde\psi_3\rangle\langle\tilde\psi_2|\tilde\psi_2\rangle}=\frac{\langle\tilde\psi_2|\tilde\psi_2\rangle}{\langle\tilde\psi_3|\tilde\psi_3\rangle}.
\label{eq: 2.15}
\end{equation}

Now consider the submatrix $A_{BD}$ of $A(\t,\t')$ with indices restricted to region $BD$. Then $\det(A_{BD})$ equals the squared volume of the parallelepiped spanned by the vectors $\{|\chi\rangle\}$ in $BD$\footnote{Recall the volume element in general relativity $dV=\sqrt{\det g}\,d^n x$, where $g_{\mu\nu}$ is the inner-product matrix of line element $dx^\mu$.}. In \cref{fig:  determinant formula}(b), we schematically draw this parallelepiped in green. The orange shaded parallelepiped is spanned by $|\chi\rangle$ in region $B$ and has squared volume $\det(A_B)$. Since $\langle\tilde\psi_3|\tilde\psi_3\rangle$ is the squared height of the $BD$-parallelepiped with respect to $B$-parallelepiped, it is given by the ratio of squared volumes:
\begin{equation}\langle\tilde\psi_3|\tilde\psi_3\rangle=\det(A_{BD})/\det(A_B).\end{equation} 
The same reasoning applies to $\langle\tilde\psi_2|\tilde\psi_2\rangle$:
\begin{equation}
\langle\tilde\psi_2|\tilde\psi_2\rangle=\det(A_{CBD})/\det(A_{CB}).
\end{equation}
Together with \eqref{eq: 2.15}, this yields
\begin{equation}
\sin^2\theta=\frac{\det(A_{CBD})\det(A_{B})}{\det(A_{BD})\det(A_{CB})}.
\label{eq: general determinant formula from region}
\end{equation}

In \cref{app: Determinant formula for multi-flavor fermions}, we provide an alternative way of deriving the above determinant formula without using parallelepiped, instead using elementary properties of block matrix determinant. The benefit of the alternative method is that it directly generalizes to the case where fermions have multi-flavor, see \eqref{eq: mutiflavor det formula}.

Let $A_{[\t_1,\t_2]}$ denote the submatrix of $A(\t,\t')$ with indices $\t,\t'\in[\t_1,\t_2]$, within a time band. Then $A_{[\t,\t+(N-1)\Delta t]}$ is an $N\times N$ Toeplitz matrix independent of $\t$ by time-translation invariance; we denote it by $A_N$ and define
\begin{equation}
\tilde F_N:=\log\det A_N,\qquad N\geq1.
\end{equation}
The determinant formula for the $N^{\text{th}}$-layer gate angle $\theta_N$ becomes
\begin{equation}
\sin^2\theta_N=\exp\big[\tilde F_{N+1}+\tilde F_{N-1}-2\tilde F_N\big],\qquad N\geq2.
\label{eq: determinant formula}
\end{equation}
The first-layer angle $\theta_1$ is special and is not captured by \eqref{eq: determinant formula}; it determines the boundary condition for the continuum fields $\Psi_{L,R}(z,t)$:
\begin{equation}
(\partial_t-\mu)\Psi_L=(\partial_t+\mu)\Psi_R,\ \ z=0,\text{ with }\mu:=-\frac{\partial_tA(0,t)}{A(0,0)}\bigg|_{t\rightarrow0^+}\geq0.
\label{eq: boundary condition in main text}
\end{equation}
For more detailed derivation of this boundary condition, see \cref{app: Derive boundary condition}.

In the next subsection we take the continuum limit of \eqref{eq: determinant formula} with
\begin{equation}
N\rightarrow\infty,\Delta t\rightarrow0,\qquad (N\Delta t)\rightarrow T=2z=\text{const}.
\label{eq: limit def of toeplitz}
\end{equation}
Here $T$ is the width of the time band and $z$ is the depth of the causal wedge. In this limit, the second difference in \eqref{eq: determinant formula} becomes the second derivative appearing in \eqref{eq: main result2}.

\begin{figure}[t]
    \centering
    \includegraphics[width=0.8\linewidth]{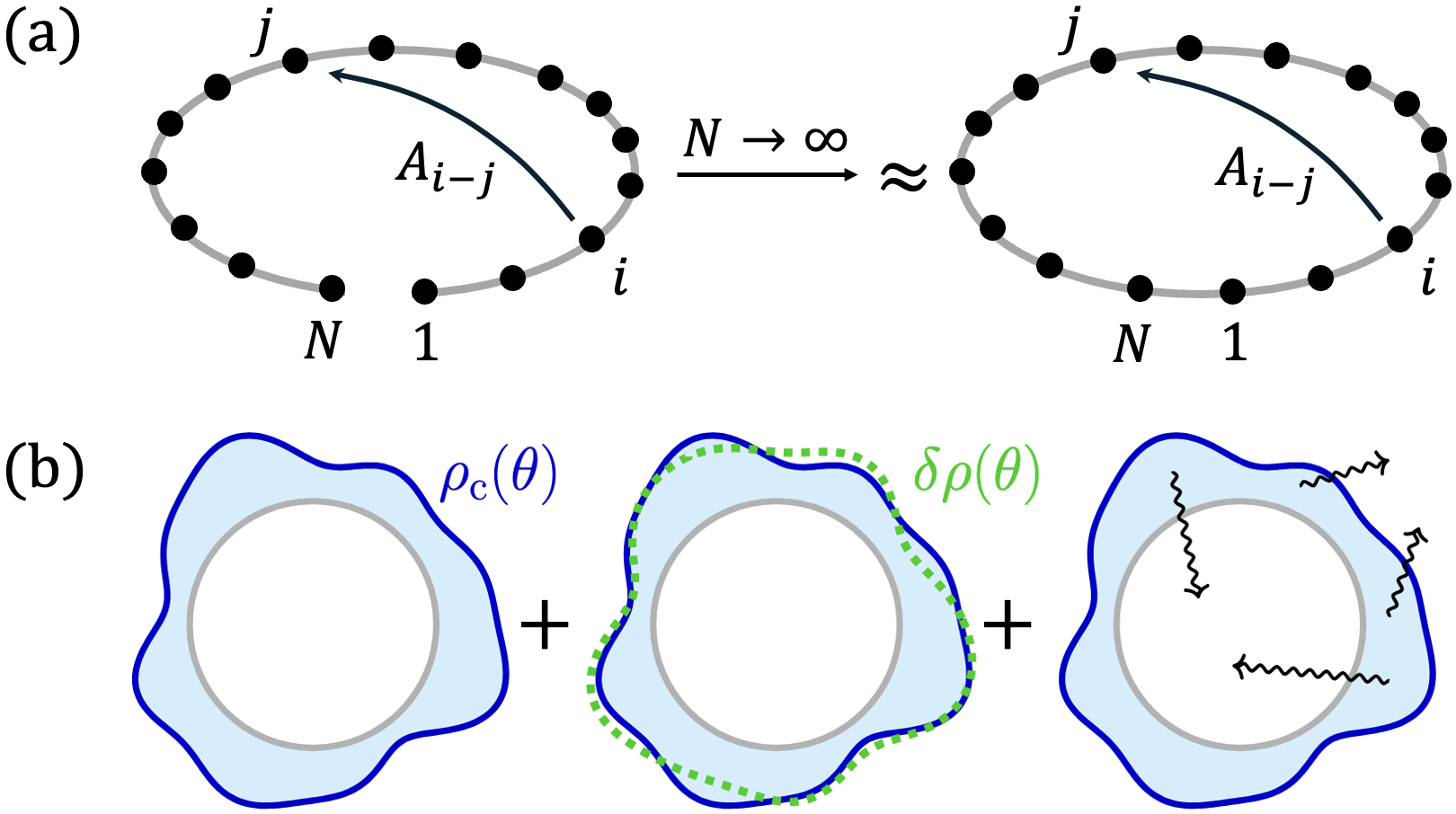}
    \caption{\textbf{(a)} Illustration of Szeg\H o's limit theorem: a Toeplitz matrix may be viewed as a single particle hopping on a one-dimensional lattice with OBC. In the large-$N$ limit, the leading spectrum is well approximated by the corresponding PBC lattice. \textbf{(b)} Illustration of the physical meaning of the three terms in \eqref{eq: free energy expansion} for the unitary matrix integral. The gray line is the unit circle on complex plane. The blue line and the shaded region represent the classical eigenvalue density $\rho_\text{c}(\theta)$. The wavy lines represent eigenvalue tunneling.}
    \label{fig: matrix integral}
\end{figure}

\subsubsection{Evaluate Toeplitz matrix determinant}
\label{sec: Evaluate Toeplitz matrix determinant}
In this section we evaluate $\tilde F_N$ using the systematic large-$N$ expansion of Toeplitz determinants. These techniques are standard in unitary matrix integral, since the partition function of a single unitary matrix can be written as a Toeplitz determinant~\cite{Murthy:2022ien,Eniceicu:2023uvd}.

To take the limit \eqref{eq: limit def of toeplitz}, we first consider $N\rightarrow\infty$ at fixed $\Delta t$. In this standard  limit, the log determinant admits an expansion
\begin{equation}
\tilde F_N=O(N)+O(1)+O(e^{-\# N}).
\label{eq: free energy expansion}
\end{equation}
The $O(N)$ and $O(1)$ terms are governed by Szeg\H o's limit theorem~\cite{wiki:SzegoLimitTheorem}. A useful intuition is to view the Toeplitz matrix as a translation-invariant hopping Hamiltonian on a 1D chain with open boundary conditions (OBC). At large $N$, most of the spectrum is insensitive to boundary effects, so one may replace open by periodic boundary conditions (PBC) and diagonalize in Fourier space, with order one eigenvalue $\eta^{\text{pbc}}(k)$, where lattice momentum $k\in[-\pi,\pi]$ are within first the Brillouin zone. This gives rise to the  $O(N)$ term in \eqref{eq: free energy expansion}:  $O(N)=N\int_{-\pi}^{\pi}dk\log\eta^{\text{pbc}}(k)$. See \cref{fig: matrix integral}(a) for illustration.

In our application of evaluating the gate angle, the $O(N)$ and $O(1)$ pieces don't contribute because \eqref{eq: determinant formula} involves a second difference. We therefore focus on the exponentially small corrections $O(e^{-\# N})$, which encode the geometric data in the continuum limit.

A systematic description of these exponentially small subleading terms is available from the theory of unitary matrix integral:
\begin{equation}
\mathcal Z_N=\int_{U(N)}dU\exp[\sum_{k=1}^{+\infty}\frac{t_k^+}{k}\Tr (U^k)+\frac{t_k^-}{k}\Tr (U^{-k})]
\label{eq: matrix model action}
\end{equation}
where $\{t_k^\pm\}$ are parameters. Physically, such an integral arises, for example, from the thermal partition function of $\mathcal N=4$ super Yang--Mills theory on $S^1\times S^{D-1}$~\cite{Aharony:2003sx}: integrating out  Kaluza--Klein modes of matter fields and gauge fields leaves an effective partition function\footnote{Following such procedure~\cite{Aharony:2003sx}, one would actually end up with a matrix integral with a double trace potential $\tilde \mZ_N=\int_{U(N)} dU\exp[\sum_{k=1}^{+\infty}\frac{g_k}{k}\Tr( U^k)\Tr (U^{-k})]$. Then, one perform a Hubbard-Stratonovich transformation~\cite{Murthy:2022ien} going from $\tilde {\mathcal Z}_N$ to $\mathcal Z_N$ in \eqref{eq: matrix model action}.} for the holonomy matrix $U$ of gauge field  around the thermal circle $S^1$.

The partition function $\mathcal Z_N$ can be written as a Toeplitz determinant (see appendix A of~\cite{Eniceicu:2024sgb} for a review). Define
$\varphi(\theta):=\exp[V(\theta)]$ with external potential $V(\theta):=\sum_{k=1}^{\infty}(\frac{t_k^+}{k}e^{\i k\theta}+\frac{t_k^-}{k}e^{-\i k\theta})$, and Fourier coefficients $\varphi(\theta)=\sum_{n\in\mathbb Z}\varphi_ne^{\i n\theta}$. The associated $N\times N$ Toeplitz matrix is $(A_N)_{ij}:=\varphi_{i-j}$, and one has: \begin{equation}\mathcal Z_N=\det(A_N)\end{equation}

In this language, the free energy $\tilde F_N=\log\mathcal Z_N$ in \eqref{eq: free energy expansion} admits a standard physical interpretation term-by-term (see \cref{fig: matrix integral}(b) for illustration):
\begin{itemize}
\item The $O(N)$ term is the action\footnote{In most matrix models the leading free energy scales as $O(N^2)$, which corresponds to inserting an additional factor of $N$ in the exponent of \eqref{eq: matrix model action}. We will not need this distinction.} of the saddle-point eigenvalue density $\rho_\text{c}(\theta)$ on the unit circle, determined by the competition between the external potential $V(\theta)$ and Coulomb repulsion from the Vandermonde determinant.
\item The $O(1)$ term captures one-loop fluctuations $\delta\rho(\theta)$ around $\rho_\text{c}(\theta)$.
\item The exponentially small corrections $O(e^{-\# N})$ (with associated asymptotic series in $N^{-1}$) describe eigenvalue instantons~\cite{Eniceicu:2023cxn,Chen:2024cvf}, in which finitely many eigenvalues tunnel off the unit circle to locations on the complex planes, inside or outside the unit circle. These contributions are equivalent to the giant-graviton expansion~\cite{Chen:2024cvf}.
\end{itemize}

We now recall a theorem, applied to unitary matrix integral first by Murthy~\cite{Murthy:2022ien}, that resums the full set of exponentially small corrections. This is the Borodin--Okounkov--Geronimo--Case (BOGC) formula, which we will use for bulk reconstruction later:
\paragraph{Theorem. }(Borodin--Okounkov--Geromino--Case)~\cite{Borodin2000,Geronimo:1979iy}. \textit{Given a $N\times N$ real symmetric Toeplitz matrix $A_N$, with matrix element $(A_N)_{ij}:=f_{|i-j|}, i,j=1,...,N$. View $\{f_n\}$ as Fourier coefficient of $f(\theta):=\sum_{n\in\mathbb Z}f_ne^{\i n\theta}$. Assume\footnote{The assumption that $f(\theta)>0, \forall\theta\in(0,2\pi]$ is equivalent to condition that $A_N$ is positive definite: $x^\intercal A_Nx=\sum_{i,j=1}^{N}\int\frac{d\theta}{2\pi}e^{-\i(i-j)\theta}f(\theta)x_ix_j=\int\frac{d\theta}{2\pi}f(\theta)|x(\theta)|^2>0$. For our purpose of bulk reconstruction, $A_N$ is always positive definite, as discussed in \cref{foot: positivity of A}.} $g(\theta):=\log f(\theta)$ exist everywhere on $\theta\in(0,2\pi]$,  with $\{g_n\}$ its Fourier coefficient given by $g(\theta)=\sum_{n\in\mathbb Z}g_ne^{\i n\theta}$. Further define $g_+(\theta):=\sum_{n=1}^{+\infty}g_n e^{\i n\theta},\ g_-(\theta):=\sum_{n=-\infty}^{-1}g_n e^{\i n\theta}$, and $b^-(\theta):=e^{g_-(\theta)-g_+(\theta)}$, with Fourier coefficient $\{b^-_n\}$ given by $b^-(\theta)=\sum_{n\in\mathbb Z}b^-_ne^{\i n\theta}$. Now, define a Hankel matrix $B_N$ on half infinite lattice $L^2(\mathbb Z_{\geq0})$ via $(B_N)_{ij}:=b^-_{i+j+N+1}$, $i,j\geq0$. We then have:  }
\begin{equation}
\log\det A_N=Ng_0+(\sum_{k=1}^{\infty}k g_kg_{-k})+\log\det[\textbf{1}-(B_N)^2]
\label{eq: original BOGC formula}
\end{equation}
In \cref{app: Elementary evaluation of determinant}, we benchmark the BOGC formula in simple cases where $A(t,t')$ admits a finite ($N_\qnm$) quasi-normal-mode expansion, see \eqref{eq: finite QNM A(t,t')}. For $N_\qnm=1,2,3$, we verify that the determinant computed from the BOGC formula agrees with an elementary evaluation (via Gauss elimination), which is technically feasible in these three cases.

We now apply the BOGC formula and show that, under the scaling limit \eqref{eq: limit def of toeplitz}, all exponentially small terms contribute at the same order:
\begin{equation}
\lim_{\eqref{eq: limit def of toeplitz}}\log\det[\textbf{1}-(B_N)^2]=\log\det[\textbf{1}-B(z)^2],
\label{eq: F(z)}
\end{equation}
with $B(z)$ as in \eqref{eq: main result1},\eqref{eq: main result2}, which we repeat it here for completeness:
\begin{equation}
B(z)_{ij}:=\frac{\beta_je^{-\lambda_j 2z}}{\lambda_i+\lambda_j},\ \text{with: }\beta_i:=2\lambda_i\times\prod_{k\neq i}\frac{\lambda_k+\lambda_i}{\lambda_k-\lambda_i}\times \prod_{n}\frac{J_n-\lambda_i}{J_n+\lambda_i}
\end{equation}
where $\{\i J_n\}$ and  $\{\i\lambda_i\}$ are simple zeros and poles of $\mA(\omega)$ in the upper half-plane.

We relegate the uninteresting technical derivation to \cref{app: technical detail of using BOGC formula}. To extract the conformal factor $\Omega(z)$, note that the continuum limit of \eqref{eq: determinant formula} is
\begin{equation}
\lim_{\eqref{eq: limit def of toeplitz}}\frac{1-\sin^2\theta_N}{(\Delta t)^2}=-\frac14\partial_z^2\log\det[\textbf{1}-B(z)^2].
\end{equation}
Combining this with \eqref{eq: relate gate angle to Omega} yields \eqref{eq: main result2}:
\begin{equation}
m^2\Omega(z)^2=-\partial_z^2\log\det[\textbf{1}-B(z)^2]
\end{equation}

\subsection{Method II: via inverse scattering}
\label{sec: Method II: via inverse scattering}
In this section we provide an alternative derivation of the key result \eqref{eq: main result1} and \eqref{eq: main result2}. Instead of starting from the discrete circuit, we directly work in the continuum and  derive \eqref{eq: main result1} and \eqref{eq: main result2} using inverse scattering. In  \cref{sec: Causal kernel decomposition and horizon modes}, starting from the boundary spectral function $\mA(\omega)$, we first extract the S-matrix $\mS(\omega)$ via causal kernel decomposition, which captures the scattering phase between modes leaving the static patch through the future horizon and modes entering the static patch  from the past horizon. Given $\mS(\omega)$, reconstructing  $m\Omega(z)$ becomes a classic inverse scattering problem~\cite{Newton1982Scattering}: determine the scattering potential from its scattering amplitude. In \cref{sec: Inverse scattering}, we generalize the standard scalar field construction in~\cite{Newton1982Scattering} to the Majorana system and re-derive \eqref{eq: main result1},\eqref{eq: main result2}.

\subsubsection{Causal kernel decomposition and horizon modes}
\label{sec: Causal kernel decomposition and horizon modes}
\begin{figure}[t]
    \centering
    \includegraphics[width=1\linewidth]{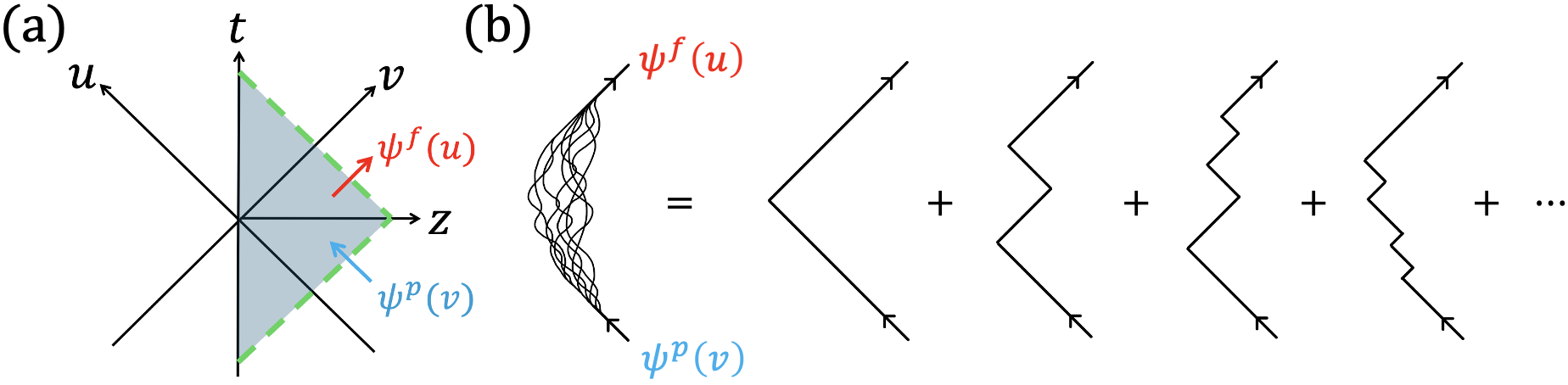}
    \caption{\textbf{(a)} Past/future horizon modes defined in \eqref{eq: define past and future mode}. The horizon is indicated by the green dashed line. \textbf{(b)} The ``forward scattering'' expansion \eqref{eq: forward scattering, zig-zag}.}
    \label{fig: inverse scattering}
\end{figure}

It is convenient to work in lightcone coordinates
\begin{equation}
u:=t-z,\ v=t+z.
\end{equation}
As in the discrete setting \eqref{eq: kernel, circuit fermion}, in the continuum a bulk fermion can be reconstructed from boundary fermions via causal kernels:
\begin{equation}
\Psi_R(u,v)=\int_{u}^vdt'K^f(z,t')\chi(t'),\ \Psi_L(u,v)=\int_{u}^vdt'K^p(z,t')\chi(t')
\end{equation}
As shown in \cref{fig: inverse scattering}, we can further define horizon `future' and `past' modes, which can be reconstructed directly in terms of boundary operators:
\begin{equation}
\begin{aligned}
&\Psi^f(u):=\lim_{v\rightarrow+\infty}\Psi_R(u,v):=\int_u^{+\infty}dt' K^f(t'-u)\chi(t')\\
&\ \Psi^p(v):=\lim_{u\rightarrow-\infty}\Psi_L(u,v):=\int_{-\infty}^{v}dt'K^p(t'-v)\chi(t')
\end{aligned}
\label{eq: define past and future mode}
\end{equation}
The kernels $K^{p,f}$ are causal:
\begin{equation}
K^f(t)\equiv K^f(t)\Theta(t),\ K^p(t)\equiv K^p(t)\Theta(-t)
\end{equation}
In Fourier space\footnote{In \cref{app: convention for Fourier}, we collect our Fourier transform conventions.}, this implies:
\begin{equation}
\mK^f(\omega)\text{ is analytic in upper half plane. } \ \ \ \  \mK^p(\omega) \text{ is analytic in lower half plane.}
\label{eq: analyticicty of K}
\end{equation}
The horizon modes satisfy the standard anti-commutation relations of chiral fermions:
\begin{equation}
\{\Psi^{f}(u),\Psi^f(u')\}=(2\pi)\delta(u-u'),\ \{\Psi^{p}(v),\Psi^p(v')\}=(2\pi)\delta(v-v')
\end{equation}
In Fourier space, this requires the continuum analogue of \eqref{eq: KAK=1}:
\begin{equation}
\mathcal K^f(\omega)\mA(\omega)\mK^f(-\omega)=1,\ \mathcal K^p(\omega)\mA(\omega)\mK^p(-\omega)=1
\label{eq: KAK=1, continuous}
\end{equation}
Due to reality condition, for real $\omega$ we have $\mA(\omega)\geq0$ and $\mK^{p,f}(-\omega)=\mK^{p,f}(\omega)^*$. Thus \eqref{eq: KAK=1, continuous} fixes only the modulus of $\mK^{p,f}(\omega)$ on the real axis. To determine the phase (and hence the analytic continuation to the complex plane), one imposes \eqref{eq: analyticicty of K} and applies a Kramers--Kronig relation~\cite{wiki:KramersKronig} to $\log\mK^f(\omega)$\footnote{If $g(\omega)$ is analytic in the upper half-plane and $\Re g(\omega)$ is known on the real axis, then $\Im g(\omega)$ is fixed by the Hilbert transform $\Im g(\omega)=-\frac{1}{\pi}\mathcal P\int_{\mathbb R}d\omega'\frac{\Re g(\omega')}{\omega'-\omega}$, and $g(\omega)$ in the upper half-plane follows from Cauchy's theorem. Here one takes $g(\omega)=\log\mK^f(\omega)$.}. Once $\mK^f$ is fixed, defining \begin{equation}\mK^p(\omega):=\mK^f(-\omega)
\end{equation} automatically satisfies the lower-half-plane analyticity of $\mK^p$.

One may also ask about the analyticity of the inverse kernels $\mR^{f,p}(\omega):=[\mK^{f,p}(\omega)]^{-1}$. In the discrete setting, the inverses automatically inherit triangularity from finite-dimensional linear algebra. In the continuum, however, analyticity of $\mR^{f,p}$ does not follow from \eqref{eq: analyticicty of K}; in particular, $\mK^f$ analytic in the upper half-plane does not by itself imply that $\mR^f$ is analytic there. $\mathcal{K}^f$ and $\mathcal{R}^f$ are both analytic only if $\mathcal{K}^f$ has no zero or pole on the upper half-plane. 

Given the causal kernels, one can compute the S-matrix:
\begin{equation}
\{\Psi^f(\omega'),\Psi^p(\omega)\}:=\mS(\omega)\delta(\omega+\omega'),\ \mS(\omega)=\frac{\mK^f(-\omega)}{\mK^f(\omega)}
\label{eq: S in fourier space}
\end{equation}
For real $\omega$, $\mS(\omega)$ is a pure phase.

In \cref{app: Matching with bulk QFT}, we re-derive this causal kernel decomposition directly in the bulk QFT. We show that if
\begin{equation}
\int_0^{+\infty}dz\cdot \Omega(z)<\infty
\end{equation}
then, comparing with \eqref{eq: C22}, the causal kernel $\mathcal K^f(\omega)$ exists and is given by the J\"ost function $\mathcal J(\omega)$~\cite{Festuccia:2005pi,Festuccia:2008zx,Dodelson:2023vrw}. In \cref{app: Analytic property of Jost function}, we further show that zeros of $\mJ(\omega)$ in the upper half plane correspond to bound states. Assuming there are no such bound states for the Majorana equation \eqref{eq: eom of QFT in frequency space}, the inverse kernel $\mR^f(\omega):=\frac{1}{\mJ(\omega)}$ is also analytic in the upper half plane.

\subsubsection{Inverse scattering}
\label{sec: Inverse scattering}
From the Majorana equation \eqref{eq: eom of QFT in frequency space}, the conformal factor $m\Omega(z)$ plays the role of a scattering potential. Given scattering amplitude $\mS(\omega)$, reconstructing $\Omega(z)$ is therefore an inverse scattering problem.

Before addressing the inverse problem, it is useful to recall the corresponding `forward scattering' problem: determine $\mS(\omega)$ from a given $\Omega(z)$. A convenient approach is the interaction picture. From the action \eqref{eq: action of QFT} we read off the Hamiltonian $H$ and decompose it into a free part $H_0$ describing a massless Majorana fermion (with decoupled left/right movers)\footnote{Left/right scattering arises from both the mass term $m\Omega(z)$ and the boundary at $z=0$; we include both effects in $V$. More precisely, $H_0$ describes a massless Majorana fermion on the full line $z\in(-\infty,+\infty)$.} and an interaction $V$ that couples left and right movers. Near the horizon, $\Omega(z)\to0$ exponentially as $z\to+\infty$, so the left/right movers become asymptotically non-interacting; the horizon modes $\Psi^f(u)$ and $\Psi^p(v)$ furnish the out- and in-states of the scattering problem.

We denote by $\Psi^{(0)}(u,v)$ the interaction-picture fields evolved by $H_0$. They satisfy the standard chiral fermion anti-commutation relations:
\begin{equation}
\{\Psi^{(0)}_L(u,v),\Psi^{(0)}_L(u',v')\}=(2\pi)\delta(v-v'),\ \{\Psi^{(0)}_R(u,v),\Psi^{(0)}_R(u',v')\}=(2\pi)\delta(u-u')
\end{equation}
In the interaction picture, the potential is time dependent: $V^{(0)}(t):=e^{\i H_0(t-t_0)}Ve^{-\i H_0(t-t_0)}$, where $t_0$ is a reference time. We can now compute the scattering matrix at finite location:
\begin{equation}
S(u_f,v_f|u_i,v_i):=\{\Psi_R(u_f,v_f),\Psi_L(u_i,v_i)\}
\end{equation}
We choose the interaction-picture reference time $t_0$ to coincide with the initial operator $\Psi_L(u_i,v_i)$, so that $\Psi_L(u_i,v_i)=\Psi_L^{(0)}(u_i,v_i)$. We therefore only need to expand the final operator $\Psi_R(u_f,v_f)$ in a Dyson series:
\begin{equation}
\Psi_R(u_f,v_f)=\Psi^{(0)}_R(u_f,v_f)+\sum_{n=1}^{+\infty}\i^n\int dt_1\cdots dt_n[V^{(0)}(t_1),\cdots[V^{(0)}(t_n),\Psi^{(0)}_R(u_f,v_f)],
\end{equation}
with integration region $\frac{v_i+u_i}{2}<t_1<\cdots<t_n<\frac{v_f+u_f}{2}$. The resulting finite-location S-matrix can be resummed into an integral equation:
\begin{equation}
S(u_f,v_f|u_i,v_i)/(2\pi)=m\Omega(\frac{v_i-u_f}{2})-\int_{u_i}^{u_f}du'\int_{v_i}^{v_f}dv'm\Omega(\frac{v'-u'}{2})m\Omega(\frac{v'-u_f}{2})S(u',v'|u_i,v_i)
\label{eq: finite location S matrix}
\end{equation}
The asymptotic S-matrix
\begin{equation}
S(u,v):=S(u,+\infty|-\infty,v)
\end{equation}
is obtained by taking the horizon limit of the finite-location quantity:
\begin{equation}
S(u,v):=\{\Psi^f(u),\Psi^p(v)\}
\end{equation}
By time-translation invariance, $S(u,v)$ depends only on the radial coordinate $z=\frac{v-u}{2}$. Using \eqref{eq: S in fourier space}, its Fourier representation is:
\begin{equation}
S(u,v)/(2\pi)=S(2z),\ S(z):=\int \frac{d\omega}{2\pi}e^{\i\omega z}\mS(\omega)
\label{eq: S(z)}
\end{equation}
Taking the limit of \eqref{eq: finite location S matrix}, we obtain a formal relation between the S-matrix and the conformal factor\footnote{This formula holds for $z=\frac{v-u}{2}>0$; for $z<0$ the S-matrix includes the boundary scatterer at $z=0$.}:
\begin{equation}
\begin{aligned}
&S(u,v)/(2\pi)=m\Omega(\frac{v-u}{2})\\
&\ \ \ \ +\sum_{n=1}^{+\infty}(-1)^nm^{2n+1}\int dv_1\cdots dv_n\int du_1\cdots du_n\prod_{k=1}^{n}\left[\Omega(\frac{v_k-u_k}{2})\right]\prod_{k=1}^{n+1}\left[\Omega(\frac{v_{k-1}-u_k}{2})\right]
\end{aligned}
\label{eq: forward scattering, zig-zag}
\end{equation}
with $v_0:=v$ and $u_{n+1}:=u$, and integration region $v<v_1<\cdots <v_n<+\infty$ and $-\infty<u_1<\cdots<u_n<u$. Equation \eqref{eq: forward scattering, zig-zag} is a small-mass expansion in which a timelike trajectory is resolved into a sum over zig-zag null trajectories, as in \cref{fig: inverse scattering}(b).

Although \eqref{eq: forward scattering, zig-zag} does not admit a simple recursive form like \eqref{eq: finite location S matrix}, inverse scattering methods still allow one to invert it and  reconstruct $m\Omega(z)$ from $S(z)$. We defer the derivation to \cref{app: Technical detail of deriving inverse scattering formula} and record the result:
\begin{equation}
m^2\Omega(z)^2=-\partial_z^2\log\det[1-(P_zhP_z)^2]
\label{eq: mOmega in terms of P_zhP_z}
\end{equation}
\begin{equation}
\text{with: }h(z_1,z_2):=S(z_1+z_2), \ P_{z}(z_1,z_2):=\delta(z_1-z_2)\Theta(z_1-z)
\end{equation}
Here $h$ and $P_z$ are operators on $L^2(\mathbb R)$, with $P_z\equiv \textbf{1}_{[z,+\infty)}$ the projector onto $L^2(\mathbb R_{\geq z})$.

To connect with the explicit expressions \eqref{eq: main result1}--\eqref{eq: main result2}, we specialize to spectral functions admitting a quasi-normal-mode expansion, assuming $\mA(\omega)$ has no branch cut:
\begin{equation}
A(t,t')=\sum_{i}c_ie^{-J_i|t-t'|},\ \ \mA(\omega)=\sum_{i}\frac{c_iJ_i}{J_i^2+\omega^2}
\end{equation}
where $\{\i J_i\}$ are poles of $\mA(\omega)$ in the upper half plane. If $\mA(\omega)$ has zeros at $\{\i \lambda_k\}$ on the upper half plane, we can combine these into the rational form 
\begin{equation}
\mA(\omega)=C\frac{\prod_k(\lambda_k^2+\omega^2)}{\prod_i(J_i^2+\omega^2)}
\label{eq: A(omega) decomposed into poles and zeros}
\end{equation}
where $C$ is an irrelevant positive constant.

Assuming there are no bound states in the upper half-plane, both $\mK^f(\omega)$ and its inverse $\mR^f(\omega)$ are analytic there. This uniquely fixes the causal factorization:
\begin{equation}
\mK^f(\omega)=\sqrt{C}\frac{\prod_i(\omega+\i J_i)}{\prod_{k}(\omega+\i\lambda_k)}
\end{equation}
Then \eqref{eq: S in fourier space} gives the S-matrix:
\begin{equation}
\mS(\omega)=\prod_{k}\frac{\i\lambda_k+\omega}{\i\lambda_k-\omega}\times\prod_i\frac{\i J_i-\omega}{\i J_i+\omega}
\label{eq: S(omega) decomposed into poles and zeros}
\end{equation}
In the upper half-plane, $\mS(\omega)$ has poles at $\{\i\lambda_k\}$ with residues
\begin{equation}
\operatorname{Res}\mS(\omega\rightarrow \i\lambda_k)=(-\i)\beta_k
\label{eq: beta from residue}
\end{equation}
with $\beta_k$ defined in \eqref{eq: main result1}. In real space, the S-matrix $S(z)$ defined in \eqref{eq: S(z)} can be written as a sum over poles,
\begin{equation}
S(z)=\sum_i\beta_ie^{-\lambda_i z},\ z>0
\label{eq: 2.47}
\end{equation}
We defer the remaining technical steps to \cref{app: Technical detail of deriving inverse scattering formula}, where we show that \eqref{eq: mOmega in terms of P_zhP_z} combined with \eqref{eq: 2.47} reduces to \eqref{eq: main result1}--\eqref{eq: main result2}.

Thus the parameters $\{\i\lambda_k,\beta_k\}$ in \eqref{eq: main result1}--\eqref{eq: main result2} are determined entirely by the pole and residue data of the S-matrix. As shown in \cref{app: An alternative inverse formula via contour integral of S-matrix}, one can also write an equivalent formula for $m\Omega(z)$ directly as a contour integral of $\mS(\omega)$, using techniques familiar from giant graviton expansion in unitary matrix integral~\cite{Eniceicu:2023uvd}, without introducing $\{\i\lambda_k,\beta_k\}$ explicitly:
\begin{equation}
m^2\Omega(z)^2=-\partial_z^2\log\left[1+\sum_{k=1}^{+\infty}G^{(k)}(z)\right]
\label{eq: main result3, contour integral of S}
\end{equation}
\begin{equation}
G^{(k)}(z):=\frac{1}{(k!)^2}\int\prod_{i=1}^{k}\frac{d\omega_i}{2\pi}\frac{d\tilde\omega_i}{2\pi}\times \frac{\prod_{1\leq i<j\leq k}(\omega_i-\omega_j)^2(\tilde\omega_i-\tilde\omega_j)^2}{\prod_{1\leq i,j\leq k}(\omega_i-\tilde\omega_j)^2}\times\prod_{i=1}^{k}e^{\i 2z(\omega_i-\tilde\omega_i)}\times \prod_{i=1}^{k}\frac{\mS( \omega_i)}{\mS(\tilde \omega_i)}
\label{eq: main result4, contour integral of S}
\end{equation}
where the integral contour of $\omega_i$ is along $\int_{\mathbb R+\i0^+}$, while   $\tilde\omega_i$ along $\int_{\mathbb R-\i0^+}$.

Finally, although the BOGC approach in \cref{sec: Method I: via unitary matrix integral} and the inverse scattering approach here look different, they are closely related. In fact, Geronimo and Case~\cite{Geronimo:1979iy} (1979) derived what is now called the BOGC formula using inverse scattering ideas. 
\section{Curvature near horizon and AdS universality}
\label{sec: Curvature near horizon and AdS universality}
In this section we use \eqref{eq: main result1} and \eqref{eq: main result2} to study the near-horizon geometry. From the perspective of entanglement renormalization~\cite{Swingle:2009bg}, the deep bulk region $z\rightarrow+\infty$ probes IR physics of the boundary theory, where one may expect universality to emerge, independent of microscopic details. We focus on the near-horizon curvature $R_\text h$ and its dependence on the boundary correlator $A(t,t')$. 

Let's order $\{\i\lambda_i,i\geq0\}$, the zeros of $A(\omega)$ in the upper half plane, in increasing order of their real part: $0<\Re\lambda_0\leq\Re\lambda_1\leq\Re\lambda_2\leq\cdots$. Then, the formula \eqref{eq: main result2} for conformal factor $\Omega(z)$ admits a near horizon expansion where the $\log\det[\textbf{1}-B(z)^2]$ term is expanded into $-\sum_{k=1}^{+\infty}\frac{1}{k}\Tr[B(z)^{2k}]$, further organized into a sum over the exponential decays in $z$\footnote{We further assume $\lambda_0,\lambda_1,\lambda_2$ are real and nondegenerate, with $0<\lambda_0<\lambda_1<\lambda_2$.}:
\begin{equation}
m^2\Omega(z)^2=4\beta_0^2e^{-4\lambda_0z}+8\beta_0\beta_1e^{-2(\lambda_0+\lambda_1)z}+2\frac{\beta_0^4}{\lambda_0^2}e^{-8\lambda_0z}+\text{(subleading)}
\label{eq: 3.4}
\end{equation}
The first two terms and the third term come from $\Tr[B(z)^2]$ and $\Tr[B(z)^4]$ respectively.

As a starting point, we first determine the temperature $T_\text {bh}$ of this blackhole geometry, which only needs the first exponentially decaying term of $\Omega(z)$, controlled by $\lambda_0$:
\begin{equation}
ds^2=\Omega^2(z)(-dt^2+dz^2),\ m^2\Omega(z)^2=4\beta_0^2e^{-4\lambda_0z}+\text{(subleading)}
\end{equation}
This immediately gives the Hawking temperature:
\begin{equation}
T_\text {bh}=\frac{\lambda_0}{\pi}
\label{eq: temperature of geometry}
\end{equation}
At this order, the geometry is Rindler. To extract the near-horizon curvature, we may either evaluate the exact Ricci scalar of the conformal metric \eqref{eq: conformal coordinate},
\begin{equation}
R(z)=-\Omega(z)^{-2}\partial_z^2[\log\Omega(z)^2]
\end{equation}
and take the limit $z\rightarrow+\infty$, or equivalently, directly read off the horizon curvature after changing to Rindler coordinates,
\begin{equation}
\rho:=e^{-2\lambda_0z}, \ \tilde t=2\lambda_0t
\end{equation}
for which the metric becomes
\begin{equation}
ds^2=\frac{\beta_0^2}{\lambda_0^2m^2}\left[1+\frac{2\beta_1}{\beta_0}\rho^{\frac{\lambda_1}{\lambda_0}-1}+\frac{\beta_0^2}{2\lambda_0^2}\rho^2+(\text{subleading})\right](-\rho^2d\tilde t^2+d\rho^2)
\label{eq: 3.6}
\end{equation}
We compare this with a maximally symmetric $(1+1)$-dimensional spacetime of constant Ricci scalar $R$, written in Rindler coordinates as
\begin{equation}
ds^2_0=\frac{-\rho^2d\tilde t^2+d\rho^2}{(1+\frac{R}{8}\rho^2)^2}=\left[1-\frac{R}{4}\rho^2+(\text{subleading})\right](-\rho^2d\tilde t^2+d\rho^2)
\end{equation}
The horizon curvature is therefore fixed by the coefficient of $\rho^2$. Applying this to \eqref{eq: 3.6}, we must determine whether the fractional-power term $\rho^{\frac{\lambda_1}{\lambda_0}-1}$ is subleading compared to, comparable to, or dominant over the $\rho^2$ term. This leads to three cases:
\begin{itemize}
\item \textbf{Case I (AdS universality):} $\lambda_1>3\lambda_0$. The near-horizon geometry is universally AdS, with dimensionless curvature $R_\text h/m^2=-2$, independent of any microscopic details of boundary model:
\begin{equation}
R(z)/m^2=-2+\text{(subleading)}
\label{eq: curvature near horizon, Case I}
\end{equation}
\item \textbf{Case II (tunable curvature):} $\lambda_1=3\lambda_0$. The near-horizon curvature approaches a tunable constant:
\begin{equation}
R(z)/m^2=-2(1+ \frac{4\lambda_0^2\beta_1}{\beta_0^3})+\text{(subleading)}
\label{eq: near horizon curvature, general formula}
\end{equation}
The sign and magnitude of $R_\text h$ are tunable through the boundary correlator. In \cref{sec: Three QNMs and a concrete de Sitter model} we construct a simple model with $\lambda_1=3\lambda_0$ and tune $R_\text h>0$, yielding a de Sitter near-horizon region. Thus dS behavior is less generic than AdS, occupying a codimension-one locus in parameter space. In \cref{sec: Examples: large-q SYK} we will find that large-$q$ SYK at any finite temperature satisfies $\lambda_1=3\lambda_0$ and exhibits a nontrivial temperature dependence of $R_\text h$ via \eqref{eq: near horizon curvature, general formula}. For the ``purely dissipative Gaussian Lindbladian'' class later discussed in \cref{sec: A Gaussian Lindbladian boundary model}, one can show that all $\beta_i$ have the same sign, so the geometry remains AdS even in Case II.

A final comment is that the condition $\lambda_1=3\lambda_0$ need not represent fine tuning. For a broad class of models whose Wightman function $\mathcal G_\text W(\omega)$ has no zeros in the complex frequency plane, KMS implies $\mA(\omega)=(1+e^{\beta\omega})\mathcal G_\text W(\omega)$ (see \cref{app: convention for Fourier}), and the relation $\lambda_1=3\lambda_0$ follows automatically. More generally, this ``no-zero'' property~\cite{Dodelson:2023vrw} is a consequence of the geometry that is analytic at horizon to all order, for which the zeros of $\mA(\omega)$ sit at $\lambda_n=(2n+1)\lambda_0, n\in\mathbb{Z}_{\geq0}$. See \cref{app: Analytic property of Jost function}.
\item \textbf{Case III (non-smooth horizon):} $\lambda_1<3\lambda_0$. The near-horizon curvature diverges:
\begin{equation}
R(z)/m^2=-2\frac{\beta_1}{\beta_0^3}(\lambda_0-\lambda_1)^2e^{2(3\lambda_0-\lambda_1)z}+\text{(subleading)}
\label{eq: curvature near horizon, Case III}
\end{equation}
From \eqref{eq: 3.6} the fractional-power term $\rho^{\frac{\lambda_1}{\lambda_0}-1}$ dominates over $\rho^2$ term, leading to divergent curvature. This divergence reflects non-analyticity of the metric. From \eqref{eq: main result2}, if the metric is analytic at horizon to all order, the set $\{\lambda_n, n\geq0\}$ should satisfy that $\lambda_0\in\mathbb R_+$ and  $\{\lambda_n, n\geq1\}$ being a subset (finite or infinite) of $(\mathbb Z_{>0}+\frac12)(2\lambda_0)$. For a generic set of $\{\i\lambda_i\}$, the metric is generically non-analytic. The Ricci scalar can only detect possible non-analyticity from $\lambda_1$, while higher-derivative geometric quantities will be sensitive to non-analyticity induced from $\lambda_2,\lambda_3,...$ etc.

As a final comment, whether $R_\text h$ diverges to $+\infty$ or $-\infty$ depends on the relative signs of $\beta_0$ and $\beta_1$. For the ``purely dissipative Gaussian Lindbladian'' class later discussed in \cref{sec: A Gaussian Lindbladian boundary model}, one can show that all $\beta_i$ share the same sign, implying $R_\text h\to-\infty$ in Case III.
\end{itemize}

\section{Examples: finite QNMs}
\label{sec: Examples: finite QNMs}
In this section we study boundary correlators $A(t,t')$ with finitely many ($N_\qnm$) quasi-normal modes (QNMs)~\cite{Horowitz:1999jd,Berti:2009kk}:
\begin{equation}
A(t,t')=\sum_{i=1}^{N_\qnm}c_ie^{-J_i|t-t'|}
\label{eq: finite QNM A(t,t')}
\end{equation}
In frequency space, $\mA(\omega)$ has $N_\qnm$ poles in the upper half-plane, at $\{\i J_i\}_{i=1}^{N_\qnm}$:
\begin{equation}
\mA(\omega)=\sum_{i=1}^{N_\qnm}\frac{c_iJ_i}{J_i^2+\omega^2}
\label{eq: A(omega) finite qnm}
\end{equation}
For generic parameters $c_i,J_i$, $\mA(\omega)$ has $N_\qnm-1$ zeros in the upper half plane, at $\{\i\lambda_k\}_{k=1}^{N_\qnm-1}$.

The simplest case is $N_\qnm=1$, where $A(t,t')$ is a single exponential. In \cref{app: Elementary evaluation of determinant}, we show that the dual bulk geometry has $m\Omega(z)=0$ for all $z$, which we interpret as a massless Majorana fermion on the half-line $z\geq0$. From a standard QFT perspective one might ask how a massless bulk theory can induce exponential decay in the boundary correlator. We address this in \cref{app: One QNMs model from bulk QFT}; the effect arises from the boundary condition \eqref{eq: boundary condition in main text} at $z=0$.

In \cref{sec: Two QNMs}, we study the simplest nontrivial case $N_\qnm=2$ and find that the metric is always constant-curvature AdS$_2$. From the bulk-QFT perspective, these correspond to `discrete series' representations of the AdS$_2$ isometry with integer mass; see \cref{app: Two QNMs model from bulk QFT}.

To obtain geometries beyond AdS, in \cref{sec: Three QNMs and a concrete de Sitter model} we study $N_\qnm=3$ and exhibit a family of boundary models with $R_\text h/m^2>0$, realizing a de Sitter near-horizon region.

Finally, in \cref{sec: A Gaussian Lindbladian boundary model} we show that finite-QNM spectral functions $A(t,t')$ can be realized by Gaussian Lindbladian dynamics, in particular we give an explicit family of boundary models realizing the $N_\qnm=3$ de Sitter example.

\subsection{Two QNMs}
\label{sec: Two QNMs}
For $N_\qnm=2$, the boundary anti-commutator is
\begin{equation}
A(t,t')=c_1e^{-J_1|t-t'|}+c_2e^{-J_2|t-t'|}
\end{equation}
Using \eqref{eq: main result1} and \eqref{eq: main result2}, we obtain a closed-form expression for the conformal factor:
\begin{equation}
m\Omega(z)=\frac{2\lambda_0}{\sinh[\lambda_0(2z+z_0)]},\ \lambda_0:=\sqrt{\frac{J_1J_2(c_2J_1+c_1J_2)}{c_1J_1+c_2J_2}},\ z_0:=-\frac{1}{2\lambda_0}\log\left[\frac{(J_1-\lambda_0)^2(J_2-\lambda_0)^2}{(J_1+\lambda_0)^2(J_2+\lambda_0)^2}\right]
\label{eq: 2 QNM geometry}
\end{equation}
where $\i\lambda_0$ is the unique zero of $\mA(\omega)$ in the upper half plane. This is the AdS$_2$ Rindler metric, with constant negative curvature:
\begin{equation}R(z)/m^2=-2
\label{eq: 2 qnm curvature}
\end{equation}

For real parameters $c_1,c_2,J_1,J_2$, the boundary correlator $A(t,t')$ decays monotonically, and the dual geometry naturally contains a horizon. A notable feature is that \eqref{eq: 2 QNM geometry} also holds for complex parameters. Requiring $A(t,t')$ to remain real forces $c_1,c_2$ and $J_1,J_2$ to come in complex-conjugate pairs; then $A(t,t')$ is an exponentially decaying envelope modulated by oscillations\footnote{Let $c_1=ce^{i\theta},c_2=ce^{-i\theta},J_1=J+i\nu, J_2=J-i\nu$, with $c,\theta,J,\nu\in\mathbb R$. Then $A(t,t')=ce^{-J|t-t'|}\cos(\nu |t-t'|+\theta)$.}. Such modulated decay also yields a horizon.

\begin{figure}[t]
    \centering
    \includegraphics[width=0.9\linewidth]{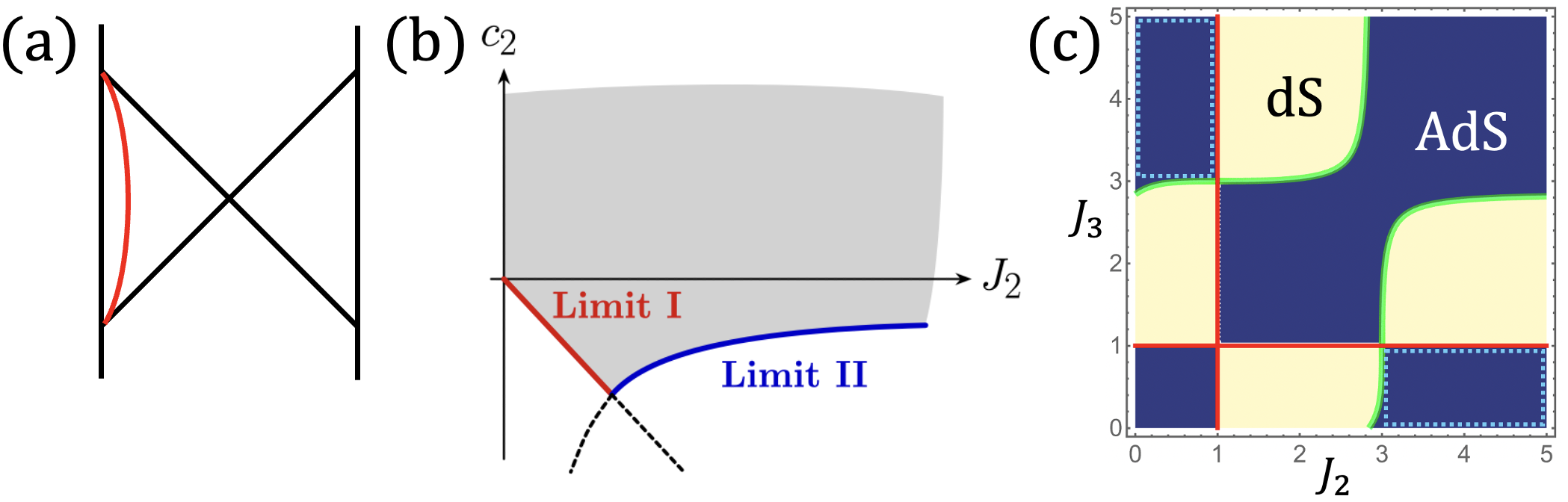}
    \caption{\textbf{(a)} For generic parameters, the boundary sits at finite radial location rather than at the asymptotic AdS$_2$ boundary. \textbf{(b)} Parameter-space ``phase diagram'' of the $N_\qnm=2$ model; the shaded region is physically allowed and its boundaries are discussed in the text. \textbf{(c)} Phase diagram of the $N_\qnm=3$ model of \cref{sec: Three QNMs and a concrete de Sitter model} at $J_1=2$, showing the sign of $R_\text h$ \eqref{eq: three QNM near horizon curvature} (AdS in blue, dS in yellow). The green line marks $R_\text h=0$ and the red line marks $R_\text h=\infty$. The region inside the dotted blue box has $c_i\geq0$ and corresponds to purely dissipative Gaussian Lindbladian models.}
    \label{fig: 2QNM phase diagram}
\end{figure}

Another key feature is that for all physical\footnote{\label{foot: positivity constraint.}Positivity of the Hilbert space requires $A(t,t')$, viewed as a matrix in $(t,t')$, to be positive semidefinite. This constrains its eigenvalues $A(\omega)$ to be nonnegative.} choices of parameters, one finds $z_0\geq0$. Thus the boundary at $z=0$ does not lie at the asymptotic AdS$_2$ boundary (\cref{fig: 2QNM phase diagram}(a)). Intuitively, a generic $A(t,t')$ does not exhibit conformal symmetry; conformal boundary models are special (e.g. SYK at low temperature~\cite{Sachdev:1992fk,Kitaev:2015talks,Maldacena:2016hyu} or conformal quantum mechanics~\cite{deAlfaro:1976vlx,Anninos:2011af}). For generic parameters the construction naturally realizes finite-cutoff holography~\cite{Coleman:2021nor,Batra:2024kjl,Anninos:2024xhc}.

A natural question is when the boundary approaches the asymptotic AdS$_2$ boundary. For simplicity, we address this in the case of real $c_1,c_2,J_1,J_2$.

For real parameters, decay requires $J_1,J_2\geq0$. Without loss of generality we set $J_1=c_1=1$ to fix the time scale and normalize the Majorana operator. The remaining parameters $c_2$ and $J_2$ are constrained by positivity (\cref{foot: positivity constraint.}), which requires $\mA(\omega)\geq0$ for all real $\omega$. This implies $c_2\geq -J_2$ and $c_2\geq -J_2^{-1}$; see \cref{fig: 2QNM phase diagram}(b). We now study the two boundary limits of the allowed region:
\begin{itemize}
\item \textbf{Limit I, zero temperature blackhole. }We set $c_2=-J_2+J_2^{-1}\varepsilon^2$ with $0<J_2<1$. Expand in $0<\varepsilon\ll1$, we find:
\begin{equation}
\lambda_0=\frac{\varepsilon}{\sqrt{1-J_2^2}}+O(\varepsilon^3),\ \ z_0=2(1+J_2^{-1})+O(\varepsilon^2), \ m\Omega(z)=\frac{1}{z+(1+J_2^{-1})}
\end{equation}
Since $\lambda_0\to0$ in this limit, \eqref{eq: temperature of geometry} implies that the black hole temperature approaches zero. The metric approaches the Poincar\'e patch of AdS$_2$, while the boundary ($z=0$) remains at finite radial location.
\item \textbf{Limit II, asymptotic boundary. }We set $c_2=-J_2^{-1}+J_2^{-1}\varepsilon^2$ with $J_2>1$. Expand in $0<\varepsilon\ll1$, we find:
\begin{equation}
\lambda_0=\sqrt{J_2^2-1}\varepsilon^{-1}+O(\varepsilon),\ z_0=\frac{2}{J_2-1}\varepsilon^2+O(\varepsilon^4)
\end{equation}
Since $\lambda_0$ diverges, it is useful to rescale $z:=\varepsilon\tilde z$ and $t:=\varepsilon\tilde t$, keeping $\tilde z,\tilde t=O(1)$ as $\varepsilon\to0^+$. Physically, because $\lambda_0\to\infty$, the temperature measured in the original $t$ coordinate diverges by \eqref{eq: temperature of geometry}; a finite limit requires zooming into time scales of order the inverse temperature. The rescaled coordinates probe the UV of the original variables, and the geometry approaches:
\begin{equation}
ds^2=\tilde\Omega(\tilde z)^2(-d\tilde t^2+d\tilde z^2),\ \ m\tilde\Omega(\tilde z)=\frac{2\sqrt{J_2^2-1}}{\sinh(2\sqrt{J_2^2-1}\tilde z)}
\end{equation}
In these coordinates, the boundary at $\tilde z=0$ lies at the asymptotic AdS$_2$ boundary.
\end{itemize}

Since the geometry is constant-curvature AdS$_2$, one might ask why the bulk QFT does not exhibit the usual infinite tower of QNMs of AdS$_2$ Rindler~\cite{Berti:2009kk}. The reason is that the two-QNM model corresponds to a special mass. Using \eqref{eq: 2 qnm curvature} to set units by the AdS length $\ell_{\text{AdS}}$, one finds $m=1$, i.e. an integer mass. Integer masses furnish discrete series representations of SL(2,$\mathbb R$)\footnote{Strictly speaking, our finite cutoff at $z=0$ breaks SL(2,$\mathbb R$) to time translations, so the single-particle Hilbert space is not organized into full SL(2,$\mathbb R$) irreps. Here ``discrete series'' refers to the corresponding theory on the full AdS$_2$ Rindler background.}. Technically, the mode functions are hypergeometric functions for generic non-integer $m$; at integer $m$ the hypergeometric functions truncate to polynomials, leaving only finitely many poles. Equivalently, as one tunes $m$ to an integer, most poles and zeros of $\mA(\omega)$ collide and cancel, leaving finitely many uncanceled. More detailed discussion will be reserved to \cref{app: Two QNMs model from bulk QFT}.



\subsection{Three QNMs and a concrete de Sitter model}
\label{sec: Three QNMs and a concrete de Sitter model}
The goal of this section is to find an explicit $N_\qnm=3$ model whose near-horizon region is de Sitter, since we have seen that $N_\qnm=1,2$ cannot realize $R_\text h>0$.

The $N_\qnm=3$ ansatz has six boundary parameters $c_{1\sim3},J_{1\sim3}$. In the upper half plane, $\mA(\omega)$ has three poles $\{\i J_i\}_{i=1}^3$ and two zeros $\{\i\lambda_0,\i\lambda_1\}$. As discussed in \cref{sec: Curvature near horizon and AdS universality}, avoiding AdS universality requires tuning $\lambda_1=3\lambda_0$. We therefore begin by extracting $\lambda_0,\lambda_1$ from the zeros of $\mA(\omega)$:
\begin{equation}
\mA(\omega)=\sum_{i=1}^{3}\frac{c_iJ_i}{J_i^2+\omega^2}\equiv\frac{P(\omega)}{\prod_{i=1}^{3}(J_i^2+\omega^2)},\ P(\omega):=a_1\omega^4+a_2\omega^2+a_3
\end{equation}
\begin{equation}
\text{with: }a_1:=(\sum_{i=1}^{3}c_iJ_i),\ a_2:=a_1(\sum_{i=1}^3J_i^2)-(\sum_{i=1}^3c_iJ_i^3),\ a_3:=(\prod_{i=1}^{3}J_i^2)(\sum_{i=1}^3c_iJ_i^{-1})
\label{eq: a1a2a3}
\end{equation}
The zeros of $P(\omega)$ give $\lambda_0^2=\frac{a_2-\sqrt{a_2^2-4a_1a_3}}{2a_1}$ and $\lambda_1^2=\frac{a_2+\sqrt{a_2^2-4a_1a_3}}{2a_1}$. Imposing $\lambda_1=3\lambda_0$ (with $\lambda_0>0$) yields $a_2=10a_1\lambda_0^2$ and $a_3=9a_1\lambda_0^4$. Without loss of generality we set $\lambda_0=1$ (fixing the time unit) and $a_1=1$ (fixing the Majorana operator normalization), so $a_2=10$ and $a_3=9$. We then take $(J_1,J_2,J_3)$ as free dimensionless parameters, with $c_i$ determined by \eqref{eq: a1a2a3}:
\begin{equation}
c_i=\frac{(J_i^2-1)(J_i^2-9)}{J_i}\prod_{j=1,j\neq i}^{3}\frac{1}{J_i^2-J_j^2}
\label{eq: 3QNM c_i}
\end{equation}
From \eqref{eq: near horizon curvature, general formula}, we obtain the curvature near horizon:
\begin{equation}
R_\text h/m^2=-\frac{1}{2}\left[1-\frac{3}{4}p(J_1)p(J_2)p(J_3)\right],\ p(x):=\frac{(x-3)(x+1)^3}{(x+3)(x-1)^3}
\label{eq: three QNM near horizon curvature}
\end{equation}
\begin{figure}[t]
    \centering
    \includegraphics[width=0.99\linewidth]{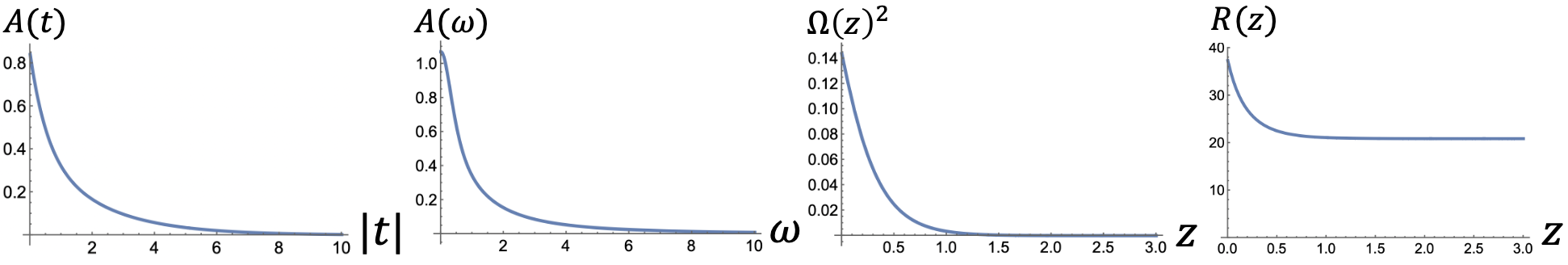}
    \caption{$A(t,t')$, $\mA(\omega)$, $\Omega(z)^2$, and $R(z)$ for a representative $N_\qnm=3$ model with a de Sitter near-horizon region (parameters in \eqref{eq: parameter of a concrete dS model}).}
    \label{fig: 3QNM}
\end{figure}
\noindent We now identify parameters with $R_\text h>0$. For simplicity we take $J_i\in\mathbb R_{>0}$. In \cref{fig: 2QNM phase diagram}(c), we fix $J_1=2$ and plot the sign of $R_\text h$ in the $(J_2,J_3)$ plane: AdS (blue) and dS (yellow). The green curve marks $R_\text h=0$, while the red curve marks a divergence of $R_\text h$. As a check, the region with $c_i\geq0$ (inside the dashed light-blue box) corresponds to purely dissipative Gaussian Lindbladian models and is always AdS, as discussed in \cref{sec: A Gaussian Lindbladian boundary model}.

Away from the horizon the full $z$-dependence is not particularly illuminating, but for completeness we record the explicit expression from \eqref{eq: main result2}:
\begin{equation}
m^2\Omega(z)^2=-\partial_z^2\log\left[1-\frac{\beta_0^2}{4}e^{-4z}-\frac{\beta_0\beta_1}{8}e^{-8z}-\frac{\beta_1^2}{36}e^{-12z}+\frac{\beta_0^2\beta_1^2}{2304}e^{-16z}\right]
\end{equation}
with $\beta_0=4\prod_{i=1}^3\frac{J_i-1}{J_i+1}$ and $ \beta_1=-12\prod_{i=1}^3\frac{J_i-3}{J_i+3}$ from~\eqref{eq: main result1}.

In \cref{fig: 3QNM}, we plot $A(t,t'), \mA(\omega), \Omega(z)^2, R(z)$ for a concrete de Sitter model near horizon, with parameters $
J_1=2,J_2=2.9, J_3=0.5
\label{eq: parameter of a concrete dS model}$. In the next section we construct an explicit boundary model that realizes this boundary two-point function.

\subsection{A Gaussian Lindbladian boundary model}
\label{sec: A Gaussian Lindbladian boundary model}
To provide some more physical motivation for the finite QNM model studied in this section, now we construct a simple Gaussian Lindbladian model that realizes a GFF whose two-point function decomposes into finitely many QNMs, as in \eqref{eq: finite QNM A(t,t')}. 

Let's start with $N_\qnm$ Majorana fermions (with $2^{N_\qnm/2}$ dimensional Hilbert space) with canonical commutation $\{\chi_i,\chi_j\}=2\delta_{ij}$. The operator dynamics is governed by Lindbladian equation\footnote{We are interested in the case where $O$ and $L_a$ are fermionic. In this case, unlike usual Lindbladian equation where at least one of $O,L_a$ is bosonic, the sign in front of $L_a^\dagger OL_a$ is minus, see \cite{Liu:2022god}.}:
\begin{equation}
\frac{dO}{dt}=\i[H,O]+\sum_a(-2L_a^\dagger OL_a-\{L_a^\dagger L_a,O\})
\label{eq: lindbladian eq}
\end{equation}
with:
\begin{equation}
H=\frac{1}{4}\sum_{i,j}\chi_i h_{ij}\chi_j,\ L_a=\frac{1}{2}\sum_iM_{ai}\chi_i
\end{equation}
where $h_{ij}$ is a purely imaginary anti-symmetric matrix, and $M_{ai}$ is unconstrained. Choosing $H$ quadratic in $\chi$ and $L_a$ linear in $\chi$ keeps the dynamics Gaussian, so correlators satisfy Wick's theorem and the boundary fermion furnishes a GFF.

The finite-time evolution implied by \eqref{eq: lindbladian eq} is
\begin{equation}
\chi_{i}(t)=\sum_{j=1}^{N_\qnm}\left[e^{-\mathcal Lt}\right]_{ij}\chi_j(0),\ t>0,\ \ \ \text{with }  \mathcal L:=\i h+\operatorname{Re}M^\dagger M
\end{equation}
Let's use a particular fermion component $\chi_1(t)$ to do bulk reconstruction. Thus the anti-commutator correlator is
\begin{equation}
A(t,t'):=\frac{1}{2}2^{-N/2}\Tr(\{\chi_1(t),\chi_1(t')\})=\left[e^{-\mathcal L|t-t'|}\right]_{11}=\sum_{i=1}^{N_\qnm}V_{1i}(V^{-1})_{i1}e^{-J_i|t-t'|}
\end{equation}
where we assume $\mathcal L$ is diagonalizable and write $\mathcal L:=VDV^{-1}$, with $D=\mathrm{diag}(\{J_i\}_{i=1}^{N_\qnm})$ and $V$ invertible. This matches \eqref{eq: finite QNM A(t,t')}, with $c_i:=V_{1i}(V^{-1})_{i1}$ (no sum).

A special case is $H=0$, i.e. pure dissipation, which we call a ``purely dissipative Gaussian Lindbladian model''. Then $\mathcal L=\Re M^\dagger M$ is real symmetric and positive semidefinite\footnote{Write $M^\dagger M=\Re M^\dagger M+\i\Im M^\dagger M$, with $\Im M^\dagger M$ real antisymmetric. Then for all $x\in\mathbb R^{N}$, $x^\intercal (\Re M^\dagger M)x=x^\intercal (M^\dagger M)x\geq0$.}, so its eigenvalues satisfy $J_i\geq0$ and its eigenvectors form a real orthogonal matrix $V$, implying $c_i=V_{1i}^2\geq0$. We can then ask what $c_i\geq0$ and $J_i\geq0$ imply for the bulk geometry. From \eqref{eq: A(omega) finite qnm}, one can show\footnote{We view $A(\i\lambda)=\sum_{i=1}^{N_\qnm}\frac{c_iJ_i}{J_i^2-\lambda^2}$ as a function of $\lambda^2$, it will have $N_\qnm-1$ zeros in general. Since $A(\omega)$ is a real function on the imaginary axes,  and notice that $A(\i\lambda)\rightarrow-\infty$ as $\lambda^2\rightarrow J_k^2+0^+$ and $A(\i\lambda)\rightarrow+\infty$ as $\lambda^2\rightarrow J_{k+1}^2-0^+$, so there must be at least one zero in the interval between $J_{k}$ and $J_{k+1}$. Since the number of interval already equals the number of zeros, so there could be one and only one zero in each interval.} that the $N_\qnm-1$ zeros $\{\i\lambda_k\}$ interlace with the poles: ordering $J_i$ increasingly, one has $J_k<\lambda_k<J_{k+1}$. This ordering implies (via \eqref{eq: main result1}) that all $\beta_i$ have the same sign. Then the near-horizon curvature is negative in all three cases of \cref{sec: Curvature near horizon and AdS universality}. Interestingly, purely dissipative Gaussian Lindbladians are always AdS near the horizon and cannot realize de Sitter near-horizon curvature.

Next, we proceed to construct an explicit Gaussian Lindbladian model that realizes the $N_\qnm=3$ de Sitter example of \cref{sec: Three QNMs and a concrete de Sitter model} by finding a concrete matrix $\mathcal L$. Our procedure is randomized for two reasons: (i) for fixed $\{c_i,J_i\}$ there are generically many admissible $\mathcal L$, and it suffices to find one example; (ii) the physical constraint $\mathcal L+\mathcal L^\intercal=2\Re M^\dagger M\succeq0$ is nontrivial, so a simple generate-and-check strategy is effective. The algorithm is:
\paragraph{Pseudo-code. }\textit{Given a set of $\{c_i,J_i\}_{i=1}^{N_\qnm}$ (for simplicity we focus on real $c_i$ and positive $J_i$), the code back-engineers a physical $N_\qnm$ dimensional matrix  $\mathcal L$ satisfying: (1), the spectrum of $\mathcal L$ is $\{J_i\}$; (2) its decomposition $\mathcal L=VDV^{-1}$ satisfy $V_{1i}(V^{-1})_{i1}=c_i$; (3) its symmetric part $\mathcal L+\mathcal L^{\intercal}$ is positive semi-definite. It contains the following five steps.}

\noindent $\bullet$ \textit{Input list of $\{J_i\}$ and $\{c_i\}$. Normalize $\{c_i\}$ such that $\sum_i c_i=1$. }

\noindent $\bullet$ \textit{Generate $N_\qnm$ non-zero random real numbers $x_i$ and assemble into a column vector $x$, used as $V_{1i}:=x_i$.} 

\noindent $\bullet$ \textit{Define a $N_\qnm\times1$ column vector $y_i=\frac{c_i}{V_{1i}}$. Define $N_\qnm\times N_\qnm$ projector $P=I-\frac{|y\rangle\langle y|}{\langle y|y\rangle}$.}

\noindent $\bullet$ \textit{Generate an $(N_\qnm-1)\times N_\qnm$ random real rectangular matrix $Q$. Then let $V:=\begin{bmatrix}x^\intercal\\
QP
\end{bmatrix}$, and  let $\mathcal L:=V\operatorname{diag}(\{J_i\})V^{-1}$.}

\noindent $\bullet$ \textit{Check if $\mathcal L+\mathcal L^\intercal$ is positive semi-definite. If yes, code ends and output $\mathcal L$; if not, go back to the second step.}

\hfill

\noindent Running this algorithm on the concrete $N_\qnm=3$ de Sitter example of \cref{sec: Three QNMs and a concrete de Sitter model}, with parameters \eqref{eq: parameter of a concrete dS model} and coefficients $c_1=0.453515$, $c_2=-0.0418932$, $c_3=0.428922$ (from \eqref{eq: 3QNM c_i}), we find one realization of $\mathcal L$:
\begin{equation}
\mathcal L=\begin{bmatrix}
1.18971&
-0.530428&
-0.984871\\
 -1.11738& 1.31535& -1.55849\\
 0.143654& -0.0742184& 2.89494
\end{bmatrix}
\end{equation}
Repeating the run yields many realizations.

\section{Examples: large-$q$ SYK}
\label{sec: Examples: large-q SYK}
In this section we apply the above formalism to the bulk geometry of the large-$q$ Sachdev--Ye--Kitaev (SYK) model~\cite{Sachdev:1992fk,Kitaev:2015talks,Maldacena:2016hyu,Polchinski:2016xgd}. At zero temperature, SYK is dual to JT gravity on AdS$_2$ Rindler wedge~\cite{Maldacena:2016upp,Almheiri:2014cka,Jensen:2016pah,Mertens:2022irh}. The dual geometry of the finite temperature SYK model has been discussed\footnote{In the double scaled limit of SYK~\cite{Berkooz:2018jqr,Berkooz:2018qkz}, the model is conjectured~\cite{Susskind:2021esx,Susskind:2022bia,Susskind:2022dfz,Rahman:2024vyg,Narovlansky:2023lfz,Verlinde:2024znh} to dual to de Sitter spacetime at finite or infinite temperature.} but remains an open question.  Here we use the reconstruction developed above to study the bulk geometry. Interestingly, we found that the near-horizon curvature has an unexpected temperature dependence.

\subsection{General setup}\label{sec:syk_setup}

We begin by recalling the basic aspects of SYK relevant for our purposes, following \cite{Maldacena:2016hyu}. The model is $(0+1)$d quantum mechanics of $N$ Majorana fermions with $\{\chi_i,\chi_j\}=\delta_{ij}$, interacting via a $q$-local random Hamiltonian:
\begin{equation}
H_\syk=\i^{q/2}\sum_{i_1<\cdot<i_q}J_{i_1...i_q}\chi_{i_1}\cdots\chi_{i_q},\ \ \ \ \overline{J_{i_1...i_q}^2}=\frac{2^{q-1}}{q}\frac{\mJ^2(q-1)!}{N^{q-1}}
\end{equation}
We set $\mJ\equiv1$ to fix the energy scale.

In the large-$N$ limit, connected correlators vanish and the fermions obey Wick's theorem, so $\chi_i$ furnishes a GFF. In addition, if we take the large-$q$ limit with $q\rightarrow \infty, q^2/N\rightarrow 0$, the anti-commutator two-point function takes an analytic form~\cite{Maldacena:2016hyu}:
\begin{equation}
A(t,t')\equiv\frac{1}{N}\sum_{i=1}^{N}\lr{\{\chi_i(t),\chi_i(t')\}}=\Re\left[\left(\frac{\cos\theta}{\cosh(\i\theta+(t-t')\cos\theta)}\right)^{2\Delta}\right],\ \ \ \beta_\syk=\frac{2\theta}{\cos\theta},\ \Delta=\frac1q
\label{eq: syk 2pt}
\end{equation}
Here $\theta\in(0,\pi/2)$ parametrizes the inverse temperature $\beta_\syk$: $\theta=0$ corresponds to infinite temperature and $\theta=\pi/2$ to zero temperature.

Unlike the finite-QNM examples of \cref{sec: Examples: finite QNMs}, the SYK two-point function has infinitely many QNMs:
\begin{equation}
A(t,t')=\sum_{n=0}^{+\infty}c_ne^{-J_n|t-t'|}
\end{equation}
\begin{equation}
\text{with: }c_n=(\cos\theta)^{-2\Delta}(-1)^n\frac{2^{2\Delta}\Gamma(2\Delta+n)}{n!\Gamma(2\Delta)}\cos[2\theta(\Delta+n)]
\end{equation}
\begin{equation}
\text{and: }J_n=2\cos\theta(n+\Delta)=2\pi T_\fake(n+\Delta)
\label{eq: J of syk}
\end{equation}
where $T_\fake$ is the ``fake disk'' temperature~\cite{Maldacena:2016hyu,Lin:2023trc}, which controls the submaximal chaos exponent $\lambda_{\text{Lyapunov}}=2\pi T_\fake$ at finite temperature.

To extract the near-horizon data as in \cref{sec: Curvature near horizon and AdS universality}, we work in Fourier space and determine the zeros of the spectral function:
\begin{equation}
\mA(\omega)=\frac{(\cos\theta)^{-2\Delta-1}2^{2\Delta -2}}{\Gamma(2\Delta)}\Gamma(\Delta+\frac{\i\omega}{2\cos\theta})\Gamma(\Delta-\frac{\i\omega}{2\cos\theta})\operatorname{cosh}
(\frac{\theta \omega}{\cos\theta})
\end{equation}
The zeros of $\mA(\omega)$ arise from the $\cosh$ factor and lie at $\{\pm\i\lambda_k\}_{k\geq0}$:
\begin{equation}
\lambda_k=\frac{\pi\cos\theta}{\theta}(k+\frac{1}{2})=2\pi T_\syk(k+\frac12),\ k\geq0
\label{eq: lambda of syk}
\end{equation}
Comparing \eqref{eq: J of syk} and \eqref{eq: lambda of syk}, we see that the poles $\{\i J_n\}$ and zeros $\{\i\lambda_k\}$ form equally spaced lattices on the imaginary axis, with spacings set by the fake and physical temperatures, respectively. Collisions between poles and zeros can therefore occur when the two spacings are incommensurate.

Next we perform the causal kernel decomposition of \cref{sec: Causal kernel decomposition and horizon modes}. Using the Gamma function reflection formula, $\mA(\omega)$ can be rewritten as:
\begin{equation}
\mA(\omega)=A_0\frac{\Gamma(\Delta+\frac{\i\omega}{2\cos\theta})\Gamma(\Delta-\frac{\i\omega}{2\cos\theta})}{\Gamma(\frac12+\frac{\i\theta\omega}{\pi\cos\theta})\Gamma(\frac12-\frac{\i\theta\omega}{\pi\cos\theta})},\ A_0:=\frac{\pi(\cos\theta)^{-2\Delta-1}2^{2\Delta -2}}{\Gamma(2\Delta)}
\end{equation}
From \eqref{eq: KAK=1, continuous}, the modulus of the future kernel on the real axis is
\begin{equation}
\left|\mK^f(\omega)\right|=\sqrt{A_0}\left|\frac{\Gamma(\frac12-\frac{\i\theta\omega}{\pi\cos\theta})}{\Gamma(\Delta-\frac{\i\omega}{2\cos\theta})}\right|
\end{equation}
Since the causal factorization requires both $\mK^f(\omega)$ and $\mR^f(\omega)=1/\mK^f(\omega)$ to be analytic in the upper half plane, one might naively guess
\begin{equation}
\mK_\naive^f(\omega)=\sqrt{A_0}\frac{\Gamma(\frac12-\frac{\i\theta\omega}{\pi\cos\theta})}{\Gamma(\Delta-\frac{\i\omega}{2\cos\theta})}
\label{eq: naive Kf for syk}
\end{equation}
which has no poles or zeros in the upper half-plane. However, this naive choice fails because the corresponding real-time kernel (or its inverse) is not causal: $K^f(t)$ and $R^f(t)$ do not vanish for $t<0$. The obstruction is that at any nonzero temperature $\mK^f_\naive(\omega)$ grows factorially at large complex $\omega$\footnote{Using Stirling's formula, for \eqref{eq: naive Kf for syk} one finds at large $\omega$ that $\log\mK_\naive^f(\omega)=(1-\eta)(\i\nu)\log(\i\nu)+(\i\nu)(\eta-1-\eta\log\eta)+(\frac{1}{2}-\Delta)\log(\i\nu)+O(1)$, with $\nu:=\frac{\omega}{2\pi T_\syk}$ and $\eta:=\frac{T_\syk}{T_\fake}\geq1$. At $T_\syk=0$ (so $\eta=1$), the $O(\omega\log\omega)$ and $O(\omega)$ terms vanish and $\mK_\naive^f(\omega)\sim\omega^{\frac{1}{2}-\Delta}$.} so the Fourier contour cannot be closed in the upper half-plane, even though there are no poles. For example, at infinite temperature one finds
\begin{equation}
R_\naive^f(t)=\int\frac{d\omega}{2\pi}e^{-\i\omega t}\frac{1}{\sqrt{\pi A_0}}\Gamma(\Delta-\frac{\i\omega}{2})=\frac{2}{\sqrt{\pi A_0}}e^{-2\Delta t-e^{-2t}}
\label{eq: 510}
\end{equation}
Thus $R^f_\naive(t)$ is nonzero for $t<0$, though doubly exponentially suppressed.

In fact, one can show that a strictly causal $R^f(t)$ does not exist at any finite temperature. The reason is that for large real $\omega$ one has exponential decay:
\begin{equation}
\log|\mR^f(\omega)|=-\frac{|\omega|}{4}(\frac{1}{T_\fake}-\frac{1}{T_\syk})-(\frac12-\Delta)\log|\omega|+O(1)
\end{equation}
It follows that $R^f(t)$ is analytic in the strip $|\Im t|<\frac{1}{4}(\frac{1}{T_\fake}-\frac{1}{T_\syk})$. By analyticity, if $R^f(t)$ vanished on an open interval $t<0$, it would vanish identically in the strip. Physically, $R^f(t)$ is ``too analytic'' to support a sharp causal discontinuity at $t=0$.

We expect the UV-sensitive behavior discussed above not to affect the near-horizon physics, which corresponds to IR physics on the boundary. We therefore evaluate the near-horizon geometry by introducing a UV regulator. To this end, define a regulated Gamma function by truncating the infinite lattice of poles of $\Gamma(z)$ to the first $(\Lambda+1)$ poles:
\begin{equation}
\Gamma_\Lambda(z):=(\Lambda^z\Lambda!)P_\Lambda(z),\ P_\Lambda(z):=\frac{1}{z(z+1)(z+2)\cdots(z+\Lambda)}
\end{equation}
where $\Lambda$ is a large positive integer. Note that
\begin{equation}
\lim_{\Lambda\rightarrow+\infty}\Gamma_\Lambda(z)=\Gamma(z)
\end{equation}
The regulated spectral function is
\begin{equation}
\mA_\Lambda(\omega):=A_0\frac{\Gamma_\Lambda(\Delta+\frac{\i\omega}{2\cos\theta})\Gamma_\Lambda(\Delta-\frac{\i\omega}{2\cos\theta})}{\Gamma_\Lambda(\frac12+\frac{\i\theta\omega}{\pi\cos\theta})\Gamma_\Lambda(\frac12-\frac{\i\theta\omega}{\pi\cos\theta})}=(A_0\Lambda^{2\Delta-1})\frac{P_\Lambda(\Delta+\frac{\i\omega}{2\cos\theta})P_\Lambda(\Delta-\frac{\i\omega}{2\cos\theta})}{P_\Lambda(\frac12+\frac{\i\theta\omega}{\pi\cos\theta})P_\Lambda(\frac12-\frac{\i\theta\omega}{\pi\cos\theta})}
\end{equation}
Since $\mA_\Lambda(\omega)$ is a finite product of the form \eqref{eq: A(omega) decomposed into poles and zeros}, the corresponding kernel is well defined:
\begin{equation}
\mK_\Lambda^f(\omega):=\sqrt{A_0\Lambda^{2\Delta-1}}\frac{P_\Lambda(\frac12-\frac{\i\theta\omega}{\pi\cos\theta})}{P_\Lambda(\Delta-\frac{\i\omega}{2\cos\theta})},\ \mR^{f}_\Lambda(\omega):=1/\mK^f_\Lambda(\omega)
\end{equation}
With this choice, the time-domain kernels $\mK^f_\Lambda(t)$ and $\mR^f_\Lambda(t)$ are causal. Using \eqref{eq: S in fourier space}, the S-matrix is
\begin{equation}
\mS_\Lambda(\omega)=\frac{P_\Lambda(\frac12+\frac{\i\theta\omega}{\pi\cos\theta})}{P_\Lambda(\frac12-\frac{\i\theta\omega}{\pi\cos\theta})}\frac{P_\Lambda(\Delta-\frac{\i\omega}{2\cos\theta})}{P_\Lambda(\Delta+\frac{\i\omega}{2\cos\theta})}=\Lambda^{\frac{\i\omega}{\pi}(\frac{1}{T_\syk}-\frac{1}{T_\fake})}\hat{\mS}_\Lambda(\omega)
\end{equation}
\begin{equation}
\text{with: }\hat{\mS}_\Lambda(\omega):=\frac{\Gamma_\Lambda(\frac12+\frac{\i\theta\omega}{\pi\cos\theta})}{\Gamma_\Lambda(\frac12-\frac{\i\theta\omega}{\pi\cos\theta})}\frac{\Gamma_\Lambda(\Delta-\frac{\i\omega}{2\cos\theta})}{\Gamma_\Lambda(\Delta+\frac{\i\omega}{2\cos\theta})}
\end{equation}
where in the last step we rewrote $P_\Lambda(z)$ in terms of $\Gamma_\Lambda(z)$. Thus, up to the phase factor $\Lambda^{\frac{\i\omega}{\pi}(\frac{1}{T_\syk}-\frac{1}{T_\fake})}$, the remaining factor $\hat{\mS}_\Lambda(\omega)$ has a finite limit as $\Lambda\rightarrow+\infty$.

We now extract the near-horizon geometry. Determining the geometry further requires the residues $\{\beta_{\Lambda,k}\}_{k\geq0}$ defined in \eqref{eq: beta from residue}:
\begin{equation}
\beta_{\Lambda,k}=e^{-2(\alpha_0\log\Lambda)\lambda_k }\hat{\beta}_{\Lambda,k},\ \ \alpha_0:=\frac{1}{2\pi}(\frac{1}{T_\syk}-\frac{1}{T_\fake})
\end{equation}
\begin{equation}
\text{with: }\lim_{\Lambda\rightarrow+\infty}\hat{\beta}_{\Lambda,k}:=\beta_k=(2\pi T_\syk)\frac{(-1)^k}{(k!)^2}\frac{\Gamma[\Delta+\frac{\pi}{2\theta}(k+\frac{1}{2})]}{\Gamma[\Delta-\frac{\pi}{2\theta}(k+\frac{1}{2})]}
\label{eq: beta in syk}
\end{equation}
For large but finite $\Lambda$, this is well behaved. A potential concern is that $\beta_{\Lambda,k}$ has no nonzero limit as $\Lambda\rightarrow+\infty$: while $\hat{\beta}_{\Lambda,k}$ converges, the prefactor $e^{-2(\alpha_0\log\Lambda)\lambda_k}$ drives all $\beta_{\Lambda,k}\to0$.

As we now show, this does not affect near-horizon geometric quantities (e.g. the Ricci scalar), since they depend only on ratios of $\beta_{\Lambda,k}$. From \eqref{eq: temperature of geometry} and \eqref{eq: lambda of syk}, the black-hole temperature is independent of $\Lambda$:
\begin{equation}
T_\text{bh}=T_\syk
\end{equation}
To obtain the horizon curvature, note from \cref{sec: Curvature near horizon and AdS universality} that large-$q$ SYK falls into Case II with $\lambda_1=3\lambda_0$. The near-horizon curvature is then given by \eqref{eq: near horizon curvature, general formula}, which we repeat for convenience:
\begin{equation}
R_{\text h,\Lambda}/m^2=-2(1+\frac{4\lambda_0^2\beta_{\Lambda,1}}{\beta_{\Lambda,0}^3})=-2(1+\frac{4\lambda_0^2\hat\beta_{\Lambda,1}}{\hat\beta_{\Lambda,0}^3})
\label{eq: 512}
\end{equation}
In the last step, we used that in the combination $\beta_{\Lambda,1}/\beta_{\Lambda,0}^3$ the ``un-wanted'' exponential factors $e^{-2(\alpha_0\log\Lambda)\lambda_k}$ cancel precisely because of  $\lambda_1=3\lambda_0$. We therefore obtain a finite limit for horizon curvature,
\begin{equation}
R_\text{h}/m^2=\lim_{\Lambda\rightarrow+\infty}R_{\text h,\Lambda}/m^2=-2(1+\frac{4\lambda_0^2\beta_{1}}{\beta_{0}^3})
\end{equation}

One may also ask whether the geometry at finite proper distance from the horizon admits a finite limit as $\Lambda\rightarrow+\infty$. It does. In \eqref{eq: main result2}, the residues $\beta_{\Lambda,k}$ always appear multiplied by $e^{-2\lambda_k z}$, so the extra factor $e^{-2(\alpha_0\log\Lambda)\lambda_k}$ can be absorbed into a coordinate shift, $\tilde z:=z+\alpha_0\log\Lambda$. Working with the renormalized coordinate $\tilde z$, only $\hat\beta_{\Lambda,k}$ appears in \eqref{eq: main result2}, and it has a finite limit given by \eqref{eq: beta in syk}. This yields a well-defined limit of geometry, $ds^2=\Omega(\tilde z)^2(-dt^2+d\tilde z^2)$.

We now discuss the zero and finite temperature cases.

\subsection{Zero temperature limit}
\label{sec: zero temperature syk}
In the zero temperature limit, we don't need $\Lambda$-regularization and the causality of $K^{p,f}(t), R^{p,f}(t)$ works just fine. The relevant coefficients to extract geometry behave as
\begin{equation}
\beta_0=-(2\pi T_\syk)(\frac{1}{2}-\Delta)+O(T_\syk^2),\ \beta_1=-(2\pi T_\syk)(\frac{1}{2}+\Delta)(\frac{1}{2}-\Delta)(\frac{3}{2}-\Delta)+O(T_\syk^2)
\end{equation}
Since $\lambda_0$, $\beta_0$, and $\beta_1$ are all $O(T_\syk)$, \eqref{eq: 512} implies that $R_\text h$ remains $O(1)$ as $T_\syk\to0$:
\begin{equation}
R_\text h/m^2=-\frac{2}{(\frac12-\Delta)^2},\ \ \ \text{when }T_\syk=0
\end{equation}
Writing the curvature as $R_\text h=-2\ell_{\text{AdS}}^{-2}$, the fermion mass is $m=(\frac12-\Delta)\ell_{\text{AdS}}^{-1}$, matching the standard AdS$_2$ relation between mass and scaling dimension for a Majorana field.

One can in fact reconstruct the full geometry, not just the near-horizon expansion. Using \eqref{eq: main result1} and \eqref{eq: main result2}, one finds\footnote{One first notice that in zero temperature, $\beta_k$ can be expressed in terms of elementary functions as $\beta_k=(2\pi T_\syk)\frac{(-1)^k}{(k!)^2}\prod_{j=0}^{2k}(\Delta-k-\frac{1}{2}+j)$. Also, notice that since $\lambda_k$ are integer spaced, then from \eqref{eq: main result2}, $\Omega^2$ can be written as Taylor series of $\varepsilon:=e^{-(2\pi T_\syk)2z}$, since no fractional power. Of course, the $B(z)$ matrix in \eqref{eq: main result2} is infinite dimensional, so we used Mathematica to obtain its first twenty Taylor coefficients, and find that it satisfy $\frac{1}{4}m^2\Omega(z)^2=(\frac{1}{2}-\Delta)^2(\varepsilon+2\varepsilon^2+3\varepsilon^3+4\varepsilon^4+5\varepsilon^5+6\varepsilon^6+\cdots)$. Therefore, though we cannot prove it, it is convincing to believe that the full answer is $\frac{1}{4}m^2\Omega(z)^2=(\frac12-\Delta)^2\sum_{n=1}^{+\infty}n\varepsilon^n$, which leads to \eqref{eq: zero temeprature syk Omega(z)}.} that the conformal factor is:
\begin{equation}
m\Omega(z)=\frac{\frac{1}{2}-\Delta}{\sinh[(2\pi T_\syk)z]}
\label{eq: zero temeprature syk Omega(z)}
\end{equation}
This reproduces AdS$_2$ Rindler, with the boundary $z=0$ at the asymptotic boundary.

\subsection{Finite temperature geometry}
\label{sec: finite temperature syk}
We first record the explicit horizon-curvature formula obtained from \eqref{eq: 512} and \eqref{eq: beta in syk}:
\begin{equation}
R_\text h/m^2=(-2)\left[1-\pi^2\frac{\sin(\pi(\Delta-\frac32\eta))}{\sin^3(\pi(\Delta-\frac12\eta))}\frac{\Gamma(\Delta+\frac32\eta)\Gamma(1-\Delta+\frac32\eta)}{\Gamma(\Delta+\frac12\eta)^3\Gamma(1-\Delta+\frac12\eta)^3}\right],\ \eta:=\frac{T_\syk}{T_\fake}
\label{eq: syk horizon curvature}
\end{equation}
We have used the reflection formula to ensure that the arguments of the Gamma functions are positive. Therefore, possible divergences of $R_\text h$ arise from the sine factor in the denominator, i.e. from $\beta_0=0$. This corresponds to a collision between the first pole $\i\lambda_0$ and one of the zeros $\{\i J_n\}$ due to the incommensurability of the physical and fake temperature lattices. Explicitly, the divergences occur at:
\begin{equation}
\frac{T_\syk}{T_\fake}=2\Delta+2n\Longrightarrow T_\syk=\frac{2}{\pi}(n+\Delta)\cos(\frac{\pi/4}{n+\Delta}) \ \ \ n=0,1,2,3,...
\end{equation}
\begin{figure}[t]
    \centering
    \includegraphics[width=1\linewidth]{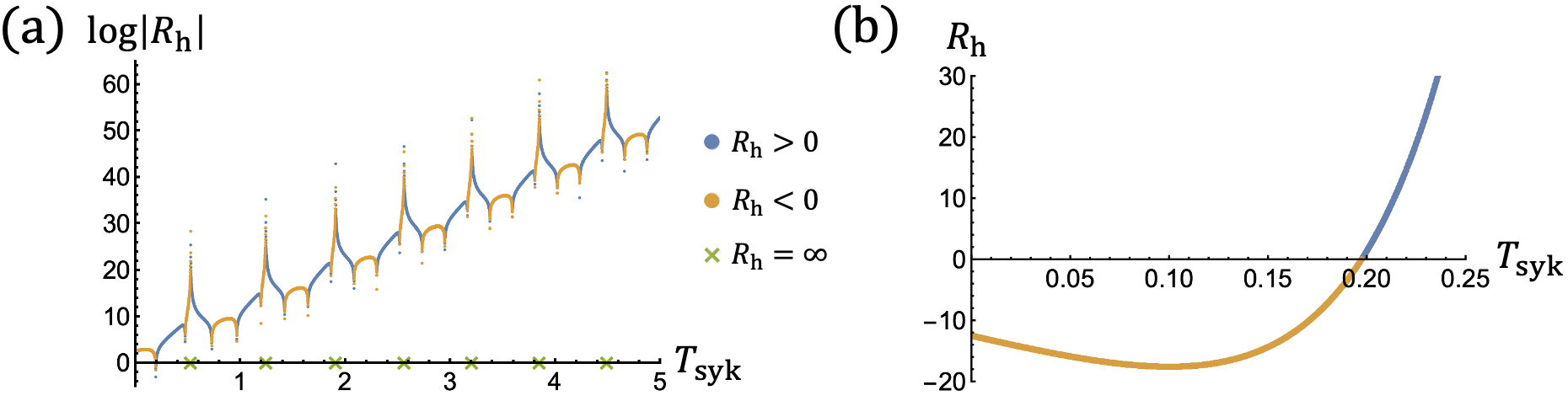}
    \caption{\textbf{(a)} The horizon curvature from \eqref{eq: syk horizon curvature} as a function of temperature for large-$q$ ($\Delta=1/10$) SYK. Colors indicate the sign of $R_\text h$. \textbf{(b)} Zoom into low temperature regime of (a).}
    \label{fig: syk curvature vs temperature}
\end{figure}
At high temperature, $T_\fake$ approaches the constant $\pi^{-1}$, so the curvature diverges almost periodically at $T_\syk\approx\frac{2}{\pi}(\Delta+n)$. The infinite-temperature limit~\cite{Lin:2022nss} of the reconstructed geometry is therefore not manifestly well-defined in our set up. Using Stirling's approximation, the ratio of Gamma functions grows as $\sim e^{(3\log 3)\frac{T_\syk}{T_\fake}}\sim e^{(3\pi\log 3)T_\syk}$, so $|R_\text h|$ grows exponentially with temperature. As shown in \cref{fig: syk curvature vs temperature}, the sign of $R_\text h$ alternates, so AdS and dS near-horizon behavior appear almost periodically as $T_\syk$ increases. At low but nonzero temperature, the near-horizon region remains AdS.

\section{Algebra near the horizon}
\label{sec: algebra near horizon}
Recent progress in AdS/CFT has emphasized that the appropriate notion of a ``subregion'' in QFT/gravity is an operator algebra rather than a tensor factor, with modular theory providing the intrinsic dynamics~\cite{Chen:2025aen,Furuya:2023fei,Ouseph:2023juq,Lashkari:2024lkt}. In particular, the large-$N$ algebra program~\cite{Leutheusser:2021frk,Leutheusser:2021qhd,Leutheusser:2022bgi} and related crossed-product/observer viewpoints~\cite{Witten:2021unn,Witten:2023qsv,Witten:2023xze,Chandrasekaran:2022cip,Chandrasekaran:2022eqq,Penington:2023dql,Kudler-Flam:2023qfl,Chen:2024rpx} clarify how semiclassical $N^{-1}$ effects can promote a strictly type-III structure to a type-I or type-II setting, in which generalized entropy and reconstruction admit a precise algebraic formulation. These developments are largely structural: the relevant generators are typically specified abstractly (via their automorphism action or universal properties) rather than constructed explicitly from boundary operators.

Here we take a step towards a constructive realization by giving boundary expressions for an approximate algebra, generated by two  null translations and a boost, which becomes exact when acting on the modes at the bifurcate horizon.

In a maximally symmetric $(1+1)$d spacetime, there are three generators: null translations $P^\pm$ and a boost $B$. They obey $[B,P^\pm]=\pm\i P^\pm$ and $[P^-,P^+]\propto-\i R B$, with the algebra closing to sl(2,$\mathbb R$), so(1,2), or the Poincar\'e algebra depending on the sign of the Ricci scalar $R$.

In a general $(1+1)$d spacetime with metric \eqref{eq: conformal coordinate}, only $B$ is an exact isometry. Nevertheless, one can define  ``null translations'' $P^\pm$ in a natural way so that $[B,P^\pm]=\pm\i P^\pm$ holds exactly. While $[P^-,P^+]$ is not proportional to $-\i R B$ when $R$ varies, we will show that $[P^-,P^+]\propto-\i R_\text h B$ still holds when acting on fields at the bifurcate horizon, where $R_\text h$ is the local horizon curvature; see \eqref{eq: action of [P,P] is B on bifurcate horizon modes}.

We first introduce Kruskal coordinates near the horizon. Using the leading near-horizon behavior $e^{-4\lambda_0 z}$ in \eqref{eq: 3.4}, we define $(U,V)$ by:
\begin{equation}
ds^2=\frac{4\beta_0^2}{m^2}\tilde\Omega^2\times e^{-4\lambda_0z}(-dt^2+dz^2)=\frac{\beta_0^2\tilde\Omega^2}{m^2\lambda_0^2}(-dUdV)
\end{equation}
\begin{equation}
\text{with: }U:=e^{2\lambda_0u}\geq0,\  V:=-e^{-2\lambda_0v}\leq0\text{,  and }\tilde\Omega^2:=m^2\Omega^2e^{4\lambda_0z}/(4\beta_0^2)
\end{equation}
Next we define $P^\pm$ and $B$ on the past/future horizons. Using the horizon modes in \eqref{eq: define past and future mode}, define the rescaled fields $\Upsilon^p(V)$ and $\Upsilon^f(U)$ on the horizons:
\begin{equation}
\Upsilon^f(U):=\frac{1}{\sqrt{2\lambda_0U}}\Psi^f(u),\ \Upsilon^p(V):=\frac{1}{\sqrt{-2\lambda_0V}}\Psi^p(v)
\label{eq: from Psi to Upsilon}
\end{equation}
The rescale factor is chosen so that the canonical anti-commutators are preserved under coordinate change from $(u,v)$ to $(U,V)$:
\begin{equation}
\{\Upsilon^f(U),\Upsilon^f(U')\}=(2\pi)\delta(U-U'),\ \{\Upsilon^p(V),\Upsilon^p(V')\}=(2\pi)\delta(V-V')
\end{equation}
The mixed anti-commutator between $\Upsilon^f$ and $\Upsilon^p$ is controlled by the S-matrix:
\begin{equation}
\tilde S(U,V):=\{\Upsilon^f(U),\Upsilon^p(V)\}=\frac{1}{2\lambda_0\sqrt{-UV}}S(u,v)
\label{eq: S-matrix of Upsilon}
\end{equation}
Equivalently, the future mode can be expanded in the past modes as:
\begin{equation}
\Upsilon^f(U)=\frac{1}{2\pi}\int dV\tilde S(U,V)\Upsilon^p(V)
\end{equation}

With these preparations, we define $P^+$ (null translation along $V$), $P^-$ (null translation along $U$), and the boost $B$ on the horizons:
\begin{equation}
\begin{aligned}
&P^+:=\frac{\i}{4\pi}\int dV\Upsilon^p(V)\partial_V\Upsilon^p(V)\\
&P^-:=\frac{\i}{4\pi}\int dU\Upsilon^f(U)\partial_U\Upsilon^f(U)\\
&B:=-\frac{\i}{4\pi}\int dV\Upsilon^p(V)V\partial_V\Upsilon^p(V)\equiv \frac{\i}{4\pi}\int dU\Upsilon^f(U)U\partial_U\Upsilon^f(U)
\end{aligned}
\label{eq: defintion of three generators}
\end{equation}
The equivalence of the two expressions for $B$ follows from unitarity of the S-matrix:
\begin{equation}
\int dU \tilde S(U,V_1)\tilde S(U,V_2)=(2\pi)^2\delta(V_1-V_2)
\end{equation}
By construction, these generators act on the horizon fields as translations and boosts:
\begin{equation}
[\i P^+,\Upsilon^p(V)]=\partial_V\Upsilon^p(V),\ [\i P^-,\Upsilon^f(U)]=\partial_U\Upsilon^f(U)
\end{equation}
\begin{equation}
[\i B,\Upsilon^p(V)]=(-\frac12-V\partial_V)\Upsilon^p(V),\ [\i B,\Upsilon^f(U)]=(\frac12+U\partial_U)\Upsilon^f(U)
\label{eq: action of boost}
\end{equation}
The factor $\frac12$ reflects the fermion scaling dimension. Their action on $\Psi_{L/R}(u,v)$ away from the horizon follows from the equations of motion.

In \cref{app: sl2R algebra in AdS2 Rindler}, we will benchmark that the definition in \eqref{eq: defintion of three generators} matches the exact sl(2,$\mathbb R$) generator of AdS$_2$-Rindler wedge with massive Majorana fermion, whose boundary is at asymptotic boundary with usual normalizable boundary condition.

We now examine the commutators. From \eqref{eq: defintion of three generators}, one checks directly that
\begin{equation}
[B,P^\pm]=\pm \i P^\pm
\end{equation}
The remaining commutator $[P^-,P^+]$ is:
\begin{equation}
[P^-,P^+]=\frac{1}{16\pi^3}\int dV_1dV_2\Upsilon^p(V_1)\Upsilon^p(V_2)\left[\int dU\cdot \tilde S(U,V_1)\partial_U\partial_{V_2}\tilde S(U,V_2)-(V_1\leftrightarrow V_2)\right]
\label{eq: kernel of PP commutator}
\end{equation}
Even on the past horizon, $[P^-,P^+]$ maps a local operator $\Upsilon^p(V)$ to a nonlocal one, so it is not generally a geometric flow. In special cases---notably AdS$_2$-Rindler (\cref{app: benchmark on AdS2 Rindler case})---the kernel becomes local and $[P^-,P^+]$ is exactly proportional to $B$.

Nevertheless, in \cref{app: Technical details of near horizon algebra} we show that the action of $[P^-,P^+]$ on the bifurcate-horizon modes $\mathcal O\in\{\Upsilon^f(0),\Upsilon^p(0)\}$ is local and related to the boost generator as expected:
\begin{equation}
\left[[P^-,P^+],\mathcal O\right]=-\i \frac{\beta_0^2}{(2\lambda_0)^2} (R_\text h/m^2)\times [B,\mathcal O] 
\label{eq: action of [P,P] is B on bifurcate horizon modes}
\end{equation}
This provides an algebraic interpretation of the near-horizon curvature studied in \cref{sec: Curvature near horizon and AdS universality}. It also suggests an emergent IR symmetry for any boundary system.

We should comment that, even away from the bifurcate horizon, the Majorana fermion dynamics along the horizon is universal if the boundary is in a thermal state~\cite{Nebabu:2023iox}. In this work, we focused on bulk reconstruction from the fermion anti-commutator two-point function $\mA(\omega)$, which is sufficient to determine the bulk geometry. To identify which quantum state in the QFT Hilbert space the bulk fermions occupy, one needs additional input, namely the fermion commutator two-point function $\mathcal G_\text{C}(\omega)$; see \cref{app: convention for Fourier}. Assuming the Kubo–Martin–Schwinger (KMS) condition holds, $\mathcal G_\text{C}(\omega)$ is related to $\mA(\omega)$ in a universal way; see \eqref{eq: relate commutator to anti-commutator}. Therefore, the bulk quantum state is the two-sided thermofield-double state. The commutator of the horizon modes satisfies universal CFT$_2$ behaviour~\cite{Nebabu:2023iox}.

\section{Discussion}
\label{sec: discussion}

In this paper, building on~\cite{Nebabu:2023iox}, we obtained an explicit analytic formula for the semiclassical bulk geometry dual to an arbitrary boundary Majorana generalized free field. We derived the formula in two complementary ways: via unitary matrix integral techniques and via inverse scattering. We then used it to study the near-horizon curvature as a probe of IR universality of boundary theory. For generic boundary data, the near-horizon region is AdS. For correlators with finitely many QNMs, the $N_\qnm=2$ case yields AdS$_2$ Rindler, while $N_\qnm=3$ admits families with de Sitter near-horizon curvature realizable by Lindbladian boundary dynamics. We also applied the construction to large-$q$ SYK, recovering AdS$_2$ Rindler at zero temperature and finding an unusual, nearly periodic pattern of sign changes and divergences in the near-horizon curvature at finite temperature, driven by the incommensurability between the physical and ``fake'' temperatures. Finally, we constructed boundary expressions for an approximate algebra generated by null translations and a boost, which becomes exact acting on bifurcate-horizon modes.

We conclude with a list of open questions and directions:
\begin{itemize}
\item \textit{Scalar field. }Here we focused on Majorana fields. It would be natural to study scalar fields as an even simpler testbed. Ref.~\cite{Nebabu:2023iox} discussed that there is a scalar analogue of the bulk quantum circuit, where each SO(2) gate is replaced by a symplectic matrix, describing the discrete evolution of a scalar GFF.
\item \textit{Higher dimensions. }Our analysis focused on the duality between $(1+1)$d bulk and $(0+1)$d boundary. A higher-dimensional extension should be straightforward under (say) spherical symmetry of the transverse geometry, where reconstruction can be performed in each angular-momentum sector and reduced to an effective two-dimensional problem. Scalar fields would again be a natural starting point, since fermions acquire additional components in higher dimensions.

\item \textit{Euclidean version. }Our approach reconstructs a Lorentzian bulk geometry from real-time boundary data. It would be interesting to start instead from the Euclidean two-point function on the thermal circle and reconstruct a Euclidean disk geometry with U(1)-symmetric metric $ds^2=d\rho^2+G(\rho)^2d\theta^2$. A preliminary step was performed in~\cite{Goel:2023svz,Almheiri:2024xtw}. A technical obstacle to this approach is the absence of a direct Euclidean analogue of the Nebabu--Qi circuit, since Euclidean signature lacks causal structure. One possible route is to work directly in the continuum and formulate the corresponding inverse scattering problem.

\item \textit{Relation to large-mass expansions\footnote{We thank Henry Lin for asking this question.}. }Our relation between the S-matrix $\mS(\omega)$ and the conformal factor $\Omega(z)$ is organized as a small-mass expansion (e.g.\ \eqref{eq: forward scattering, zig-zag}). In the opposite large-mass limit, boundary correlators are often approximated by a worldline action dominated by a (possibly complex) geodesic anchored at the boundary, so geometric data can in principle be inferred from the dependence of geodesic length on boundary endpoints~\cite{Goel:2023svz,Almheiri:2024xtw}. This viewpoint is closely related to differential entropy and kinematic space~\cite{Czech:2015qta,Balasubramanian:2013lsa,Czech:2015kbp}. It would be interesting to connect these approaches.

\item \textit{A more microscopic de Sitter model. }We found de Sitter near-horizon curvature in a class of dissipative Gaussian Lindbladian boundary models. An unsatisfactory aspect is that the effective Lindbladian dynamics is non-unitary, describing an open subsystem coupled to a Markovian bath (if one include the bath, the infinitely large whole system is still unitary though). It would be preferable to find a fully microscopic unitary model whose large-$N$ limit yields positive near-horizon curvature.
\item \textit{Entanglement entropy. }We used only boundary anti-commutators to reconstruct the bulk geometry, but did not determine the bulk quantum state. The bulk quantum state is determined by the commutator of fermion $\chi(t)$ on the boundary~\cite{Nebabu:2023iox}. It would be interesting to compute bulk entanglement entropy in this framework.

\item \textit{Relation to Krylov space. }Krylov space and Krylov complexity~\cite{Parker:2018yvk,Nandy:2024evd}, which encode two-point-function data, provide a useful diagnostic of chaos and scrambling. Recent work by Dodelson~\cite{Dodelson:2024atp,Dodelson:2025rng} proposes a putative asymptotic ``bulk geometry'' dual to Krylov space as an intermediate structure for computing QNMs and complex-time analyticity. It would be interesting to relate that proposal to our reconstructed bulk geometry. In the main text, the continuum limit \eqref{eq: definition of continuous limit} keeps both space and time continuous while preserving a $45^\circ$ lightcone, yielding a relativistic theory. By contrast, Krylov space is obtained by a nonrelativistic limit of the Nebabu--Qi circuit, in which time becomes continuous while space remains discrete:
\begin{equation}
\Delta t\rightarrow0\text{ and }\t\rightarrow\infty\text{, while }(\t\Delta t)\rightarrow t\in\mathbb R,\text{ and }\z:=n\in\mathbb Z_{+} \text{ fixed.}
\label{eq: krylov limit}
\end{equation}
The discrete spatial coordinate $\z$ is identified with the Krylov lattice index $n$. One then finds that the gate angle $\theta(\z)$ is related to the Krylov coefficient $b_n$ by:
\begin{equation}
\lim_{\Delta t\rightarrow 0\text{~as~}\eqref{eq: krylov limit}}\frac{\theta(\z=n)}{\Delta t}=b_n
\label{eq: krylov coeffciient and angle}
\end{equation}
We will report these results and further details on the dual geometry of Krylov space in future work.
\item \textit{Fluctuations of geometry. }Our reconstruction assumes a boundary GFF, i.e. a strict large-$N$ limit. It would be interesting to include perturbative $N^{-1}$ corrections. Since connected four-point functions can be viewed as fluctuations of two-point data, and two-point data maps to geometry, one expects the connected four-point function to determine the bulk metric fluctuations.

\item \textit{More on algebras. }Our discussion of bulk algebras focused on the bifurcate horizon. Recent developments in von Neumann algebra methods suggest many further structures to explore, including subregion algebras~\cite{Lashkari:2024lkt,Ouseph:2023juq,Chandrasekaran:2026pnc}, algebras of local Lorentz symmetry in the bulk, and half-sided modular inclusions.

\item \textit{Puzzles at finite-temperature large-$q$ SYK. }In \cref{sec: Examples: large-q SYK}, the reconstructed horizon curvature exhibits an unusual temperature dependence. What is its physical interpretation? Our derivation used a subtle $\Lambda$-regularization to enforce causality of the inverse kernel; one might instead advocate using the naive (noncausal) inverse kernel, suggesting that the minimal ansatz of a local Majorana field coupled to a metric is insufficient. The correct bulk description might be nonlocal or receive ``stringy'' corrections~\cite{Shenker:2014cwa} or higher derivative corrections, as expected for submaximal chaos~\cite{Shenker:2014cwa,Lin:2023trc}. One hint is that at finite temperature the S-matrix $\mS(\omega)$ grows factorially at high energy for complex $\omega$, which is unusual for particle scattering. It is also possible that the correct dual theory does not have a sharp lightcone.

\item \textit{Physical origin of AdS universality. }In \cref{sec: Curvature near horizon and AdS universality}, we found that AdS behavior is generic near the horizon over a large region of boundary parameter space. Is this tied to an IR universality class of the boundary theory? We do not yet have a satisfactory answer.

\item \textit{Distinguishing chaos from dissipation. }Finite-QNM correlators are naturally realized by dissipative boundary dynamics (\cref{sec: Examples: finite QNMs}), while SYK provides a unitary example (\cref{sec: Examples: large-q SYK}). Both can reconstruct bulk geometries with horizons. The horizon indicates information loss in the space of simple operators, but does not by itself distinguish dissipation (operators leak into a bath) from chaos (operators scramble into complex ones). Is there a bulk criterion that differentiates the two mechanisms?

\item \textit{Horizonless geometries. }Although all examples in this paper have horizons, the construction does not require them. The scattering problem remains well defined as long as $\Omega(z)\to0$ as $z\to+\infty$; horizons correspond to exponential decay. What boundary properties lead to horizonless geometries?

\item \textit{A new scaling limit of unitary matrix integrals. }In \cref{sec: Method I: via unitary matrix integral}, we identified a scaling limit in which all subleading eigenvalue-instanton contributions enter at the same order due to the additional parameter $\Delta t$, with $N\to\infty$, $\Delta t\to0$, and $N\Delta t\to z=\text{const}$. Effectively, the discrete large parameter $N$ is traded for a continuous parameter $z>0$. It would be interesting to study this limit directly from the matrix model perspective.


\item \textit{``Measuring'' the bulk geometry from boundary experiments. }We provided an explicit formula showing that the dual bulk geometry is determined by boundary two-point functions. Since two-point functions are among the simplest observables accessible experimentally (compared to, e.g., entanglement entropies or out-of-time-ordered correlators~\cite{Garttner2017,PhysRevX.7.031011,Xu:2022vko}, which may require multiple copies or post-selection), it would be interesting to propose concrete experimental protocols and estimate the precision required in measuring the two-point function to reliably reconstruct the bulk geometry.

\end{itemize}

\section*{Acknowledgements}
We thank Shoaib Akhtar, Yiming Chen, Matthew Dodelson, Anatoly Dymarsky, Dan Eniceicu, Jonah Kudler-Flam, Nima Lashkari, Albert Law, Hong Liu, Hideo Mabuchi, Juan Maldacena, Steve Shenker, Douglas Stanford, Xinyu Sun, and Zimo Sun for helpful discussions. We thank the LITP lunch for feedback. HT especially thanks Yiming Chen for valuable comments and encouragement during the  long development of this work, Matthew Dodelson for pointing out relevant literature, and Dan Eniceicu for illuminating discussions on unitary matrix integral. Part of this work was completed while XLQ and HT were visiting the Institute for Advanced Study, and we thank IAS for hospitality. This work is supported by the Shoucheng Zhang Graduate Fellowship Program (HT), the Simons Foundation (TN, XLQ), the Eric and Wendy Schmidt AI in Science (TN), and Knight-Hennessy Scholars (HW).
\appendix
\section{Convention for the Fourier transformation}
\label{app: convention for Fourier}
Here we collect our Fourier transform conventions for the boundary fermion $\chi(t)$; the bulk horizon modes $\Psi^f(u),\Psi^p(v)$; the bulk-to-boundary kernels $K^{p,f}(t)$ and their inverses $R^{p,f}(t)$; the spectral function $A(t,t')$; and the S-matrix $S(z)$:
\begin{equation}
\begin{aligned}
&\chi(t):=\int d\omega e^{\i\omega t}\chi(\omega)\\
&\Psi^f(u):=\int d\omega e^{\i\omega u}\Psi^f(\omega),\ \Psi^p(v):=\int d\omega e^{\i\omega v}\Psi^p(\omega)\\
&K^p(t):=\int \frac{d\omega}{2\pi}e^{-\i\omega t}\mK^p(\omega),\ \ K^f(t):=\int \frac{d\omega}{2\pi}e^{-\i\omega t}\mK^f(\omega)\\
&R^p(t):=\int \frac{d\omega}{2\pi}e^{-\i\omega t}\mR^p(\omega),\ \ R^f(t):=\int \frac{d\omega}{2\pi}e^{-\i\omega t}\mR^f(\omega)\\
&A(t,t'):=\int d\omega e^{\i\omega (t-t')}\mA(\omega)\\
&S(z)=\int\frac{d\omega}{2\pi}e^{\i\omega z}\mS(\omega)
\end{aligned}
\end{equation}
We also record the relations among the Wightman function $G_\text{W}(t)$, the two-sided correlator $G_\text{12}(t)$, the spectral function $G_\text A(t)$, and the commutator $G_\text{C}(t)$:
\begin{equation}
\begin{aligned}
G_\text{W}(t)&=\Tr[e^{-\beta H}O(t)O(0)]\\
G_\text{12}(t)&=\Tr[e^{-\frac{\beta H}{2}}O(t)e^{-\frac{\beta H}{2}}O(0)]\\
G_\text A(t)&=\Tr[e^{-\beta H}\{O(t),O(0)\}]\\
G_\text C(t)&=\Tr(e^{-\beta H}[O(t),O(0)])
\end{aligned}
\end{equation}
The KMS condition is $G_\text{W}(t-\i\beta)=G_\text{W}(-t)$. We use the same Fourier-transform convention for all four: $G_{\cdots}(t)=\int d\omega\, e^{\i\omega t}\mathcal G_{\cdots}(\omega)$. Their Fourier-space relations are:
\begin{equation}
\mathcal{G}_\text A(\omega)=(1+e^{\beta \omega})\mathcal G_\text W(\omega),\ \mathcal{G}_\text C(\omega)=(1-e^{\beta \omega})\mathcal G_\text W(\omega),\ \mathcal{G}_\text {12}(\omega)=e^{\frac{\beta \omega}{2}}\mathcal G_\text W(\omega)
\label{eq: KMS relation}
\end{equation}
In particular, these KMS relation imply that the commutator two-point function is related to anti-commutator two-point function in a universal way:
\begin{equation}
\mathcal G_\text{C}(\omega)=-\tanh(\frac{\beta\omega}{2})\mathcal G_\text A(\omega)
\label{eq: relate commutator to anti-commutator}
\end{equation}

\section{Determinant formula for multi-flavor fermions}
\label{app: Determinant formula for multi-flavor fermions}
In this appendix, we generalize the determinant formula \eqref{eq: general determinant formula from region} to the case in which the fermions $\chi_\alpha(t)$ carry an additional flavor index $\alpha\in\{1,\dots, N_\text f\}$.

The anticommutator two-point function is
\begin{equation}
A(\alpha,t,\alpha',t')=\langle\{\chi_\alpha(t),\chi_{\alpha'}(t')\}\rangle
\end{equation}
with the normalization $A(\alpha,t,\alpha',t)=\delta_{\alpha\alpha'}$. We follow the vector notation of \cref{sec: A determinant formula for gate angle} and use the same setup as in \cref{fig:  determinant formula}(a), additionally assuming that there are $N_B$ time points in region $B$. We use $|\chi_D\rangle$ and $|\chi_C\rangle$ to represent the $N_\text f$ fermions in regions $D$ and $C$ (each corresponding to a single time point but containing $N_\text f$ flavors), and similarly $|\chi_B\rangle$ to represent the $N_B N_\text f$ fermions in region $B$.

Following \cref{sec: A determinant formula for gate angle}, we use $A_{B}$ to denote the $N_BN_\text f\times N_BN_\text f$ submatrix of $A(\alpha,\t,\alpha',\t')$ whose indices lie in region $B$; similarly, $A_{D}$ and $A_C$ are $N_\text f\times N_\text f$ matrices, $A_{DB}$ and $A_{BC}$ are $(N_\text f+N_BN_\text f)\times (N_\text f+N_BN_\text f)$ matrices, and $A_{DBC}$ is a $(2N_\text f+N_BN_\text f)\times (2N_\text f+N_BN_\text f)$ matrix. For convenience, we parametrize these matrices in terms of four intermediate matrices $W_{1\sim4}$:
\begin{equation}
A_{DBC}=\begin{bmatrix}
\textbf{1} & W_2^\intercal &W_4\\
W_2 & W_1 & W_3\\
W_4^\intercal & W_3^\intercal & \textbf{1}
\end{bmatrix},\ A_{DB}=\begin{bmatrix}
\textbf{1} & W_2^\intercal \\
W_2 & W_1
\end{bmatrix},\ 
A_{BC}=\begin{bmatrix}
 W_1 & W_3\\
W_3^\intercal & \textbf{1}
\end{bmatrix}, \ A_B=W_1
\label{eq: BB0}
\end{equation}

In \cref{fig:  determinant formula}, the circuit implements a step-by-step orthonormalization: $|\psi_3\rangle$ is the component of $|\chi_D\rangle$ orthonormal to $|\chi_B\rangle$, while $|\psi_1\rangle$ is the component of $|\chi_C\rangle$ orthonormal to $|\chi_B\rangle$. We denote by $|\tilde\psi_3\rangle$ and $|\tilde\psi_1\rangle$ the corresponding unnormalized vectors,
\begin{equation}
|\tilde\psi_3\rangle=|\chi_D\rangle-W_2^\intercal W_1^{-1}|\chi_B\rangle,\ |\tilde\psi_1\rangle=|\chi_D\rangle-W_3^\intercal W_1^{-1}|\chi_B\rangle
\end{equation}
which are fixed by the orthogonality conditions $\langle{\tilde\psi_3|\chi_B}\rangle=0$ and $\langle{\tilde\psi_1|\chi_B}\rangle=0$. The inner-product matrix of $|\tilde\psi_3\rangle$, $|\chi_B\rangle$, and $|\tilde\psi_1\rangle$ is then
\begin{equation}
\tilde A_{DBC}=\begin{bmatrix}
\textbf{1}-W_2^\intercal W_1^{-1}W_2 & 0 & W_4-W_2^\intercal W_1^{-1}W_3\\
0&W_1&0\\
W_4^\intercal-W_3^\intercal W_1^{-1}W_2&0&\textbf{1}-W_3^\intercal W_1^{-1}W_3
\end{bmatrix}
\label{eq: BB1}
\end{equation}
Note that $\tilde A_{DBC}$ can be obtained from $A_{DBC}$ by Gaussian elimination: add $[(-W_3W_1^{-1})\times (2^{\text{nd}}\text{ row})]$ to the third row and add $[(-W_2W_1^{-1})\times (2^{\text{nd}}\text{ row})]$ to the first row, and similarly for the corresponding columns. Therefore, $\tilde A_{DBC}$ and $A_{DBC}$ have the same determinant,
\begin{equation}
\det(A_{DBC})=\det(\tilde A_{DBC})
\label{eq: BB2}
\end{equation}
To obtain $|\psi_{1,3}\rangle$ from $|\tilde\psi_{1,3}\rangle$, we only need to normalize them. Define
\begin{equation}
\Lambda_3\Lambda_3^\intercal:=\textbf{1}-W_2^\intercal W_1^{-1}W_2,\ \ \ \ \Lambda_1\Lambda_1^\intercal:=\textbf{1}-W_3^\intercal W_1^{-1}W_3
\label{eq: BB3}
\end{equation}
so that
\begin{equation}
|\psi_3\rangle=\Lambda_3^{-1}|\tilde\psi_3\rangle,\ \ \ \ |\psi_1\rangle=\Lambda_1^{-1}|\tilde\psi_1\rangle
\end{equation}
For later use, we compute the inner product between $\psi_3$ and $\psi_1$:
\begin{equation}
\langle\psi_3|\psi_1\rangle=\Lambda_3^{-1}(W_4-W_2^\intercal W_1^{-1}W_3)\Lambda_1^{-\intercal}
\label{eq: BB4}
\end{equation}
Each gate in the circuit is an SO($2N_\text f$) matrix. For the specific gate $U$ in \cref{fig:  determinant formula}(a), we write
\begin{equation}
\begin{bmatrix}
\psi_3\\
\psi_4
\end{bmatrix}=\begin{bmatrix}
U_{31} & U_{32}\\
U_{41} & U_{42}
\end{bmatrix}
\begin{bmatrix}
\psi_1\\
\psi_2
\end{bmatrix}
\end{equation}
where each block $U_{ij}$ is an $N_\text f\times N_\text f$ matrix. It follows that
\begin{equation}
\langle\psi_3|\psi_1\rangle=U_{31}^\intercal
\label{eq: BB5}
\end{equation}
Combining \eqref{eq: BB0}--\eqref{eq: BB5} and using standard determinant identities for block matrices, one finds
\begin{equation}
\begin{aligned}
\det(A_{DBC})&=\det(\Lambda_1)^2\det(\Lambda_3)^2\det(A_B)\det(\textbf{1}-U_{31}U_{31}^\intercal)\\
\det(A_{DB})&=\det(\Lambda_3)^2\det(A_B)\\
\det(A_{BC})&=\det(\Lambda_1)^2\det(A_B)\\
\end{aligned}
\end{equation}
Eliminating $\det(\Lambda_{1,3})$, we obtain
\begin{equation}
\det(\textbf{1}-U_{31}U_{31}^\intercal)=\frac{\det(A_{DBC})\det(A_B)}{\det(A_{DB})\det(A_{BC})}
\end{equation}
This is a direct generalization of \eqref{eq: general determinant formula from region}. Indeed, when $N_\text f=1$, $U_{31}=\cos\theta$ is a scalar, and $\det(\textbf{1}-U_{31}U_{31}^\intercal)$ reduces to $\sin^2\theta$.

To elucidate the physical meaning of the factor $\det(\textbf{1}-U_{31}U_{31}^\intercal)$, we adopt an ansatz for the bulk QFT. For multi-flavor fermions, the bulk theory naturally admits the action
\begin{equation}
S=\frac{1}{4\pi}\int dtdz\cdot \i\Psi_L^\alpha\partial_t\Psi_L^\alpha+\i\Psi_R^\alpha\partial_t\Psi_R^\alpha-\i\Psi_L^\alpha\partial_z\Psi_L^\alpha+\i\Psi_R^\alpha\partial_z\Psi_R^\alpha+2\i \hat m_{\alpha\beta}(z,t)\Omega(z,t)\Psi_L^\alpha \Psi_R^\beta
\label{eq: action of multiflavor QFT}
\end{equation}
which is a direct generalization of \eqref{eq: action of QFT}. Here $\hat m(z,t)$ is an $N_\text f\times N_\text f$ real symmetric matrix.

We next relate the discrete-circuit gate data to the bulk QFT. From \eqref{eq: relate gate angle to Omega}, for $N_\text f=1$ and in the small-$\Delta t$ limit we have $\cos\theta\approx(\Delta t)\frac{1}{2}m\Omega(z)$. The corresponding multi-flavor generalization is
\begin{equation}
U_{31}\approx (\Delta t)\frac{1}{2}\hat m(z,t)\Omega(z,t)
\end{equation}
which implies
\begin{equation}
\log\det(\textbf{1}-U_{31}U_{31}^\intercal)=\Tr\log(\textbf{1}-U_{31}U_{31}^\intercal)\approx-\frac14(\Delta t)^2\Tr(\hat m(z,t)^2)\Omega(z,t)^2
\label{eq: BB15}
\end{equation}
As in \cref{sec: A determinant formula for gate angle}, let $A_{[\t_1,\t_2]}$ denote the submatrix of $A(\alpha,\t,\alpha',\t')$ with indices $\t,\t'\in[\t_1,\t_2]$ (a time band). In the continuum limit, this becomes $A_{[t_1,t_2]}$. Moreover, a boundary time band $[t_1,t_2]$ determines a bulk point with lightcone coordinates $(u,v)$ via $t_1=u$ and $t_2=v$, and we therefore also write $A_{[t_1,t_2]}\equiv A_{[u,v]}$. In this notation, the matrices $A_{DBC},A_{DB},A_{BC},A_{B}$ become
\begin{equation}
A_{B}\equiv A_{[u,v]},\ A_{DB}\equiv A_{[u,v+\Delta t]},\ A_{BC}\equiv A_{[u-\Delta t,v]},\ A_{DBC}\equiv A_{[u-\Delta t,v+\Delta t]}
\end{equation}
Consequently,
\begin{equation}
\log\left[\frac{\det(A_{DBC})\det(A_B)}{\det(A_{DB})\det(A_{BC})}\right]\approx-(\Delta t)^2\partial_u\partial_v\log\det(A_{[u,v]})
\end{equation}
Combining this with \eqref{eq: BB15}, we obtain in the continuum limit the determinant formula
\begin{equation}
\Tr(\hat m(z,t)^2)\Omega(z,t)^2=4\partial_u\partial_v\log\det(A_{[u,v]})
\label{eq: mutiflavor det formula}
\end{equation}
This is consistent with \eqref{eq: main result2}, since when $N_\text f=1$ and the setup is time-translation invariant, we have $4\partial_u\partial_v=-\partial_z^2$.

An interesting question is under what conditions one can obtain a spacetime-independent mass matrix, $\hat m(z,t)\equiv\hat m$. Only in this case does it make sense to interpret $\Omega(z,t)$ as a geometry that is experienced universally by all fermion flavors (and only then is ``geometry'' a meaningful notion). This requires the matrix $U_{31}U_{31}^\intercal$ to be proportional across different spacetime locations. It would be interesting to translate this requirement into a condition on the boundary fermions' anticommutator $A(\alpha,t,\alpha',t')$.

\section{Technical details of applying BOGC formula}
\label{app: technical detail of using BOGC formula}
In this appendix we record the details of applying the BOGC formula \eqref{eq: original BOGC formula} to obtain \eqref{eq: F(z)}. We first restate the formula:
\paragraph{Theorem. }(Borodin--Okounkov--Geromino--Case)~\cite{Borodin2000,Geronimo:1979iy}. \textit{Given a $N\times N$ real symmetric Toeplitz matrix $A_N$, with matrix element $(A_N)_{ij}:=f_{|i-j|}, i,j=1,...,N$. View $\{f_n\}$ as Fourier coefficient of $f(\theta):=\sum_{n\in\mathbb Z}f_ne^{in\theta}$. Assume $g(\theta):=\log f(\theta)$ exist everywhere on $\theta\in(0,2\pi]$,  with $\{g_n\}$ its Fourier coefficient given by $g(\theta)=\sum_{n\in\mathbb Z}g_ne^{in\theta}$. Further define $g_+(\theta):=\sum_{n=1}^{+\infty}g_n e^{in\theta},\ g_-(\theta):=\sum_{n=-\infty}^{-1}g_n e^{in\theta}$, and $b^-(\theta):=e^{g_-(\theta)-g_+(\theta)}$, with Fourier coefficient $\{b^-_n\}$ given by $b^-(\theta)=\sum_{n\in\mathbb Z}b^-_ne^{in\theta}$. Now, define a Hankel matrix $B_N$ on half infinite lattice $L^2(\mathbb Z_{\geq0})$ via $(B_N)_{ij}:=b^-_{i+j+N+1}$, $i,j\geq0$. We then have:  }
\begin{equation}
\log\det A_N=Ng_0+(\sum_{k=1}^{\infty}k g_kg_{-k})+\log\det[\textbf{1}-(B_N)^2]
\end{equation}

\hfill

Since this appendix is very technical, we summarize the procedure as follows:
\begin{itemize}
\item \textbf{Step one,} factorize $f(\theta)$ into $e^{g_0}e^{g_+(\theta)}e^{g_-(\theta)}$.
\item \textbf{Step two,} calculate $b^-(\theta)$ and extract its Fourier coefficient $\{b_n^-\}$.
\item \textbf{Step three,} evaluate $\Tr(B_N^{2m})$, since in $\log\det[\textbf{1}-(B_N)^2]=-\sum_{m=1}^{\infty}\frac{1}{m}\Tr(B_N^{2m})$, we  only need the information of trace of the moment of $B_N$.
\item \textbf{Step four,} take continuous limit of $\log\det[\textbf{1}-(B_N)^2]$, following \eqref{eq: limit def of toeplitz}, and obtain \eqref{eq: F(z)}.
\end{itemize}

As a preparation, recall that around~\eqref{eq: main result1}, we assume the anti-commutator two-point function of GFF $A(t,t')$ in upper half frequency plane $\mA(\omega)$ have simple poles at $\{\i J_k\}$. Further assume time reversal symmetry, then $\mA(\omega)$ can be written as:
\begin{equation}
\mA(\omega)=\sum_{k}\frac{c_kJ_k}{J_k^2+\omega^2}
\label{eq: A(omega)}
\end{equation}
where the coefficients $\{c_k\}$ encode the residues. For notational convenience we assume finitely many poles $\{\i J_k\}_{k=1}^{N_\qnm}$; the case of infinitely many poles can be recovered by taking $N_\qnm\to\infty$ at the end.

In the time domain, $A(t,t')$ admits the quasi-normal-mode (QNM) expansion:
\begin{equation}
A(t,t')=\sum_{k=1}^{{N_\qnm}}c_ke^{-J_k|t-t'|}
\label{eq: QNM expansion of A(t,t')}
\end{equation}
Recall that $A_N:=A_{[\t,\t+(N-1)\Delta t]}$ is defined by discretizing time; its Toeplitz entries $(A_N)_{ij}:=f_{|i-j|}$ are:
\begin{equation}
f_n=\sum_{k=1}^{{N_\qnm}}c_ke^{-J_k(\Delta t)n}:=\sum_{k=1}^{{N_\qnm}}c_kr_k^n,\ \ r_k:=e^{-J_k\Delta t}
\label{eq: A4, f_n}
\end{equation}
where we introduced $r_k$ for notational simplicity.

\noindent$\bullet$ \textbf{Step one. }View $\{f_n\}$ as the Fourier coefficients of $f(\theta)=\sum_{n\in\mathbb Z}f_ne^{\i n\theta}$. It is convenient to use the complex variable $\xi:=e^{\i\theta}$:
\begin{equation}
f(\xi):=\sum_{n\in\mathbb Z}f_n\xi^n=\sum_{i=1}^{{N_\qnm}}\frac{c_i(1-r_i^2)}{(1-r_i\xi)(1-r_i\xi^{-1})}=\frac{\sum_{i=1}^{{N_\qnm}}c_i(1-r_i^2)\prod_{j\neq i}(1+r_j^2-r_j(\xi+\xi^{-1}))}{\prod_{i=1}^{N_\qnm}(1-r_i\xi)(1-r_i\xi^{-1})}
\label{eq: A3 f(xi)}
\end{equation}
The next step is to factorize $f(\theta)$ as $e^{g_0}e^{g_+(\theta)}e^{g_-(\theta)}$, equivalently to factorize $f(\xi)$ as:
\begin{equation}
f(\xi)=f_0 f_+(\xi)f_-(\xi),\ \text{ with } g_0=\log f_0,\ g_\pm=\log f_\pm
\label{eq: A4, factorize f(xi)}
\end{equation}
where $f_+(\xi)-1$ contains only non-negative powers of $\xi$ and $f_-(\xi)-1$ contains only non-negative powers of $\xi^{-1}$. In our case, the symmetry $f(\xi)=f(\xi^{-1})$ implies $f_-(\xi)=f_+(\xi^{-1})$. To determine $f_+(\xi)$, we factorize the numerator of \eqref{eq: A3 f(xi)} as a polynomial $P(x)$ in $x:=\frac12(\xi+\xi^{-1})$:
\begin{equation}
P(x):=\sum_{i=1}^{{N_\qnm}}c_i(1-r_i^2)\prod_{j\neq i}(1+r_j^2-2r_jx)
\label{eq: s_i in terms of x_i}
\end{equation}
Then for generic choice of parameter, $P(x)$ is an $({N_\qnm}-1)$ degree polynomial of $x$, with roots $\{x_i\}_{i=1}^{{N_\qnm}-1}$. Therefore $P=C_1\prod_{i=1}^{{N_\qnm}-1}(x-x_i)$ with $C_1$ some constant. Now, we can further factorize $(x-x_i)$ into $-(2s_i)^{-1}(1-s_i\xi)(1-s_i\xi^{-1})$ with \begin{equation}s_i:=x_i-\sqrt{x_i^2-1}.
\label{eq: B8}
\end{equation} Finally, we obtain a factorized re-writing of $f(\xi)$:
\begin{equation}
f(\xi)=f_0\times\frac{\prod_{i=1}^{{N_\qnm}-1}(1-s_i\xi)}{\prod_{i=1}^{{N_\qnm}}(1-r_i\xi)}\times\frac{\prod_{i=1}^{{N_\qnm}-1}(1-s_i\xi^{-1})}{\prod_{i=1}^{{N_\qnm}}(1-r_i\xi^{-1})}
\end{equation}
\begin{equation}
\text{with: } f_0:=(\prod_{i=1}^{{N_\qnm}-1}s_i^{-1})(\prod_{i=1}^{N_\qnm}r_i)(\sum_{i=1}^{N_\qnm}c_i(1-r_i^2)r_i^{-1})
\label{eq: f0}
\end{equation}
Comparing with \eqref{eq: A4, factorize f(xi)} determines $f_\pm(\xi)$.

\noindent$\bullet$ \textbf{Step two. }Compute $b^-(\xi):=f_-(\xi)/f_+(\xi)$ and its Laurent coefficients $\{b_n^-\}$:
\begin{equation}
b^-(\xi)=\sum_{n\in\mathbb Z}b^-_n\xi^n
\end{equation}
One finds:
\begin{equation}
\begin{aligned}
b^-(\xi)&=\prod_{i=1}^{{N_\qnm}-1}\frac{1-s_i\xi^{-1}}{1-s_i\xi}\times \prod_{j=1}^{{N_\qnm}}\frac{1-r_j\xi}{1-r_j\xi^{-1}}=-\left[\frac{\prod_{j=1}^{N_\qnm}r_j}{\prod_{i=1}^{{N_\qnm}-1}s_i}\right]\xi+\sum_{i=1}^{{N_\qnm}-1}\frac{\alpha_i}{1-s_i \xi}+\sum_{j=1}^{{N_\qnm}}\frac{\tilde\alpha_j}{\xi-r_j}
\end{aligned}
\label{eq: A8, b-(xi)}
\end{equation}
with:
\begin{equation}
\alpha_i=(1-s_i^2)\prod_{k=1,k\neq i}^{{N_\qnm}-1}\frac{1-s_ks_i}{1-s_ks_i^{-1}}\times \prod_{j=1}^{{N_\qnm}}\frac{1-r_js_i^{-1}}{1-r_js_i}
\label{eq: alpha_i}
\end{equation}
Since $(B_N)_{ij}=b^-_{i+j+N+1}$ with $i,j\geq0$ and $N\geq1$, we only need $b_n^-$ for $n\geq2$. Expanding $\frac{1}{1-s_i\xi}=\sum_{n\geq0}s_i^n\xi^n$ contains only nonnegative powers, while $\frac{1}{\xi-r_j}=-\sum_{n\geq0}r_j^n\xi^{-n-1}$ contains only negative powers. Hence, in \eqref{eq: A8, b-(xi)} only the second term contributes to $b_n^-$ for $n\geq2$:
\begin{equation}
b^-_n=\sum_{i=1}^{{N_\qnm}-1}\alpha_is_i^n,\ n\geq2
\end{equation}

\noindent$\bullet$ \textbf{Step three. }Construct the Hankel matrix $B_N$ on $L^2(\mathbb Z_{\geq0})$ with entries $(B_N)_{ij}:=b^-_{i+j+N+1}$ for $i,j\geq0$. It is convenient to use the orthonormal basis $\{|i\rangle\}_{i=0}^{\infty}$ with $\lr{i|j}=\delta_{ij}$. Then:
\begin{equation}
(B_N)_{ij}=\sum_{k}^{{N_\qnm}-1}\alpha_k s_k^{N+1}s_k^{i+j}\longrightarrow B_N=\sum_{k=1}^{{N_\qnm}-1}\alpha_ks_{k}^{N+1}|u_k\rangle\langle u_k|,\ \text{with }|u_k\rangle:=\sum_{i=0}^{+\infty}s_k^i|i\rangle
\end{equation}
To compute $\Tr(B_N^{2m})$, define the $(N_\qnm-1)\times (N_\qnm-1)$ Gram matrix $G$ of $\{|u_k\rangle\}$:
\begin{equation}
G_{kl}:=\lr{u_k|u_{l}}=\sum_{i=0}^{+\infty}s_k^is_{l}^i=\frac{1}{1-s_ks_{l}}
\end{equation}
Also define the diagonal matrix $\Lambda$ of the same dimension:
\begin{equation}
\Lambda_{kl}:=\alpha_ks_k^{N+1}\delta_{kl}
\end{equation}
Then:
\begin{equation}
\Tr(B_N^{2m})=\Tr[(G\Lambda)^{2m}]
\end{equation}
It is therefore convenient to define the finite matrix $\tilde B_N$:
\begin{equation}
(\tilde B_N)_{kl}:=(G\Lambda)_{kl}=\frac{\alpha_l s_l^{N+1}}{1-s_ks_l}
\label{eq: A18}
\end{equation}
Then we have:
\begin{equation}
\log\det[\textbf{1}-(B_N)^2]=\log\det[\textbf{1}-(\tilde B_N)^2]
\label{eq: A19}
\end{equation}

\noindent$\bullet$ \textbf{Step four. }We now take the continuum limit \eqref{eq: limit def of toeplitz} of $\tilde B_N$. We first solve for the roots of $P(x)$ at small $\Delta t$. It is convenient to parametrize $x:=e^{\frac12\lambda^2(\Delta t)^2}$. To leading order in $\Delta t$:
\begin{equation}
P(x)=2(\Delta t)^{2{N_\qnm}-1}\times\prod_{i=1}^{{N_\qnm}}(J_i^2-\lambda^2)\times\left[\sum_{i=1}^{{N_\qnm}}\frac{c_iJ_i}{J_i^2-\lambda^2}\right]+O((\Delta t)^{2{N_\qnm}})
\end{equation}
Comparing with \eqref{eq: A(omega)}, the roots of $P(x)$ are determined by the zeros of $\mA(\omega\equiv\i\lambda)$. Thus, if $\mA(\omega)$ has zeros $\{\i\lambda_k\}_{k=1}^{N_\qnm-1}$ in the upper half-plane, then \eqref{eq: s_i in terms of x_i} and \eqref{eq: alpha_i} imply:
\begin{equation}
s_i=1-\lambda_i\Delta t+O((\Delta t)^2), \ \alpha_i=(\Delta t)\times2\lambda_i\times\prod_{k=1,k\neq i}^{{N_\qnm}-1}\frac{\lambda_k+\lambda_i}{\lambda_k-\lambda_i}\times \prod_{j=1}^{{N_\qnm}}\frac{J_j-\lambda_i}{J_j+\lambda_i}+O((\Delta t)^2)
\end{equation}
Therefore, using \eqref{eq: A18}, $\tilde B_N$ has a finite continuum limit:
\begin{equation}
B(z)_{ij}:=\lim_{\eqref{eq: limit def of toeplitz}}(\tilde B_N)_{ij}=\frac{\beta_je^{-\lambda_j2z}}{\lambda_i+\lambda_j}
\end{equation}
which reproduces \eqref{eq: main result1} and \eqref{eq: main result2}. Together with \eqref{eq: A19}, this yields \eqref{eq: F(z)}.

\section{An elementary calculation of determinants in one and two QNM models}
\label{sec: An elementary calculation of determinant of one and two QNM model}
In \cref{app: technical detail of using BOGC formula} we used the BOGC formula to evaluate Toeplitz determinants. Here we benchmark that computation in  simple cases, $N_\qnm=1$, $N_\qnm=2$ and $N_\qnm=3$ in \eqref{eq: QNM expansion of A(t,t')}, and show that the BOGC result (\cref{app: BOGC evaluation of determinant}) agrees with a direct elementary computation of determinant (\cref{app: Elementary evaluation of determinant}).
\subsection{BOGC evaluation of determinant}
\label{app: BOGC evaluation of determinant}
\paragraph{Example: $N_\qnm=1$. }This case is particularly simple. From \eqref{eq: A3 f(xi)} we have the factorization $f(\xi)=c_1(1-r_1^2)\,(1-r_1\xi)^{-1}(1-r_1\xi^{-1})^{-1}$, from which $f_0$ and $f_\pm(\xi)$ follow via \eqref{eq: A4, factorize f(xi)}. The Fourier coefficients satisfy $g_k=g_{-k}=\frac{1}{k}r_1^k$ for $k>0$. Since there is no subleading $\log\det[\textbf{1}-(\tilde B_N)^2]$ term when $N_\qnm=1$, \eqref{eq: original BOGC formula} yields:
\begin{equation}
\det A_N=[c_1(1-r_1^2)]^N(1-r_1^2)^{-1}
\label{eq: det of M=1 QNMs via BOGC}
\end{equation}
As a check, for $N=1$ this gives $\det A_1=c_1$, as expected.

Using \eqref{eq: determinant formula}, one finds $\sin^2\theta_N=1$ for all $N\geq2$, i.e. all gates beyond the first layer are exact SWAPs. Then \eqref{eq: relate gate angle to Omega} implies that in the continuum limit the conformal factor satisfies $m\Omega(z)\equiv0$.

\paragraph{Example: $N_\qnm=2$. }In this case there is a single root $s_1$. It is convenient to introduce two intermediate quantities $Q_1$ and $Q_2$:
\begin{equation}
Q_2=c_1(1-r_1^2)(1+r_2^2)+c_2(1-r_2^2)(1+r_1^2),\ \  Q_1=c_1(1-r_1^2)r_2+c_2(1-r_2^2)r_1
\label{eq: B2, Q1,Q2}
\end{equation}
Then:
\begin{equation}
s_1=\frac{Q_2-\sqrt{Q_2^2-4Q_1^2}}{2Q_1}
\end{equation}
We also record the following intermediate relations:
\begin{equation}
f_0=s_1^{-1}Q_1,\ g_{k}=g_{-k}=\frac1k(r_1^k+r_2^k-s_1^k), k>0
\end{equation}
When ${N_\qnm}=2$, $\tilde B_N$ is a one-dimensional matrix, so $\log\det[\textbf{1}-(\tilde B_N)^2]=\log(1-\tilde B_N^2)$. Substituting these expressions, we find:
\begin{equation}
\det A_N=\frac{(1-r_1s_1)^2(1-r_2s_1)^2\eta_+^N-(1-r_1s_1^{-1})^2(1-r_2s_1^{-1})^2s_1^2\eta_-^N}{(1-r_1^2)(1-r_2^2)(1-r_1r_2)^2(1-s_1^2)},\ \eta_\pm:=\frac{Q_2\pm\sqrt{Q_2^2-4Q_1^2}}{2}
\label{eq: B5, eta}
\end{equation}
A quick check is to evaluate $N=1,2$ and verify (e.g. in Mathematica) agreement with the elementary results $\det A_1=c_1+c_2$ and $\det A_2=(c_1+c_2)^2-(c_1r_1+c_2r_2)^2$.
\paragraph{Example: $N_\qnm=3$.} In this case there are two roots, $s_1$ and $s_2$. It is convenient to introduce three intermediate quantities $Q_1,Q_2,Q_3$:
\begin{equation}
\begin{aligned}
\small
Q_3&:=(r_1r_2r_3)(\sum_{i=1}^{3}c_i(1-r_i^2)r_i^{-1})\\
Q_2&:=-c_1(1-r_1^2)(r_2+r_3)(1+r_2r_3)-c_2(1-r_2^2)(r_1+r_3)(1+r_1r_3)-c_3(1-r_3^2)(r_1+r_2)(1+r_1r_2)\\
Q_1&:=2Q_3+(\prod_{i=1}^{3}1+r_i^2)(\sum_{i=1}^{3}\frac{c_i(1-r_i^2)}{1+r_i^2})
\end{aligned}
\label{eq: D6}
\end{equation}
Then \eqref{eq: s_i in terms of x_i} reduces to a quadratic polynomial $P(x)=4Q_3x^2+2Q_2x+Q_1-2Q_3$, with roots $x_1$ and $x_2$. Using \eqref{eq: B8} we then obtain $s_1$ and $s_2$, and from \eqref{eq: f0} we read off $f_0=s_1^{-1}s_2^{-1}Q_3$. For $N_\qnm=3$, $\tilde B_N$ is a two-dimensional matrix. Collecting these results and applying the BOGC formula \eqref{eq: original BOGC formula}, we find
\begin{equation}
\det A_N=\sum_{i=0}^{4}\texttt{a}_i\eta_i^N
\end{equation}
with
\begin{equation}
\eta_i\in\{f_0,\ f_0s_1s_2,\ f_0s_1^2s_2^2,\ f_0s_1^2,\ f_0s_2^2\}.
\label{eq: D8 eta}
\end{equation}
Here $\texttt{a}_i$ are coefficients independent of $N$. Equivalently, $\det A_N$ is a sum of five terms, each exponential in $N$. Later we will match these exponents $\eta_i$ against a direct (elementary) computation of the determinant.

\subsection{Elementary evaluation of determinant}
\label{app: Elementary evaluation of determinant}
\paragraph{Example: ${N_\qnm}=1$ case. }The matrix $A_N$ is explicitly
\begin{equation}
\small
A_{N+1}=c_1^{N+1}\times\begin{bmatrix}
1 & r_1 & r_1^2 & \cdots &r_1^N\\
r_1 & 1 & r_1 &\cdots & r_1^{N-1}\\
r_1^2 & r_1 & 1 &\cdots &r_1^{N-2}\\
\vdots &\vdots &\vdots & & \vdots\\
r_1^N & r_1^{N-1} & r_1^{N-2}& \cdots &1
\end{bmatrix}_{N+1},\ \ r_1:=e^{-J_1\Delta t}
\end{equation}
The subindex $[\cdots]_{N+1}$ on the matrix braket serves as a  reminder  of the dimension of that matrix.  Using the identity
\begin{equation}
\small
\det\begin{bmatrix}
1 & r_1 & r_1^2 & \cdots &r_1^N\\
r_1 & 1 & r_1 &\cdots & r_1^{N-1}\\
r_1^2 & r_1 & 1 &\cdots &r_1^{N-2}\\
\vdots &\vdots &\vdots & & \vdots\\
r_1^N & r_1^{N-1} & r_1^{N-2}& \cdots &1
\end{bmatrix}_{N+1}=
\det\begin{bmatrix}
1-r_1^2 & r_1 & r_1^2 & \cdots &r_1^N\\
0 & 1 & r_1 &\cdots & r_1^{N-1}\\
0 & r_1 & 1 &\cdots &r_1^{N-2}\\
\vdots &\vdots &\vdots & & \vdots\\
0 & r_1^{N-1} & r_1^{N-2}& \cdots &1
\end{bmatrix}_{N+1}
\end{equation}
obtained by adding $\left[(-r_1)\times 2^{\text{nd}}\text{\ column}\right]$ to the $1^{\text{st}}$ column (which does not change the determinant by multilinearity), we obtain the recursion
\begin{equation}
\det A_{N+1}=c_1(1-r_1^2)\det A_{N}.
\end{equation}
Matching the initial condition at $N=1$, this recursion reproduces the BOGC result \eqref{eq: det of M=1 QNMs via BOGC}.
\paragraph{Example: ${N_\qnm}=2$ case. }Similarly, we reduce the determinant as follows:
\begin{equation}
\small
\begin{aligned}
\det(A_{N+1})&=\det\begin{bmatrix}
f_0 & f_1 & f_2 & \cdots &f_N\\
f_1 & f_0 & f_1 &\cdots & f_{N-1}\\
f_2 & f_1 & f_0 &\cdots &f_{N-2}\\
\vdots &\vdots &\vdots & & \vdots\\
f_N & f_{N-1} & f_{N-2}& \cdots &f_0
\end{bmatrix}_{N+1}
\stackrel{(1)}{=}\det
\begin{bmatrix}
Q_3 & f_1 & f_2 & \cdots &f_N\\
Q_1 & f_0 & f_1 &\cdots & f_{N-1}\\
0 & f_1 & f_0 &\cdots &f_{N-2}\\
\vdots &\vdots &\vdots & & \vdots\\
0 & f_{N-1} & f_{N-2}& \cdots &f_0
\end{bmatrix}_{N+1}\\
&\stackrel{(2)}{=}\det\begin{bmatrix}
Q_2 & Q_1 & 0 & \cdots &0\\
Q_1 & f_0 & f_1 &\cdots & f_{N-1}\\
0 & f_1 & f_0 &\cdots &f_{N-2}\\
\vdots &\vdots &\vdots & & \vdots\\
0 & f_{N-1} & f_{N-2}& \cdots &f_0
\end{bmatrix}_{N+1}
\end{aligned}
\label{eq: B9}
\end{equation}
In step $(1)$ we add $\left[(-d_1)\times 2^{\text{nd}}\text{\ column}+(-d_2)\times 3^{\text{rd}}\text{\ column}\right]$ to the $1^{\text{st}}$ column, where $d_1,d_2$ are to be determined. In step $(2)$ we add $\left[(-d_1)\times 2^{\text{nd}}\text{\ row}+(-d_2)\times 3^{\text{rd}}\text{\ row}\right]$ to the $1^{\text{st}}$ row. The quantities $Q_1,Q_2,Q_3$ are given by $Q_3=f_0-d_1f_1-d_2f_2$, $Q_1=f_1-d_1f_0-d_2f_1$, and $Q_2=Q_3-d_1Q_1$.

The coefficients $d_1,d_2$ are fixed by
\begin{equation}
f_{k}=d_1f_{k-1}+d_2f_{k-2},\forall k>0.
\end{equation}
For ${N_\qnm}=2$ with $f_k$ given in \eqref{eq: A4, f_n}, we find
\begin{equation}
d_1=r_1+r_2, \ d_2=-r_1r_2.
\end{equation}
Moreover, $Q_1$ and $Q_2$ coincide with the intermediate quantities introduced in the BOGC derivation, \eqref{eq: B2, Q1,Q2}.

We now evaluate the determinant in the second line of \eqref{eq: B9}:
\begin{equation}
\det(A_{N+1})=Q_2\det\begin{bmatrix}
f_0 & f_1  & \cdots &f_{N-1}\\
f_1 & f_0  &\cdots & f_{N-1}\\
\vdots &\vdots & & \vdots\\
f_{N-1} & f_{N-2} & \cdots &f_0
\end{bmatrix}_N
-Q_1\det\begin{bmatrix}
Q_1 & 0 & \cdots &0\\
f_1 & f_0  &\cdots & f_{N-1}\\
\vdots &\vdots & & \vdots\\
f_{N-1} & f_{N-2} & \cdots &f_0
\end{bmatrix}_N.
\end{equation}
This yields the recursion
\begin{equation}
\det(A_{N+1})=Q_2\det(A_{N})-Q_1^2\det(A_{N-1}).
\end{equation}
The characteristic roots are $\eta_\pm$ defined in \eqref{eq: B5, eta}. Matching the initial conditions at $N=1,2$, the solution of this recursion agrees with the BOGC result \eqref{eq: B5, eta}. 

\paragraph{Example: the $N_\qnm=3$ case.} Using the same procedure as above, we reduce the determinant to a finitely banded matrix:

{\small
\begin{equation}
\textstyle
\begin{aligned}
\det(A_{N+1})&=\det\begin{bmatrix}
f_0 & f_1 & f_2 & \cdots &f_N\\
f_1 & f_0 & f_1 &\cdots & f_{N-1}\\
f_2 & f_1 & f_0 &\cdots &f_{N-2}\\
\vdots & & & & \vdots\\
f_N & f_{N-1} & f_{N-2}& \cdots &f_0
\end{bmatrix}_{N+1}
\stackrel{(1)}{=}\det
\begin{bmatrix}
Q_4 & f_1 & f_2 & \cdots &f_N\\
Q_5 & f_0 & f_1 &\cdots & f_{N-1}\\
Q_3 & f_1 & f_0 &\cdots &f_{N-2}\\
\vdots & & & & \vdots\\
0 & f_{N-1} & f_{N-2}& \cdots &f_0
\end{bmatrix}_{N+1}\\
&\stackrel{(2)}{=}\det\begin{bmatrix}
Q_1 & Q_5 & Q_3 & \  &\ \\
Q_5 & f_0 & f_1 &\cdots & f_{N-1}\\
Q_3 & f_1 & f_0 &\cdots &f_{N-2}\\
\  &\vdots & & & \vdots\\
\  & f_{N-1} & f_{N-2}& \cdots &f_0
\end{bmatrix}_{N+1}\stackrel{(3)}{=}\det\begin{bmatrix}
Q_1 & Q_2 & Q_3 & \ & \ & \ &\ \\
Q_2 & Q_1 & Q_5& Q_3 & \ & \ &\ \\
Q_3 & Q_5 & f_0 & f_1 & f_2 &\cdots & f_{N-2}\\
\  & Q_3 & f_1 & f_0 & f_1 & \cdots & f_{N-3}\\
\ & \ & f_2 & f_1 & f_0 & \cdots & f_{N-4}\\
\ & \ & \vdots & \ & \ & \ & \vdots \\
\ & \ & f_{N-2} & f_{N-3} & f_{N-4} & \cdots & f_0\\
\end{bmatrix}_{N+1}\\
&=\cdots=\det\begin{bmatrix}
Q_1  & Q_2 & Q_3 &  &  &  &  &  &  \\
 Q_2 & Q_1 & Q_2 & Q_3 &  &  &  &  &  \\
 Q_3 & Q_2 & \ddots & \ddots & \ddots &  &  &  &    \\
  & Q_3 & \ddots &  &  &  &  &  &    \\
 &  & \ddots &  & Q_1 & Q_2 & Q_3 &  &    \\
  &  &  &  & Q_2 & Q_1 & Q_5 & Q_3 &    \\
  &  &  &  & Q_3 & Q_5 & f_0 & f_1 & f_2  \\
  &  &  &  &  & Q_3 & f_1 & f_0 & f_1   \\
  &  &  &  &  &  & f_2 & f_1 & f_0 \\
\end{bmatrix}_{N+1}
\label{eq: D17}
\end{aligned}
\end{equation}}
In step (1), we add $\left[(-d_1)\times 2^{\text{nd}}\text{\ column}+(-d_2)\times 3^{\text{rd}}\text{\ column}+(-d_3)\times 4^{\text{th}}\text{\ column}\right]$ to the $1^{\text{st}}$ column, where $d_1,d_2,d_3$ are parameters to be fixed. In step (2), we add $\left[(-d_1)\times 2^{\text{nd}}\text{\ row}+(-d_2)\times 3^{\text{rd}}\text{\ row}+(-d_3)\times 4^{\text{th}}\text{\ row}\right]$ to the $1^{\text{st}}$ row. These operations set most entries in the first row and first column to zero.
In step (3), we iterate the same column and row operations to eliminate most entries in the second row and second column. Repeating this procedure reduces the matrix to a five-banded form. The coefficients $d_1,d_2,d_3$ are determined by the following relation:
\begin{equation}
f_{k}=d_1f_{k-1}+d_2f_{k-2}+d_{3}f_{k-3},\forall k
\end{equation}
For the three-QNM case, using $f_k$ defined in \eqref{eq: A4, f_n}, we find:
\begin{equation}
d_1=r_1+r_2+r_3,\ d_2=-r_1r_2-r_1r_3-r_2r_3,\ d_3=r_1r_2r_3
\end{equation}
The five parameters $Q_{1\sim5}$ are given by:
\begin{equation}
\begin{aligned}
Q_4=f_0-d_1f_1-d_2f_2-d_3f_3,\quad Q_5=f_1-d_1f_0-d_2f_1-d_3f_2 \\ 
Q_3=f_2-d_1f_1-d_2f_0-d_3f_1,\quad Q_2=Q_5-d_1Q_3,\quad Q_1=Q_4-d_1Q_5-d_2Q_3
\end{aligned}
\label{eq: D20}
\end{equation}
In what follows, we will only need $Q_{1,2,3}$, since they determine the determinant recursion relation. One can directly verify that $Q_{1,2,3}$ defined in \eqref{eq: D20} agree with the intermediate variables in \eqref{eq: D6} used in the BOGC method.

We are now ready to derive the determinant recursion relation for the five-banded matrix in the last line of \eqref{eq: D17}:
{\small
\begin{equation}
\textstyle
\begin{aligned}
\det A_{N+1}&=
\det\begin{bmatrix}
Q_1  & Q_2 & Q_3 &  &  &    \\
 Q_2 & Q_1 & Q_2 & Q_3 &  &    \\
 Q_3 & Q_2 & Q_1 & Q_2 & Q_3 &      \\
  & Q_3 & Q_2 & Q_1 & Q_2 & Q_3     \\
 &  &  &  & \ddots  &      \\
\end{bmatrix}_{N+1}\\
&=Q_1\det A_{N}-Q_2\det
\begin{bmatrix}
 Q_2 & Q_3 &  &  &    \\
 Q_2 & Q_1 & Q_2 & Q_3 &      \\
 Q_3 & Q_2 & Q_1 & Q_2 & Q_3 &      \\
  & Q_3 & Q_2 & Q_1 & Q_2 & Q_3     \\
 &  &  &  & \ddots  &      \\
\end{bmatrix}_{N}+Q_3\det
\begin{bmatrix}
 Q_2 & Q_3 &  &  &    \\
 Q_1 & Q_2 & Q_3 &  &    \\
 Q_3 & Q_2 & Q_1 & Q_2 & Q_3 &      \\
  & Q_3 & Q_2 & Q_1 & Q_2 & Q_3     \\
  & & Q_3 & Q_2 & Q_1 & Q_2 & Q_3     \\
 & &  &  &  & \ddots  &      \\
\end{bmatrix}_{N}\\
&:= Q_1\det A_N-Q_2\det E^{(1)}_N+Q_3\det E^{(2)}_N
\end{aligned}
\end{equation}
}In the last line, we introduced two new matrices $E^{(1)}_N$ and $E^{(2)}_N$ as intermediate variables. Note that the matrix subscripts match the corresponding matrix dimensions. We next derive two additional relations for $E^{(1)}_N$ and $E^{(2)}_N$:
{\small
\begin{equation}
\textstyle
\begin{aligned}
\det E^{(1)}_N&=\det\begin{bmatrix}
 Q_2 & Q_3 &  &  &    \\
 Q_2 & Q_1 & Q_2 & Q_3 &      \\
 Q_3 & Q_2 & Q_1 & Q_2 & Q_3 &      \\
  & Q_3 & Q_2 & Q_1 & Q_2 & Q_3     \\
 &  &  &  & \ddots  &      \\
\end{bmatrix}_{N}=Q_2\det A_{N-1}-Q_3\det
\begin{bmatrix}
 Q_2  & Q_2 & Q_3 &      \\
 Q_3  & Q_1 & Q_2 & Q_3 &      \\
   & Q_2 & Q_1 & Q_2 & Q_3     \\
 &Q_3 & Q_2 & Q_1 & Q_2 & Q_3 &      \\
 & & Q_3 & Q_2 & Q_1 & Q_2 & Q_3     \\
&  &  &  &  & \ddots  &      \\
\end{bmatrix}_{N-1}\\
&=Q_2\det A_{N-1}-Q_3\det E^{(1)\intercal}_{N-1}\\
&=Q_2\det A_{N-1}-Q_3\det E^{(1)}_{N-1}
\end{aligned}
\end{equation}}and:
\begin{equation}
\begin{aligned}
\det E^{(2)}_N&=\begin{bmatrix}
 Q_2 & Q_3 &  &  &    \\
 Q_1 & Q_2 & Q_3 &  &    \\
 Q_3 & Q_2 & Q_1 & Q_2 & Q_3 &      \\
  & Q_3 & Q_2 & Q_1 & Q_2 & Q_3     \\
  & & Q_3 & Q_2 & Q_1 & Q_2 & Q_3     \\
 & &  &  &  & \ddots  &      \\
\end{bmatrix}_{N}\\
&=Q_2\det\begin{bmatrix}
 Q_2 & Q_3 &  &  &    \\
 Q_2 & Q_1 & Q_2 & Q_3 &      \\
 Q_3 & Q_2 & Q_1 & Q_2 & Q_3 &      \\
  & Q_3 & Q_2 & Q_1 & Q_2 & Q_3     \\
 &  &  &  & \ddots  &      \\
\end{bmatrix}_{N-1}-Q_3\det
\begin{bmatrix}
 Q_1  & Q_3 &  &    \\
 Q_3  & Q_1 & Q_2 & Q_3 &      \\
   & Q_2 & Q_1 & Q_2 & Q_3     \\
   & Q_3 & Q_2 & Q_1 & Q_2 & Q_3     \\
  &  &  &  & \ddots  &      \\
\end{bmatrix}_{N-1}\\
&=Q_2\det E^{(1)}_{N-1}-Q_3(Q_1\det A_{N-2}-Q_3^2\det A_{N-3})
\end{aligned}
\end{equation}
To summarize, we obtain the following coupled recursion relation:
\begin{equation}
\begin{aligned}
\det A_{N+1}&=Q_1\det A_N-Q_2\det E^{(1)}_{N}+Q_3\det E^{(2)}_{N}\\
\det E^{(1)}_N&=Q_2 \det A_{N-1}-Q_3\det E^{(1)}_{N-1}\\
\det E^{(2)}_N&=Q_2\det E^{(1)}_{N-1}-Q_3(Q_1\det A_{N-2}-Q_3^2\det A_{N-3})
\end{aligned}
\end{equation}
We first use the third equation to eliminate $\det E^{(2)}_N$, and then use the second equation to eliminate $\det A_N$. This yields a recursion relation involving only $\det E^{(1)}_N$:
\begin{equation}
\begin{aligned}
\det E^{(1)}_{N+1}&=(Q_1-Q_3)\det E^{(1)}_{N}+(Q_1Q_3-Q_2^2)\det E^{(1)}_{N-1} \\
&\ \ \ \ +(Q_2^2Q_3-Q_1Q_3^2)\det E^{(1)}_{N-2}+(Q_3^4-Q_1Q_3^3)\det E^{(1)}_{N-3}+Q_3^5\det E^{(1)}_{N-4}
\end{aligned}
\end{equation}
The characteristic equation yields five roots $\eta_{0,1,2,3,4}$, given by:
\begin{equation}
\begin{aligned}
\eta_0&=Q_3\\
\eta_{1\sim4}&=\frac{1}{4} \bigg(Q_1-2 Q_3\pm\sqrt{(Q_1-2 Q_2+2 Q_3) (Q_1+2 Q_2+2 Q_3)}\\&\ \ \ \ \ \ \ \ \ \ \ \ \ \ \ \ \ \ \ -\sqrt{2} \sqrt{Q_1^2-2 Q_2^2-4 Q_3^2\pm(Q_1-2 Q_3) \sqrt{(Q_1-2 Q_2+2 Q_3) (Q_1+2 Q_2+2 Q_3)}}\bigg)\\
\end{aligned}
\end{equation}
A symbolic computation in Mathematica confirms that these five roots $\eta_i$ coincide with those obtained via the BOGC method in \eqref{eq: D8 eta}.

\section{An alternative inverse formula via contour integral of S-matrix}
\label{app: An alternative inverse formula via contour integral of S-matrix}
In this appendix we derive \eqref{eq: main result3, contour integral of S} and \eqref{eq: main result4, contour integral of S}, which rewrite the inversion formula \eqref{eq: main result2} directly in terms of contour integrals of the S-matrix. We follow techniques developed by Eniceicu~\cite{Eniceicu:2023uvd}, in the context of giant graviton expansion of unitary matrix integral, where an equivalent reformulation of the BOGC formula is obtained via contour integrals:
\paragraph{Corollary. }(Eniceicu)~\cite{Eniceicu:2023uvd}. \textit{We use the same set up and notation as in BOGC formula \eqref{eq: original BOGC formula}. The subleading part of determinant can be re-written in terms of contour integral:}
\begin{equation}
\log\det\left[\textbf{1}-(B_N)^2\right]=\log\left[1+\sum_{k=1}^{+\infty}G^{(k)}_N\right]
\end{equation}
\begin{equation}
G_N^{(k)}=\frac{(-1)^k}{(k!)^2}\oint\prod_{i=1}^{k}\frac{dp_i}{2\pi \i}\frac{d\tilde p_i}{2\pi \i}\times\frac{\prod_{1\leq i<j\leq k}(p_i-p_j)^2(\tilde p_i-\tilde p_j)^2}{\prod_{1\leq i,j\leq k}(p_i-\tilde p_j)^2}\times\prod_{i=1}^{k}(\frac{p_i}{\tilde p_i})^N\times \prod_{i=1}^{k}\frac{b_-(\tilde p_i)}{b_-(p_i)}
\end{equation}
\textit{where the integral contour  for $\{p_i\}$ is a counter-clockwise circle slightly smaller than unit circle and  for $\{\tilde p_i\}$ is a counter-clockwise circle slightly larger than unit circle.}

\hfill

As in \cref{app: technical detail of using BOGC formula}, we now take the scaling limit \eqref{eq: limit def of toeplitz}. Recalling $b^-(\xi)$ from \eqref{eq: A8, b-(xi)} together with $s_k=1-\Delta t\,\lambda_k$ and $r_k=e^{-J_k\Delta t}$, and parametrizing $\xi=e^{\i\Delta t\,\omega}$, one finds that the continuum limit of $b^-(\xi)$ becomes precisely the S-matrix in \eqref{eq: S(omega) decomposed into poles and zeros}:
\begin{equation}
b^-(\xi)=\prod_{i}\frac{1-s_i\xi^{-1}}{1-s_i\xi}\times \prod_{j}\frac{1-r_j\xi}{1-r_j\xi^{-1}}=\prod_k\frac{\lambda_k+\i \omega}{\lambda_k-\i\omega}\times\prod_n\frac{J_n-\i\omega}{J_n+\i\omega}:=[\mS(\omega)]^{-1}
\end{equation}
Next set $p_i:=e^{\i\Delta t\,\omega_i}$ and $\tilde p_i:=e^{\i\Delta t\,\tilde\omega_i}$. Then \eqref{eq: limit def of toeplitz} gives $(p_i/\tilde p_i)^N=e^{\i2z(\omega_i-\tilde\omega_i)}$, so the $N$-dependence becomes $z$-dependence. The remaining $\Delta t$ factors cancel: the Jacobian contributes $(\i\Delta t)^{2k}$, the Vandermonde in the numerator gives $(\i\Delta t)^{4\cdot \frac{k(k-1)}{2}}$, and the denominator gives $(\i\Delta t)^{2k^2}$. Thus $G_N^{(k)}$ has a finite $O(1)$ continuum limit as $\Delta t\rightarrow0$, which we denote by $G^{(k)}(z)$:
\begin{equation}
G^{(k)}(z):=\frac{1}{(k!)^2}\int\prod_{i=1}^{k}\frac{d\omega_i}{2\pi}\frac{d\tilde\omega_i}{2\pi}\times \frac{\prod_{1\leq i<j\leq k}(\omega_i-\omega_j)^2(\tilde\omega_i-\tilde\omega_j)^2}{\prod_{1\leq i,j\leq k}(\omega_i-\tilde\omega_j)^2}\times\prod_{i=1}^{k}e^{\i 2z(\omega_i-\tilde\omega_i)}\times \prod_{i=1}^{k}\frac{\mS( \omega_i)}{\mS(\tilde \omega_i)}
\end{equation}
where the contours for $\omega_i$ and $\tilde\omega_i$ are inherited from those of $p_i$ and $\tilde p_i$. This reproduces \eqref{eq: main result4, contour integral of S}.

As a check, we can extract the leading contribution to $m\Omega(z)$ from \eqref{eq: main result3, contour integral of S}:
\begin{equation}
m^2\Omega(z)^2\approx-\partial_z^2G^{(1)}(z)
\end{equation}
with:
\begin{equation}
G^{(1)}(z)=\int\frac{d\omega}{2\pi}\frac{d\tilde \omega}{2\pi}\frac{e^{\i 2z(\omega-\tilde \omega)}}{(\omega-\tilde\omega)^2}\frac{\mS(\omega)}{\mS(\tilde\omega)}
\label{eq: C6}
\end{equation}
Therefore:
\begin{equation}
-\partial_z^2G^{(1)}(z)=\left[\int\frac{d\omega}{2\pi}e^{\i 2z\omega}\mS(\omega)\right]\times\left[\int\frac{d\tilde\omega}{2\pi}e^{-\i 2z\tilde\omega}\mS(-\tilde\omega)\right]=S(2z)\times S(2z)
\end{equation}
with $S(z)$ being S-matrix in real space defined in \eqref{eq: S(z)}. This predicts:
\begin{equation}
m\Omega(z)\approx S(2z)
\label{eq: B8}
\end{equation}
which is indeed the first term in the `zig-zag' formula \eqref{eq: forward scattering, zig-zag} describing forward scattering.

One can further check the equivalence between \eqref{eq: main result2} and \eqref{eq: main result3, contour integral of S}--\eqref{eq: main result4, contour integral of S} by expanding both sides:
\begin{equation}
\begin{cases}
&\log\det[\textbf{1}-B(z)^2]=-\Tr[B(z)^2]-\frac12\Tr[B(z)^4]+\cdots\\
&\log[1+\sum_{k=1}^{+\infty}G^{(k)}(z)]=G^{(1)}(z)+G^{(2)}(z)-\frac12G^{(1)}(z)G^{(1)}(z)+\cdots
\end{cases}
\end{equation}
At leading order, one finds $G^{(1)}(z)=-\Tr[B(z)^2]$. This follows directly from \eqref{eq: C6} by evaluating the frequency integrals via residues:
\begin{equation}
G^{(1)}(z)=-\sum_{i,j}\frac{\beta_i\beta_je^{-(\lambda_i+\lambda_j)2z}}{(\lambda_i+\lambda_i)^2}
\end{equation}
which matches $-\Tr[B(z)^2]$ by the definition of $B(z)$ in \eqref{eq: main result2}. At next-to-leading order, one similarly verifies $G^{(1)}(z)G^{(1)}(z)-2G^{(2)}(z)=\Tr[B(z)^4]$ by evaluating the contour integral for $G^{(2)}(z)$:
\begin{equation}
G^{(2)}(z)=\frac14\sum_{ijkl}\frac{(\lambda_i-\lambda_j)^2(\lambda_k-\lambda_l)^2\beta_i\beta_j\beta_k\beta_le^{-(\lambda_i+\lambda_j+\lambda_k+\lambda_l)2z}}{(\lambda_i+\lambda_k)^2(\lambda_j+\lambda_l)^2(\lambda_i+\lambda_l)^2(\lambda_j+\lambda_k)^2}
\end{equation}
Plug in the expression of $G^{(1)}, G^{(2)}$, one can indeed check their combination gives $\Tr[B(z)^4]$:
\begin{equation}
\Tr[B(z)^4]=\sum_{ijkl}\frac{\beta_i\beta_j\beta_k\beta_le^{-(\lambda_i+\lambda_j+\lambda_k+\lambda_l)2z}}{(\lambda_i+\lambda_k)(\lambda_j+\lambda_l)(\lambda_i+\lambda_l)(\lambda_j+\lambda_k)}
\end{equation}

\section{Match with bulk QFT}
\label{app: Matching with bulk QFT}
In this appendix we discuss several aspects of the bulk QFT of a (1+1)d Majorana field in the geometry \eqref{eq: conformal coordinate}. In \cref{app: Derive boundary condition} we derive the bulk QFT boundary condition from the Nebabu--Qi circuit. In \cref{app: Causal kernel decomposition from bulk QFT} we provide an alternative derivation of the causal-kernel decomposition discussed in \cref{sec: Causal kernel decomposition and horizon modes}. Using canonical quantization, we relate the causal kernel to the J\"ost function~\cite{Festuccia:2005pi,Festuccia:2008zx,Dodelson:2023vrw}. Finally, in \cref{app: Two QNMs model from bulk QFT} we address the question raised in \cref{sec: Two QNMs}: why the AdS$_2$--Rindler two-QNM geometry admits a bulk solution with only finitely many quasinormal modes.

\begin{figure}[t]
    \centering
    \includegraphics[width=0.85\linewidth]{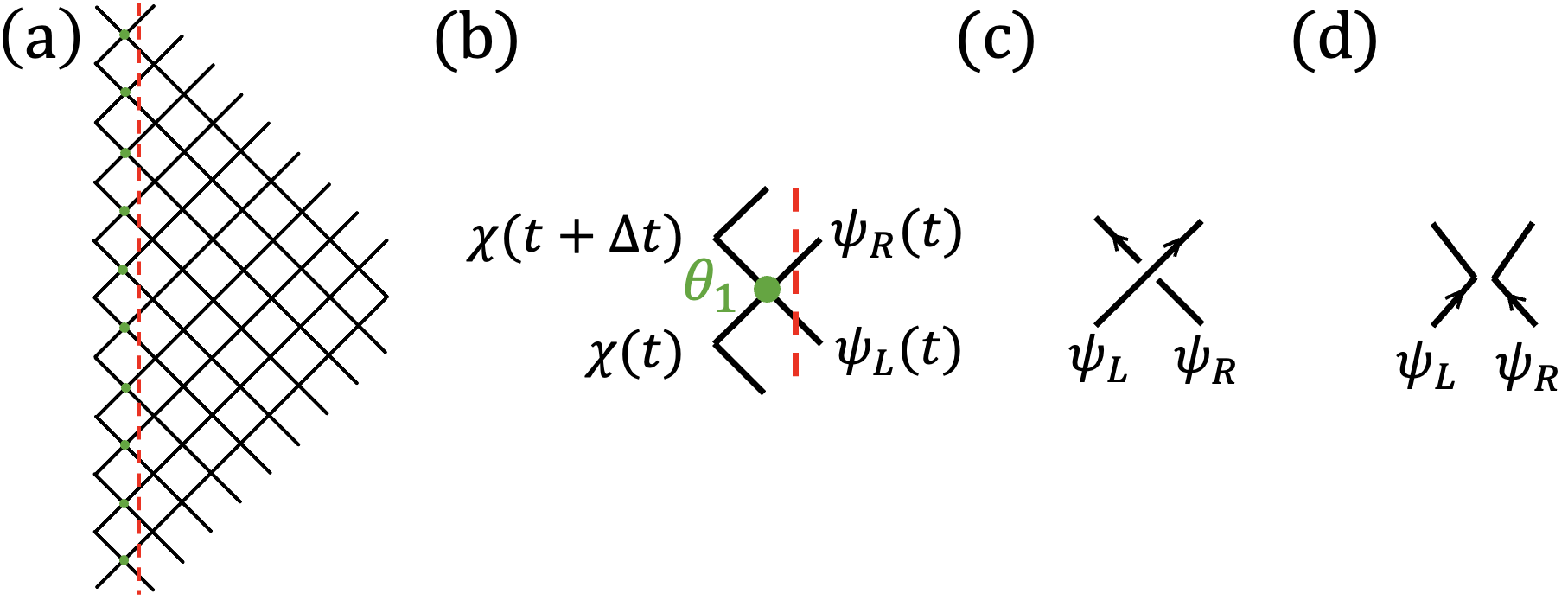}
    \caption{\textbf{(a)} An illustration of the Nebabu--Qi circuit near the boundary. The first-layer gates are marked by green dots. The red dashed line indicates the location immediately after the first layer; this corresponds to $z=0$ in the continuum limit. \textbf{(b)} Zoom-in on a representative first-layer gate from panel (a). \textbf{(c)} An almost-SWAP gate. \textbf{(d)} A gate that almost reflects an incoming left/right mover.}
    \label{fig: after first layer}
\end{figure}
\subsection{Boundary condition}
\label{app: Derive boundary condition}
From the relation between the gate angle and the conformal factor in \eqref{eq: relate gate angle to Omega}, gates in the bulk approach SWAP with $\theta_{N}\sim\frac{\pi}{2}-O(\Delta t)$. Thus $\psi_L$ and $\psi_R$ interact only weakly and essentially pass through each other; see \cref{fig: after first layer}(c). This statement holds for gates starting from the second layer ($N\geq2$). The first-layer gate (green dots in \cref{fig: after first layer}(a)) is special: its angle scales as $\theta_1\sim O(\sqrt{\Delta t})$, so the incoming $\psi_L$ and $\psi_R$ are almost reflected; see \cref{fig: after first layer}(d). The bulk field immediately after the first layer (red dashed line in \cref{fig: after first layer}(a)) sets the boundary condition for the bulk QFT, which we now derive.

Consider the region near the first-layer gate in \cref{fig: after first layer}(b). From the definition of the gate matrix in \eqref{eq: definition of gate matrix},
\begin{equation}
\psi_L(t)=(\cot\theta_1)\chi(t)-(\sin\theta_1)^{-1}\chi(t+\Delta t),\ \psi_R(t)=-(\cot\theta_1)\chi(t+\Delta t)+(\sin\theta_1)^{-1}\chi(t)
\end{equation}
The angle $\theta_1$ is fixed by requiring that $\psi_R(t)$ is normalized and orthogonal to $\chi(t+\Delta t)$, which gives
\begin{equation}
\theta_1=\sqrt{2\mu\Delta t}+O(\Delta t),\ \ \text{with: } \mu:=-\frac{\partial_tA(0,t)}{A(0,0)}\bigg|_{t\rightarrow0^+}>0
\label{eq: definition of mu}
\end{equation}
Expanding at small $\Delta t$, we find
\begin{equation}
\psi_L=\sqrt{\frac{\Delta t}{2\mu}}(-\partial_t-\mu)\chi,\ \ \psi_R=\sqrt{\frac{\Delta t}{2\mu}}(-\partial_t+\mu)\chi
\end{equation}
Using the relation between the circuit fermions $\psi_a$ and the bulk-QFT fermions $\Psi_a$ in \eqref{eq: relation between circuit fermion and QFT fermion}, we obtain
\begin{equation}
\Psi_L=\frac{1}{\sqrt{2\mu}}(-\partial_t-\mu)\chi,\ \ \Psi_R=\frac{1}{\sqrt{2\mu}}(-\partial_t+\mu)\chi
\label{eq: relate Psi with chi}
\end{equation}
Eliminating $\chi$, we arrive at the boundary condition for $\Psi$ at $z=0$:
\begin{equation}
(\partial_t-\mu)\Psi_L=(\partial_t+\mu)\Psi_R,\ \ z=0
\label{eq: boundary condition}
\end{equation}

\subsection{Causal kernel decomposition from bulk QFT}
\label{app: Causal kernel decomposition from bulk QFT}
In this section we canonically quantize the massive (1+1)d Majorana field in the background \eqref{eq: conformal coordinate}, with action~\eqref{eq: action of QFT}. For convenience, we repeat the action:
\begin{equation}
S=\frac{1}{4\pi}\int dtdz\cdot \i\Psi_L\partial_t\Psi_L+\i\Psi_R\partial_t\Psi_R-\i\Psi_L\partial_z\Psi_L+\i\Psi_R\partial_z\Psi_R+2\i m\Omega(z)\Psi_L \Psi_R
\label{eq: action of QFT, in app}
\end{equation}
We use the Fourier transform $\Psi(t,z)=\int d\omega\, e^{\i\omega t}\Psi(\omega,z)$ with reality condition $\Psi_a^\dagger(\omega,z)=\Psi_a(-\omega,z)$. The equation of motion in frequency space is
\begin{equation}
\begin{bmatrix}
\partial_z & -m\Omega(z)\\
m\Omega(z) & -\partial_z
\end{bmatrix}\begin{bmatrix}
\Psi_L\\
\Psi_R
\end{bmatrix}
=\i\omega\begin{bmatrix}
\Psi_L\\
\Psi_R
\end{bmatrix}
\label{eq: eom of QFT in frequency space}
\end{equation}
There are two linearly independent mode solutions. We denote the incoming branch by $\phi(\omega,z)$; the outgoing branch is fixed by symmetry to be $\sigma_1\phi(-\omega,z)$, where $\sigma_1$ is the Pauli-$X$ matrix. We normalize these modes near the horizon as
\begin{equation}
\phi(\omega,z)\stackrel{z\rightarrow+\infty}{=}\begin{bmatrix}
e^{\i\omega z}\\0
\end{bmatrix},\ \sigma_1\phi(-\omega,z)\stackrel{z\rightarrow+\infty}{=}\begin{bmatrix}
0\\e^{-\i\omega z}
\end{bmatrix}
\label{eq: D8}
\end{equation}
A normalized physical solution $\varphi(\omega,z)$ is a linear combination of incoming/outgoing modes,
\begin{equation}
\varphi(\omega,z)=\mathcal C_+(\omega)\phi(\omega,z)+\mathcal C_-(\omega)\sigma_1\phi(-\omega,z)
\label{eq: expand physical solution onto in and out mode}
\end{equation}
By symmetry of the Majorana equation \eqref{eq: eom of QFT in frequency space}, solutions satisfy a reality condition (e.g. $\phi(\omega,z)^*=\phi(-\omega,z)$, and similarly for $\varphi$), which implies
\begin{equation}
\mathcal C_\pm(\omega)^*=\mathcal C_\pm(-\omega)
\label{eq: reality condition of C_pm(omega)}
\end{equation}

The ratio of $\mathcal C_\pm(\omega)$ is fixed by the boundary condition \eqref{eq: boundary condition}:
\begin{equation}
\begin{bmatrix}
\mathcal C_+(\omega)\\
\mathcal C_-(\omega)
\end{bmatrix}
=\begin{bmatrix}
\phi_L(-\omega,0) & -\phi_R(-\omega,0)\\
-\phi_R(\omega,0) & \phi_L(\omega,0)
\end{bmatrix}
\begin{bmatrix}
\varphi_L(\omega,0)\\
\varphi_R(\omega,0)
\end{bmatrix}
\label{eq: C12}
\end{equation}
Here we used that the Wronskian $W(z):=\phi_L(\omega,z)\phi_L(-\omega,z)-\phi_R(\omega,z)\phi_R(-\omega,z)$ is $z$-independent by the equations of motion, and evaluating it at $z\to+\infty$ gives $W(z)=1$.

The overall normalization of $\mathcal C_\pm(\omega)$ is fixed by the canonical anti-commutator
\begin{equation}
\{\Psi_a(t,z),\Psi_b(t,z')\}=(2\pi)\delta_{ab}\delta(z-z')
\label{eq: canonical relation of QFT fermion}
\end{equation}
We expand the field operator in this mode basis as
\begin{equation}
\Psi(t,z)=\int d\omega e^{\i\omega t}\Psi(\omega,z):=\int d\omega e^{\i\omega t}c_\omega\varphi(\omega,z),\ \ \text{with: }c_\omega^\dagger=c_{-\omega},\ \{c_\omega,c_{\omega'}\}=\delta(\omega+\omega')
\end{equation}
Requiring \eqref{eq: canonical relation of QFT fermion} implies the completeness/normalization condition
\begin{equation}
\int d\omega \varphi_a(\omega,z)\varphi_b(-\omega,z')=(2\pi )\delta_{ab}\delta(z-z')
\end{equation}
Evaluating this at $z,z'\to\infty$ yields
\begin{equation}
\mathcal C_\pm(\omega)\mathcal C_\pm(-\omega)=1
\label{eq: |C|=1}
\end{equation}
Together with \eqref{eq: reality condition of C_pm(omega)}, this implies that $\mathcal C_\pm(\omega)$ are pure phases for real $\omega$.

We now compute the boundary correlator $\{\chi(t),\chi(t')\}$. Rewriting \eqref{eq: relate Psi with chi} in frequency space gives
\begin{equation}
\Psi(\omega,0)=\begin{bmatrix}
\frac{1}{\sqrt{2\mu}}(-\i\omega-\mu)\\
\frac{1}{\sqrt{2\mu}}(-\i\omega+\mu)
\end{bmatrix}
\chi(\omega):=a(\omega)\chi(\omega)
\text{, with: }
\chi(t)=\int d\omega e^{\i\omega t}\chi(\omega)
\label{eq: C16}
\end{equation}
The boundary condition \eqref{eq: boundary condition} can be written as
\begin{equation}
\frac{\varphi_L(\omega,0)}{a_L(\omega)}=\frac{\varphi_R(\omega,0)}{a_R(\omega)}=\mathcal R(\omega)
\label{eq: C17}
\end{equation}
where $\mathcal R(\omega)$ inherits the reality property of $\varphi(\omega,z)$, i.e. $\mathcal R(\omega)^*=\mathcal R(-\omega)$ for real $\omega$. Hence, from \eqref{eq: C16} we find
\begin{equation}
\chi(\omega)=\mathcal R(\omega)c_\omega
\end{equation}
and therefore the boundary two-point function is
\begin{equation}
\{\chi(t),\chi(t')\}=\int d\omega e^{\i\omega(t-t')}\mathcal R(\omega)\mathcal R(-\omega),
\end{equation}
so that
\begin{equation}
\mA(\omega)=\mathcal R(\omega)\mathcal R(-\omega).
\end{equation}
From \eqref{eq: C12} and \eqref{eq: C17} we also obtain
\begin{equation}
\begin{bmatrix}
\mathcal C_+(\omega)\\
\mathcal C_-(\omega)
\end{bmatrix}
=\mathcal R(\omega)\begin{bmatrix}
\phi_L(-\omega,0) & -\phi_R(-\omega,0)\\
-\phi_R(\omega,0) & \phi_L(\omega,0)
\end{bmatrix}
\begin{bmatrix}
a_L(\omega)\\
a_R(\omega)
\end{bmatrix}:=\mathcal R(\omega)\begin{bmatrix}
-\mathcal J(-\omega)\\
\mathcal J(\omega)
\end{bmatrix}
\label{eq: C12}
\end{equation}
where $\mathcal J(\omega)$ is the J\"ost function~\cite{Festuccia:2005pi,Festuccia:2008zx,Dodelson:2023vrw} (in analogy with the scalar case). In the last equality we used $a_L(\omega)=-a_R(-\omega)$ from \eqref{eq: C16}. Note that $\mathcal J(\omega)^*=\mathcal J(-\omega)$ for real $\omega$. The scattering amplitude is
\begin{equation}
\mS(\omega):=-\frac{\mathcal C_+(\omega)}{\mathcal C_-(\omega)}=\frac{\mathcal J(-\omega)}{\mathcal J(\omega)}
\label{eq: C22}
\end{equation}
For real $\omega$, write $\mathcal J(\omega)=\rho_{\mathcal J}(\omega)e^{\i\theta_{\mathcal J}(\omega)}$ with $\rho_{\mathcal J}(-\omega)=\rho_{\mathcal J}(\omega)$ and $\theta_{\mathcal J}(-\omega)=-\theta_{\mathcal J}(\omega)$, and similarly $\mathcal R(\omega)=\rho_{\mathcal R}(\omega)e^{\i\theta_{\mathcal R}(\omega)}$. Using \eqref{eq: |C|=1} gives $\rho_{\mathcal R}(\omega)=[\rho_{\mathcal J}(\omega)]^{-1}$. Thus the boundary spectral function and scattering phase are determined by the J\"ost function:
\begin{equation}
\mA(\omega)=\frac{1}{\rho_{\mathcal J}(\omega)^2},\ \  \mS(\omega)=e^{-2\i\theta_{\mathcal J}(\omega)}
\end{equation}

The phase $\theta_{\mathcal R}(\omega)$ is not fixed by physical data. In \cref{sec: Causal kernel decomposition and horizon modes} we identified a preferred choice from the causal-kernel decomposition: one demands that $\mathcal K^f(\omega)=\frac{1}{\mR^f(\omega)}$ is analytic in the upper half-plane.

In \cref{app: Analytic property of Jost function} we will show that, under the condition
\begin{equation}
\int_0^{+\infty}dz\cdot \Omega(z)<\infty,
\end{equation}
the J\"ost function $\mJ(\omega)$ is analytic in the upper half-plane. In this situation the causal kernel exists and can be chosen as $\mR^f (\omega)=\frac{1}{\mJ(\omega)}$, with
\begin{equation}
\mA(\omega)=\frac{1}{\mJ(\omega)\mJ(-\omega)}
\label{eq: D25}
\end{equation}

\subsection{Analytic property of J\"ost function}
\label{app: Analytic property of Jost function}
In this section we establish the following statements under the condition \eqref{eq: Omega condition}:
\begin{itemize}
\item[\textbf{(1).}] The incoming solution $\phi(\omega,z)$ is analytic in the upper half $\omega$-plane. Consequently, the J\"ost function $\mJ(\omega)$ is analytic in the upper half-plane.

\item[\textbf{(2).}] In the upper half-plane and at high energy ($\Im\omega\to+\infty$), $\phi(\omega,z)$ approaches the free solution.

\item[\textbf{(3).}] Zeros of the J\"ost function $\mJ(\omega)$ in the upper half-plane correspond to bound states.

\item[\textbf{(4).}] For an analytic metric with temperature \eqref{eq: temperature of geometry}, $\mJ(\omega)$ has simple poles at the fermionic Matsubara frequencies $\{-\i (2n+1)\lambda_0=-\i(n+\frac12)\beta\mid n=0,1,2,\dots\}$ and is otherwise analytic. As a corollary, using \eqref{eq: D25} and \eqref{eq: KMS relation}, the Wightman function $\mathcal G_\text W(\omega)$ has no zeros.

\item[\textbf{(5).}] The (unnormalized) physical solution $\tilde\varphi(\omega,z)$ is analytic on the entire complex $\omega$-plane and approaches the free physical solution at high energy.

\item[\textbf{(6).}] At high energy the S-matrix obeys $\mS(\omega)\to-1$ as $|\Im\omega|\to+\infty$.
\end{itemize}
These properties are used in deriving the inverse-scattering formula and related analyticity statements in the frequency plane.

\noindent \textbf{(1). }We first recall the definition of the J\"ost function from~\eqref{eq: C12}:
\begin{equation}
\mJ(\omega)=\phi_L(\omega,0)a_R(\omega)-\phi_R(\omega,0)a_L(\omega)
\label{eq: define Jost function}
\end{equation}
From \eqref{eq: C16}, $a_{L,R}(\omega)$ are analytic on the whole $\omega$-plane, so it suffices to show that $\phi(\omega,0)$ is analytic in the upper half-plane. In fact, we can prove the stronger statement that $\phi(\omega,z)$ is analytic in the upper half-plane for any $z\geq0$.

\paragraph{Theorem. }\textit{The incoming wavefunction $\phi(\omega,z)$ is bounded for all $\Im\omega\geq0$ and all $z\geq0$, provided that}
\begin{equation}
C_0:=\int_0^{+\infty}dz\cdot \Omega(z)<\infty.
\label{eq: Omega condition}
\end{equation}
\textit{Proof. }The incoming solution obeys the Volterra equation
\begin{equation}
\phi(\omega,z)=\begin{bmatrix}
e^{\i\omega z} \\
0
\end{bmatrix}-\int_z^{+\infty}dz'\begin{bmatrix}
0 & e^{\i\omega(z-z')}\\
e^{-\i\omega(z-z')} & 0
\end{bmatrix}
m\Omega(z')\phi(\omega,z')
\label{eq: Volterra equation for in-coming mode}
\end{equation}
which can be solved iteratively as
\begin{equation}
\phi(\omega,z):=\sum_{n=0}^{+\infty}\phi^{(n)}(\omega,z),
\end{equation}
with
\begin{equation}
\phi^{(0)}(\omega,z)=\begin{bmatrix}
e^{\i\omega z} \\
0
\end{bmatrix},\  \phi^{(n)}(\omega,z)=-\int_z^{+\infty}dz'\begin{bmatrix}
0 & e^{\i\omega(z-z')}\\
e^{-\i\omega(z-z')} & 0
\end{bmatrix}
m\Omega(z')\phi^{(n-1)}(\omega,z'),\ n\geq1.
\label{eq: D30}
\end{equation}
We bound $|\phi(\omega,z)|$ using the Euclidean norm. For any vector $v$ and matrix $M$, $|Mv|\leq\|M\|_2|v|=\sqrt{\text{max-eigval}(M^\dagger M)}|v|$. Writing $\nu:=\Im\omega$, we obtain
\begin{equation}
|\phi^{(n)}(\omega,z)|\leq\int_z^{+\infty}dz'm\Omega(z')e^{|\nu|(z'-z)}|\phi^{(n-1)}(\omega,z')|.
\end{equation}
It is convenient to define $\tilde\phi^{(n)}(\omega,z):=e^{|\nu|z}\phi^{(n)}(\omega,z)$. Iterating the bound gives
\begin{equation}
\begin{aligned}
|\tilde\phi^{(n)}(\omega,z)|&\leq\int_z^{+\infty}dz_1\int_{z_1}^{+\infty}dz_2\cdots\int_{z_{n-1}}^{+\infty}dz_n\cdot \prod_{k=1}^{n}\left[m\Omega(z_k)\right]\cdot e^{(|\nu|-\nu)z_n}\\
&=e^{(|\nu|-\nu)z}\int_z^{+\infty}dz_1\int_{z_1}^{+\infty}dz_2\cdots\int_{z_{n-1}}^{+\infty}dz_n\cdot \prod_{k=1}^{n}\left[m\Omega(z_k)\right]\cdot e^{(|\nu|-\nu)(z_n-z)}\\
&\leq e^{(|\nu|-\nu)z}\int_z^{+\infty}dz_1\int_{z_1}^{+\infty}dz_2\cdots\int_{z_{n-1}}^{+\infty}dz_n\cdot \prod_{k=1}^{n}\left[m\Omega(z_k)e^{(|\nu|-\nu)(z_k-z)}\right]\\
&=e^{(|\nu|-\nu)z}\frac{1}{n!}\left[\int_z^{+\infty}dz'\cdot m\Omega(z')e^{(|\nu|-\nu)(z'-z)}\right]^n
\label{eq: technical detail of deriving bound}
\end{aligned}
\end{equation}
Since $|\tilde\phi^{(0)}(\omega,z)|=e^{(|\nu|-\nu)z}$ and $1\leq e^{(|\nu|-\nu)(z_k-z)}$ (because $(|\nu|-\nu)(z_k-z)\geq0$), we obtain
\begin{equation}
|\phi(\omega,z)|\leq e^{-\nu z}\exp\left[\int_z^{+\infty}dz'\cdot m\Omega(z')e^{(|\nu|-\nu)(z'-z)}\right].
\end{equation}
For $\nu\geq0$ we have $|\nu|-\nu=0$, so this simplifies to
\begin{equation}
|\phi(\omega,z)|\leq e^{-\nu z}\exp\left[\int_z^{+\infty}dz'\cdot m\Omega(z')\right]\leq e^{-\nu z}e^{mC_0},\ \Im\omega\geq0.
\label{eq: bound on phi mode function}
\end{equation}
This proves the claim.\hfill$\qed$

In fact, using the asymptotic behavior $\Omega(z)\sim e^{-2\lambda_0 z}$, one can show a stronger statement: $|\phi(\omega,z)|$ is bounded as long as $\Im\omega>-\lambda_0$.

\noindent \textbf{(2). }As a corollary, we can bound the deviation from the free solution:
\begin{equation}
\begin{aligned}
|\phi(\omega,z)-\phi^{(0)}(\omega,z)|&\leq\int_z^{+\infty}dz'e^{|\nu|(z'-z)}m\Omega(z')|\phi(\omega,z')|\\
&\leq\int_z^{+\infty}dz'e^{|\nu|(z'-z)}m\Omega(z')e^{-\nu z'}e^{mC_0}\\
&\leq e^{-\nu z}mC_0e^{mC_0},\ \Im\omega\geq0
\end{aligned}
\label{eq: D35}
\end{equation}
where the first step uses the Volterra equation \eqref{eq: Volterra equation for in-coming mode}. Thus, for finite $C_0$, the incoming solution approaches the free solution in the upper half-plane.

\noindent \textbf{(3). }As another corollary of \eqref{eq: bound on phi mode function}, we show that zeros of the J\"ost function in the upper half-plane correspond to bound states. Suppose $\mJ(\omega_0)=0$ with $\Im\omega_0>0$. By \eqref{eq: define Jost function}, the incoming solution $\phi(\omega_0,z)$ by itself satisfies the boundary condition \eqref{eq: C17} and is therefore physical. The factor $e^{-(\Im\omega_0) z}$ in \eqref{eq: bound on phi mode function} implies exponential decay toward the horizon, so this mode is a bound state.

\noindent \textbf{(4). }Next, consider a conformal factor $\Omega(z)$ that is analytic to all orders near the horizon, with temperature \eqref{eq: temperature of geometry}. Equivalently, after passing to local Rindler coordinates, the metric admits a Taylor expansion with no fractional powers. The most general such conformal factor can be written as the ansatz
\begin{equation}
m\Omega(z):=\sum_{n=0}^{+\infty}a_ne^{-2(2n+1)\lambda_0z}
\end{equation}
This follows by combining the leading exponential $e^{-4\lambda_0z}(-dt^2+dz^2)$ with the local Rindler coordinate $-\rho^2dt^2+d\rho^2$, where $\rho:=e^{-2\lambda_0z}$. The remaining factor $\tilde\Omega(\rho)$ is a Taylor series in $\rho$ with only even powers; hence $\Omega(\rho)\propto \rho\tilde\Omega(\rho)$ contains only odd powers of $\rho$.

We claim that $\phi(\omega,z)$ has poles only at $\{-\i (2n+1)\lambda_0\mid n=1,2,3,\dots\}$. This is straightforward to see by tracking the $\omega$-dependence in each term of the iterative solution \eqref{eq: D30}. A low-order example is:
\begin{equation}
\phi^{(2)}(\omega,z)=\begin{bmatrix}
e^{\i\omega z}\\0    
\end{bmatrix}
\sum_{n_1,n_2=0}^{+\infty}\frac{a_{n_1}a_{n_2}e^{-2\lambda_0(m_1+m_2)z}}{(-4\i\lambda_0)(\omega+\i m_1\lambda_0)(m_1+m_2)},\ m_i:=2n_i+1
\end{equation}
The general expression for even number of iteration is:
\begin{equation}
\phi^{(2k)}(\omega,z)=\begin{bmatrix}
e^{\i\omega z}\\0    
\end{bmatrix}
(\prod_{i=1}^{2k}\sum_{n_i=0}^{+\infty})\frac{(\prod_{i=1}^{2k}a_{n_i})e^{-2\lambda_0(\sum_{i=1}^{2k}m_i)z}}{(-4\i\lambda_0)^{k}\prod_{i=1}^{k}\left[(\omega+\i\lambda_0(\sum_{j=1}^{2i-1}m_j))(\sum_{j=1}^{2i}m_j)\right]},\ m_i:=2n_i+1
\end{equation}
Thus the poles lie at $\omega=-\i (2n+1)\lambda_0$, and they are all simple. A more careful discussion of convergence can be found in~\cite{Newton1982Scattering}; see also Appendix~C of~\cite{Dodelson:2023vrw}.

Using \eqref{eq: D25} and \eqref{eq: KMS relation}, the Wightman function is
\begin{equation}
\mathcal G_\text W(\omega)=\frac{1}{(1+e^{\beta\omega})\mJ(\omega)\mJ(-\omega)},\ 
\end{equation}
so zeros of $(1+e^{\beta\omega})$ cancel the poles of $\mJ(\omega)\mJ(-\omega)$ at fermionic Matsubara frequencies. Hence $\mathcal G_\text W(\omega)$ has no zeros.

\noindent \textbf{(5). }For later convenience, we define an unnormalized physical solution $\tilde\varphi(\omega,z)$ that satisfies a Volterra equation analogous to \eqref{eq: Volterra equation for in-coming mode}:
\begin{equation}
\tilde\varphi(\omega,z)=\tilde\varphi^{(0)}(\omega,z)+\int_0^{z}dz'\begin{bmatrix}
0 & e^{\i\omega(z-z')}\\
e^{-\i\omega(z-z')} & 0
\end{bmatrix}
m\Omega(z')\tilde\varphi(\omega,z'),\ \ \ \tilde\varphi^{(0)}(\omega,z)=\begin{bmatrix}
a_L(\omega)e^{\i\omega z} \\
a_R(\omega)e^{-\i\omega z}
\end{bmatrix}
\label{eq: Volterra equation for physical mode}
\end{equation}
where $\tilde\varphi^{(0)}(\omega,z)$ is the free solution. Since $\tilde\varphi(\omega,0)=\tilde\varphi^{(0)}(\omega,0)$, $\tilde\varphi$ satisfies the boundary condition \eqref{eq: C17} with the choice $\mR(\omega)=1$. Thus $\varphi(\omega,z)=\mR(\omega)\tilde\varphi(\omega,z)$. We can also expand $\tilde\varphi$ in the incoming/outgoing basis using \eqref{eq: C12}:
\begin{equation}
\tilde\varphi(\omega,z)=-\mJ(-\omega)\phi(\omega,z)+\mJ(\omega)\sigma_1\phi(-\omega,z)
\label{eq: unnormalized physical solution}
\end{equation}
Proceeding as in \eqref{eq: technical detail of deriving bound}, we obtain the bound
\begin{equation}
|\tilde\varphi(\omega,z)|\leq e^{|\Im\omega|z}\sqrt{\mu^{-1}\left[(\Re\omega)^2+(|\Im\omega|+\mu)^2\right]}\exp\left[\int_0^{z}dz'm\Omega(z')\right]
\label{eq: D38}
\end{equation}
valid for any complex $\omega$ and $z\geq0$. In particular, $\tilde\varphi(\omega,z)$ is analytic on the entire complex $\omega$-plane. Its deviation from the free solution $\tilde\varphi^{(0)}$ can be bounded by
\begin{equation}
|\tilde\varphi(\omega,z)-\tilde\varphi^{(0)}(\omega,z)|\leq e^{|\Im\omega| z}mC_0e^{mC_0}\sqrt{\mu^{-1}\left[(\Re\omega)^2+(|\Im\omega|+\mu)^2\right]}
\label{eq: D39}
\end{equation}
for any complex $\omega$ and $z\geq0$.

\noindent \textbf{(6). }Finally, we comment on the high-energy behavior of the S-matrix. If $\Omega(z)$ is regular, then at sufficiently high energy we may neglect the mass term and study massless Majorana fermions. The only scatterer is then the boundary at $z=0$, with boundary condition~\eqref{eq: boundary condition}, which yields
\begin{equation}
\mS^{(0)}(\omega)=\frac{a_R(-\omega)}{a_R(\omega)}=\frac{\i\omega+\mu}{-\i\omega+\mu}
\end{equation}
and therefore
\begin{equation}
\mS_\infty=\mS^{(0)}_{\infty}=-1
\end{equation}

\subsection{One QNMs model from bulk QFT}
\label{app: One QNMs model from bulk QFT}
In this appendix we address a question raised in \cref{sec: Examples: finite QNMs}. There we found that the dual bulk geometry of the one-QNM model has conformal factor $m\Omega(z)=0$ for all $z$, i.e. it corresponds to a massless fermion on the (1+1)d half-space ($z\geq0$). From the bulk-QFT perspective, one may then ask how the boundary two-point function can nevertheless decay exponentially. The answer is that the exponential decay is induced by the boundary condition~\eqref{eq: boundary condition} at $z=0$.

From \eqref{eq: define Jost function} we obtain the J\"ost function
\begin{equation}
\mJ(\omega)=\frac{1}{\sqrt{2\mu}}(\i\omega+\mu)
\end{equation}
Quasinormal modes are defined as zeros of $-\mJ(-\omega)$ in the upper half-plane; this yields a single QNM at $\omega=\i\mu$. Using \eqref{eq: C22} and \eqref{eq: D25}, the spectral function and S-matrix are
\begin{equation}
\mA(\omega)=\frac{2\mu}{\omega^2+\mu^2},\ \mS(\omega)=\frac{\i\omega+\mu}{-\i\omega+\mu}
\end{equation}
We can also compute the bulk two-point function:
\begin{equation}
\{\Psi_a(u,v),\Psi_b(u',v')\}=2\pi\begin{bmatrix}
\delta(v-v') & g(u'-v)\\
g(u-v') & \delta(u-u')
\end{bmatrix},\ g(z):=\delta (z)-2\mu e^{-\mu z}\Theta(z)
\end{equation}
In the off-diagonal left/right correlator, the $\delta(z)$ term corresponds to the usual boundary reflection. The second term $2\mu e^{-\mu z}$ encodes imperfect reflection at the boundary.

\subsection{Two QNMs model from bulk QFT}
\label{app: Two QNMs model from bulk QFT}
Here we re-derive, purely from the bulk QFT, the S-matrix of the two-QNM model discussed in \cref{sec: Two QNMs}. The punchline is that the ``finite number of QNMs'' is not in tension with the fact that the background has constant AdS curvature: the usual AdS$_2$--Rindler analysis assumes the boundary sits at the asymptotic boundary with standard normalizable boundary conditions, which indeed gives an infinite tower~\cite{Horowitz:1999jd}. Once we move the boundary to a finite radial position (as dictated by the circuit), most of that tower disappears.

We start from the geometry \eqref{eq: 2 QNM geometry}:
\begin{equation}
m\Omega(z)=\frac{2\lambda_0}{\sinh[\lambda_0(2z+z_0)]},\ \ 0<z_0<z<+\infty
\end{equation}
It is convenient to use
\begin{equation}
y:=\lambda_0(2z+z_0),\ 0<y<+\infty,
\end{equation}
so that $\Omega(y)=1/\sinh y$. We keep the mass $m$ general for the moment (the two-QNM model corresponds to $m=1$ in AdS units). In terms of $y$, the Majorana equation \eqref{eq: eom of QFT in frequency space} becomes:
\begin{equation}
\begin{bmatrix}
\partial_y & -m\Omega(y)\\
m\Omega(y) & -\partial_y
\end{bmatrix}\begin{bmatrix}
\Psi_L\\
\Psi_R
\end{bmatrix}
=\i\tilde{\omega}\begin{bmatrix}
\Psi_L\\
\Psi_R
\end{bmatrix},\ \tilde\omega:=\frac{\omega}{2\lambda_0},\ \Omega(y)=\frac{1}{\sinh y},\ y>0
\end{equation}
Next, introduce the coordinate $x$ such that $x=0$ is the horizon and $x=1$ is the asymptotic boundary:
\begin{equation}
x:=e^{-2y}, \ x\in(0,1),
\label{eq: D48}
\end{equation}
In terms of $x$, the Majorana equation becomes hypergeometric. It is useful to organize solutions by their near-horizon and near-boundary behavior. We take $\Psi_1$ and $\Psi_2$ to be the incoming/outgoing solutions, normalized near the horizon as
\begin{equation}
\Psi_1\rightarrow\begin{bmatrix}
x^{-\frac{\i\tilde{\omega}}{2}}\\0
\end{bmatrix}=\begin{bmatrix}
e^{\i\tilde{\omega} y}\\0
\end{bmatrix},\ \Psi_2\rightarrow\begin{bmatrix}
0\\x^{\frac{\i\tilde{\omega}}{2}}
\end{bmatrix}=\begin{bmatrix}
0\\e^{-\i\tilde{\omega} y}
\end{bmatrix},\ \ \text{as }x\rightarrow0^+.
\label{eq: D46}
\end{equation}
We also define $\Psi_3$ and $\Psi_4$ as the normalizable and non-normalizable solutions, respectively, characterized by their behavior near the asymptotic boundary:
\begin{equation}
\Psi_3\rightarrow\begin{bmatrix}
(1-x)^{m}\\(1-x)^{m}
\end{bmatrix}=\begin{bmatrix}
(2y)^m\\(2y)^m
\end{bmatrix},\ \Psi_4\rightarrow\begin{bmatrix}
(1-x)^{-m}\\-(1-x)^{-m}
\end{bmatrix}=\begin{bmatrix}
(2y)^{-m}\\-(2y)^{-m}\end{bmatrix},\ \ \text{as }x\rightarrow1^-
\end{equation}
We will first record the result of mode functions:
\begin{equation}
\begin{aligned}
\Psi_1(\omega,y)&=\begin{bmatrix}
x^{-\frac{\i\tilde{\omega}}{2}}(1-x)^{-m}{}_2F_1(-m,\frac{1}{2}-m-\i\tilde{\omega},\frac{1}{2}-\i\tilde{\omega},x)\\
\frac{-m}{\frac{1}{2}-\i\tilde{\omega}}x^{\frac{1}{2}-\frac{\i\tilde{\omega}}{2}}(1-x)^{-m}{}_2F_1(1-m,\frac{1}{2}-m-\i\tilde{\omega},\frac{3}{2}-\i\tilde{\omega},x)
\end{bmatrix}\\
\Psi_2({\omega},y)&=
\begin{bmatrix}
\frac{-m}{\frac{1}{2}+\i\tilde{\omega}}x^{\frac{1}{2}+\frac{\i\tilde{\omega}}{2}}(1-x)^{-m}{}_2F_1(1-m,\frac{1}{2}-m+\i\tilde{\omega},\frac{3}{2}+\i\tilde{\omega},x)\\
x^{\frac{\i\tilde{\omega}}{2}}(1-x)^{-m}{}_2F_1(-m,\frac{1}{2}-m+\i\tilde{\omega},\frac{1}{2}+\i\tilde{\omega},x)
\end{bmatrix}\\
\Psi_3({\omega},y)&=
\begin{bmatrix}
x^{-\frac{\i\tilde{\omega}}{2}}(1-x)^m{}_2F_1(m,\frac{1}{2}+m-\i\tilde{\omega},1+2m,1-x)\\
x^{\frac{\i\tilde{\omega}}{2}}(1-x)^m{}_2F_1(m,\frac{1}{2}+m+\i\tilde{\omega},1+2m,1-x)
\end{bmatrix}\\
\Psi_4({\omega},y)&=\begin{bmatrix}
x^{-\frac{\i\tilde{\omega}}{2}}(1-x)^{-m}{}_2F_1(-m,\frac{1}{2}-m-\i\tilde{\omega},1-2m,1-x)\\
-x^{\frac{\i\tilde{\omega}}{2}}(1-x)^{-m}{}_2F_1(-m,\frac{1}{2}-m+\i\tilde{\omega},1-2m,1-x)
\end{bmatrix}
\end{aligned}
\label{eq: AdS2 mode functions}
\end{equation}
The outgoing and incoming solutions are related by the symmetry that exchanges $\Psi_L\leftrightarrow\Psi_R$ and simultaneously sends $\tilde{\omega}\to-\tilde{\omega}$. The normalizable and non-normalizable solutions are related by $m\to-m$ together with $\Psi_R\to-\Psi_R$.

Using hypergeometric connection formulas~\cite{}, for non-integer $m$ one can expand the normalizable and non-normalizable solutions in the incoming/outgoing basis:
\begin{equation}
\Psi_3=C_{31}({\omega})\Psi_1+C_{32}({\omega})\Psi_{2},\ C_{31}({\omega})=\frac{\Gamma(1+2m)\Gamma(\frac{1}{2}+\i\tilde{\omega})}{\Gamma(1+m)\Gamma(\frac{1}{2}+m+\i\tilde{\omega})},\ C_{32}({\omega})=\frac{\Gamma(1+2m)\Gamma(\frac{1}{2}-\i\tilde{\omega})}{\Gamma(1+m)\Gamma(\frac{1}{2}+m-\i\tilde{\omega})}
\end{equation}
\begin{equation}
\Psi_4=C_{41}({\omega})\Psi_1+C_{42}({\omega})\Psi_{2},\ C_{41}({\omega})=\frac{\Gamma(1-2m)\Gamma(\frac{1}{2}+\i\tilde{\omega})}{\Gamma(1-m)\Gamma(\frac{1}{2}-m+\i\tilde{\omega})},\ C_{42}({\omega})=-\frac{\Gamma(1-2m)\Gamma(\frac{1}{2}-\i\tilde{\omega})}{\Gamma(1-m)\Gamma(\frac{1}{2}-m-\i\tilde{\omega})}
\end{equation}

As a warm-up, place the boundary at the asymptotic boundary and impose the usual normalizability condition. Then $\Psi_3$ is the physical mode.

To extract QNMs, we look at the near-horizon decomposition into ingoing/outgoing waves. In our conventions, the ingoing piece is $\Psi_2$ (it depends on $t-z$), while the outgoing piece is $\Psi_1$. A quasinormal mode is a solution that is purely ingoing at the future horizon and decays in time; with time dependence $e^{\i\omega t}$ this means $\Im\omega>0$ and the coefficient of $\Psi_1$ vanishes. Anti-QNMs are the time-reversed modes (purely outgoing at the past horizon).

For the asymptotic-boundary problem, the QNM condition is $C_{31}(\omega)=0$ with $\Im\omega>0$, giving $\omega=\i(2\lambda_0)(n+m+\frac12)$ for $n=0,1,2,\cdots$, and similarly $C_{32}(\omega)=0$ gives the anti-QNMs at $\omega=-\i(2\lambda_0)(n+m+\frac12)$. This is the familiar infinite tower along the imaginary axis.

The corresponding S-matrix\footnote{In some literature one subtracts a ``Rindler background'' factor $\Gamma(\frac12+\i\tilde{\omega})/\Gamma(\frac12-\i\tilde{\omega})$ when discussing the density of states for real $\omega$ and static-patch thermodynamics. For our purpose---reconstructing geometry from the full S-matrix---we keep the full expression. HT thanks Albert Law and Zimo Sun for discussions of this point.} is
\begin{equation}
\mS({\omega})=-\frac{C_{31}({\omega})}{C_{32}({\omega})}=-\frac{\Gamma(\frac12+\i\tilde{\omega})}{\Gamma(\frac12-\i\tilde{\omega})}\frac{\Gamma(\frac12+m-\i\tilde{\omega})}{\Gamma(\frac12+m+\i\tilde{\omega})}
\label{eq: S matrix of normalizable mode}
\end{equation}
and indeed its zeros (poles) in the upper (lower) half-plane reproduce the QNM (anti-QNM) spectrum.

Now comes the key point. When $m$ is a positive integer, the analytic structure collapses: many of the Gamma-function poles that generate the infinite tower are paired up between numerator and denominator and cancel. Equivalently (and more geometrically), at integer $m$ the hypergeometric functions appearing in the mode solutions truncate to polynomials.

Concretely, ${}_2F_1(a,b,c,x)$ becomes a polynomial of degree $n\in\mathbb Z_{\geq0}$ whenever one of $a,b,c-a,c-b$ equals $-n$~\cite{NIST:DLMF}. For example, if $a=-n$ then
\begin{equation}
{}_2F_1(-n,b,c,x)=\sum_{k=0}^{n}(-1)^k\binom{n}{k}\frac{(b)_k}{(c)_k}x^k,\ (b)_n:=b(b+1)(b+2)...(b+n-1),\ n\in\mathbb Z_{\geq0}
\label{eq: hypergeometric truncated}
\end{equation}
The other truncation cases follow from the usual symmetry and connection identities of ${}_2F_1$. In our problem, the frequency $\omega$ enters only through the remaining hypergeometric parameters (and hence through finite Pochhammer symbols). As a result, the mode functions---and therefore the J\"ost function---become rational functions of $\omega$ of finite degree, so only finitely many poles/zeros (and hence QNMs) remain.
\begin{figure}[t]
    \centering
    \includegraphics[width=0.99\linewidth]{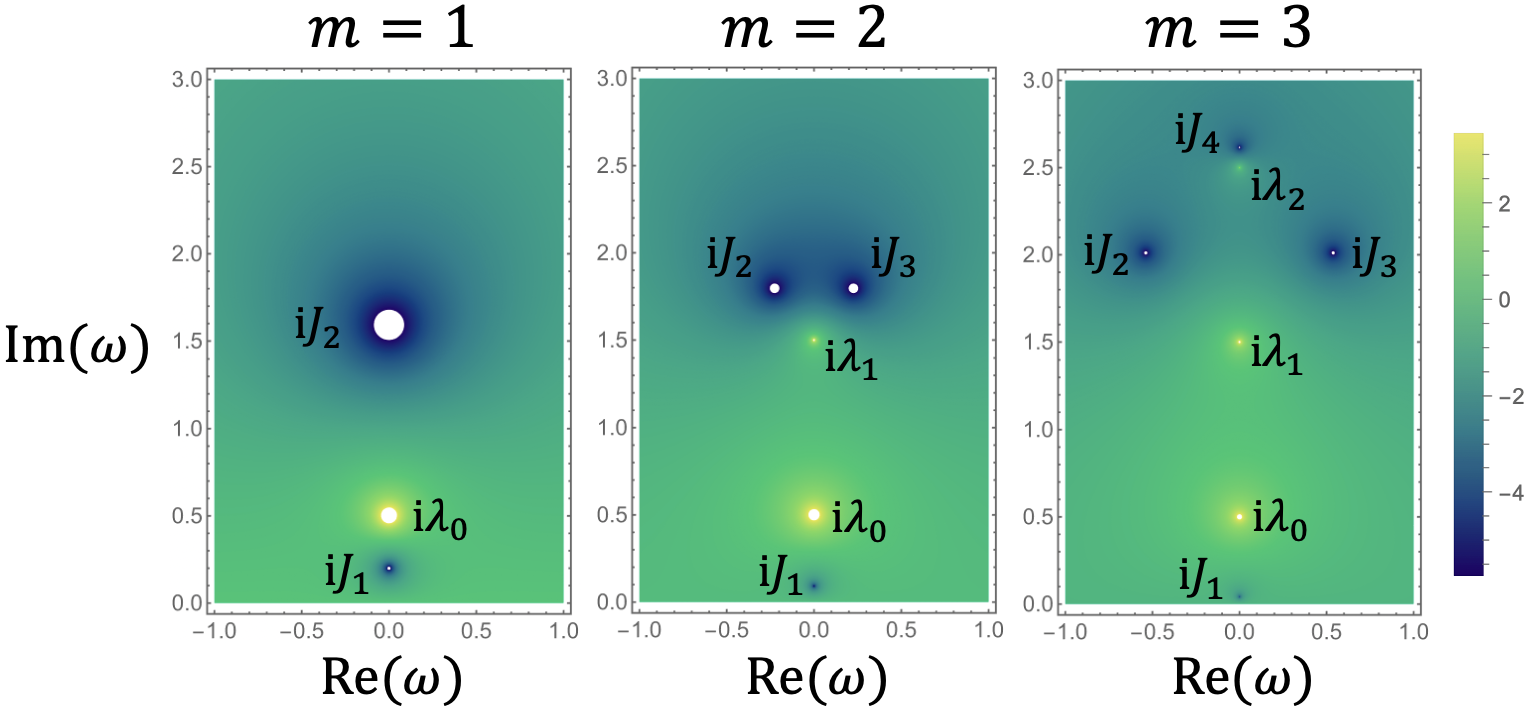}
    \caption{We numerically plot an example of $\log|\mS(\omega)|$ on the upper half plane for several integer $m=1,2,3$, and label their poles (yellow dots) and residue (blue dots). The boundary location is set at $x=0.05$ (see \eqref{eq: D48} for definition of coordinate $x$) with boundary condition parameter $\mu=1$.}
    \label{fig: integer m QNMs}
\end{figure}

To be more explicit, we now specialize to $m=1$, which corresponds to the two-QNM model studied in the main text. The mode functions become
\begin{equation}
\Psi_1({\omega},y)=\begin{bmatrix}
x^{-\frac{\i\tilde{\omega}}{2}}(1-x)^{-1}\left(1-x(\frac{\tilde{\omega}-\frac{\i}{2}}{\tilde{\omega}+\frac{\i}{2}})\right)\\
\frac{\i}{-\tilde{\omega}-\frac{\i}{2}}x^{\frac{1}{2}-\frac{\i\tilde{\omega}}{2}}(1-x)^{-1}
\end{bmatrix}
,\ \Psi_2({\omega},y)=\begin{bmatrix}
\frac{\i}{\tilde{\omega}-\frac{\i}{2}}x^{\frac{1}{2}+\frac{\i\tilde{\omega}}{2}}(1-x)^{-1}\\
x^{\frac{\i\tilde{\omega}}{2}}(1-x)^{-1}\left(1-x(\frac{-\tilde{\omega}-\frac{\i}{2}}{-\tilde{\omega}+\frac{\i}{2}})\right)
\end{bmatrix}
\end{equation}
\begin{equation}
\Psi_3=\frac{-2i}{\tilde{\omega}-\frac{\i}{2}}\Psi_1+\frac{2i}{\tilde{\omega}+\frac{\i}{2}}\Psi_2,\ \ \Psi_4=\frac{\tilde{\omega}+\frac{\i}{2}}{2i}\Psi_1+\frac{\tilde{\omega}-\frac{\i}{2}}{2i}\Psi_2
\label{eq: D54}
\end{equation}
If the boundary is placed at the asymptotic boundary, we must choose the normalizable mode $\Psi_3$ as the physical solution. In this case there are no QNMs, since $C_{31}(\omega)=\frac{-2i}{\tilde{\omega}-\frac{\i}{2}}$ has no zeros in the upper half-plane.

We now move the boundary to a finite location $x_0=e^{-2\lambda z_0}$. Using the definition of the J\"ost function in \eqref{eq: define Jost function}, we obtain
\begin{equation}
\mJ(\omega)=-\i e^{\i\omega\frac{z_0}{2}}\frac{\omega^2+\i b_1\omega+b_0}{\sqrt{2\mu}(\omega+\i\lambda_0)},\ \ \ \ b_1:=\mu+\frac{1+\sqrt{x_0}}{1-\sqrt{x_0}}\lambda,\ b_2=-\frac{1-\sqrt{x_0}}{1+\sqrt{x_0}}\lambda\mu
\end{equation}
Since the numerator is a quadratic polynomial in $\omega$, $\mJ(\omega)$ has two zeros. Substituting the relations from \eqref{eq: 2 QNM geometry} and \eqref{eq: definition of mu} to express $z_0,\lambda_0,\mu$ in terms of $\{c_i,J_i\}_{i=1}^{2}$, the J\"ost function takes the factorized form
\begin{equation}
\mJ(\omega)=-\i e^{\i\omega\frac{z_0}{2}}\frac{(\omega+\i J_1)(\omega+\i J_2)}{\sqrt{2\mu}(\omega+\i\lambda_0)}
\end{equation}
QNMs are zeros of $-\mJ(-\omega)$ in the upper half-plane, so we find two QNMs at $\{\i J_1,\i J_2\}$. The boundary spectral function and S-matrix then follow from \eqref{eq: C22} and \eqref{eq: D25}:
\begin{equation}
\mA(\omega)=(2\mu)\frac{(\omega^2+\lambda_0^2)}{(\omega^2+J_1^2)(\omega^2+J_2^2)},\ \mS(\omega)=e^{\i \omega z_0}\times\frac{\i\lambda_0+\omega}{\i\lambda_0-\omega}\times\prod_{i=1}^2\frac{\i J_i-\omega}{\i J_i+\omega}
\end{equation}
This agrees with the boundary computation in \eqref{eq: A(omega) decomposed into poles and zeros} and \eqref{eq: S(omega) decomposed into poles and zeros}. The extra phase factor $e^{\i \omega z_0}$ arises because the near-horizon normalization in \eqref{eq: D46} differs from that in \eqref{eq: D8}: near the horizon, $e^{\pm\i\tilde\omega y}=e^{\pm \i\omega z}e^{\pm \i \omega\frac{z_0}{2}}$.
\begin{figure}[t]
    \centering
    \includegraphics[width=0.99\linewidth]{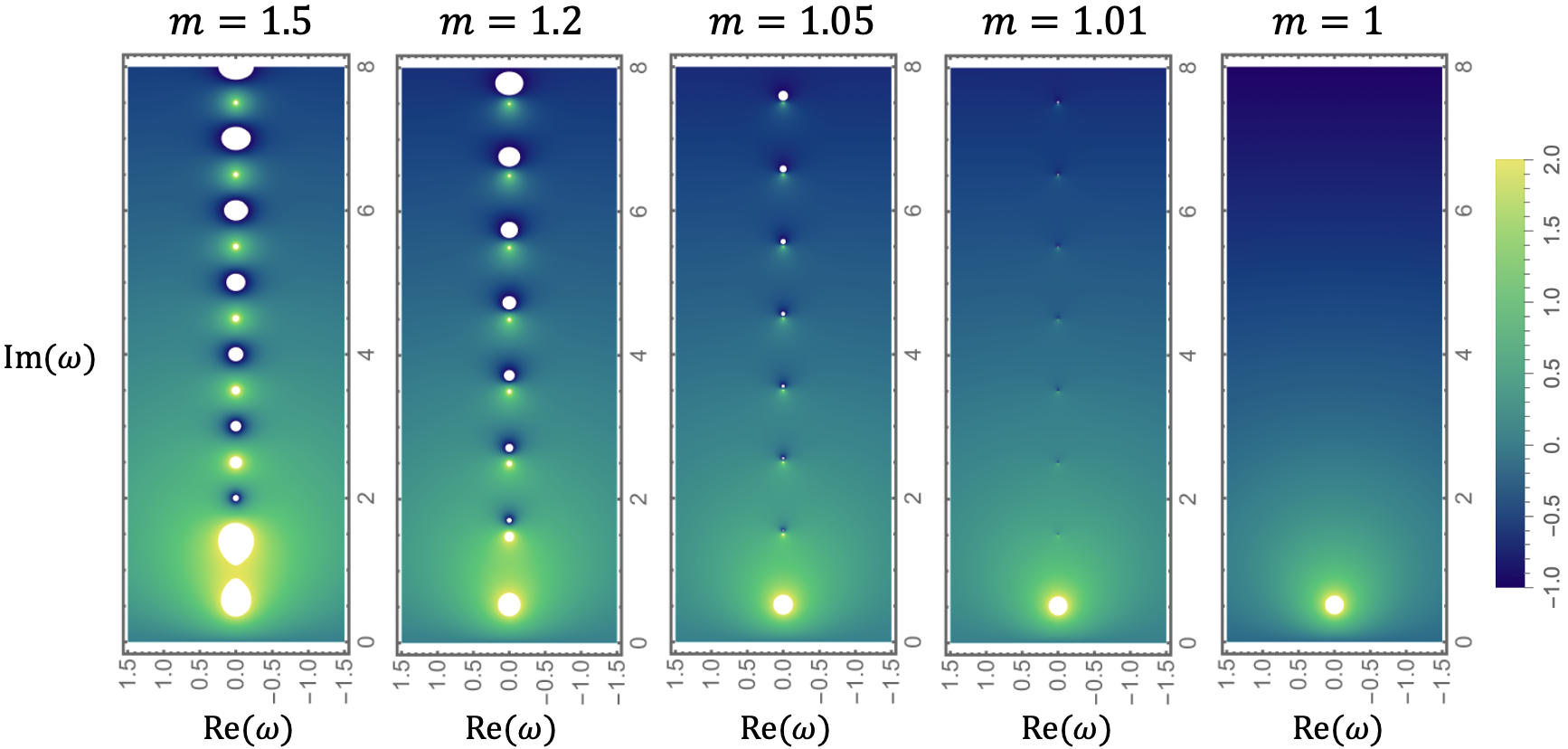}
    \caption{We numerically plot an example of $\log|\mS(\omega)|$ on the upper half plane, gradually tuning mass $m$ from non-integer ($m=1.5,1.2,1.05,1.01$) to integer($m=1$), and label their poles (yellow dots) and residue (blue dots). The boundary location is set at $x=0.9$ (see \eqref{eq: D48} for definition of coordinate $x$) with boundary condition parameter $\mu=1$. }
    \label{fig: collision of pole}
\end{figure}

Finally, we comment on the generic structure at integer mass $m$ in AdS$_2$--Rindler. In the bulk description the tunable parameters are $\lambda_0,\mu,z_0$: $\lambda_0$ sets the AdS length scale, while $\mu$ and $z_0$ fix the boundary condition and the boundary location. The J\"ost function can be written as $\mJ(\omega)=\Psi_{1,L}(\omega,z_0)a_R(\omega)-\Psi_{1,R}(\omega,z_0)a_L(\omega)$, so its poles can arise only from the mode function $\Psi_1(\omega,z)$. Using \eqref{eq: AdS2 mode functions} together with the integer-mass truncation property \eqref{eq: hypergeometric truncated}, one finds that poles can occur only at
$\omega\in\i(2\lambda_0)\{-\frac{1}{2},-\frac{3}{2},\dots,-m+\frac{1}{2}\}$.
The zeros of $\mJ(\omega)$ must in general be found numerically; one typically obtains $m+1$ zeros. Accordingly, the J\"ost function has the schematic form
\begin{equation}
\mJ(\omega)\propto e^{\i\omega\frac{z_0}{2}}\frac{\prod_{k=1}^{m+1}(\omega+\i J_k)}{\prod_{n=0}^{m-1}(\omega+\i\lambda_n)},\ \ \lambda_n:=2\lambda_0(n+\frac12)
\end{equation}

In \cref{fig: integer m QNMs} we plot $\log|\mS(\omega)|$ on the upper half-plane for several integer values of $m$. We observe poles at $\{\i\lambda_n\}_{n=0}^{m-1}$ and zeros at $\{\i J_k\}_{k=1}^{m+1}$.

In \cref{fig: collision of pole} we fix the remaining parameters and continuously tune $m$ from non-integer to integer. For non-integer $m$, there are infinitely many poles and zeros, with poles on the half-integer lattice $\{\i\lambda_n\}_{n=0}^{+\infty}$. As $m$ approaches an integer, most poles and zeros collide and cancel, leaving only finitely many.

\section{Technical detail of deriving inverse scattering formula}
\label{app: Technical detail of deriving inverse scattering formula}
In this appendix we provide the derivation of the inverse-scattering formula \eqref{eq: mOmega in terms of P_zhP_z}, which inverts the forward-scattering relation \eqref{eq: forward scattering, zig-zag}. We then show how \eqref{eq: mOmega in terms of P_zhP_z} reduces to the main results \eqref{eq: main result1} and \eqref{eq: main result2}. The derivation uses the bulk-QFT setup developed in \cref{app: Matching with bulk QFT}, and throughout we assume \eqref{eq: Omega condition}.

Rather than attempting a direct inversion of \eqref{eq: forward scattering, zig-zag}, we follow the standard Marchenko inverse-scattering route, adapted to the Majorana system. Our steps closely parallel the scalar-field treatment in Chapter~20 of~\cite{Newton1982Scattering}.

We start from the Majorana equation \eqref{eq: eom of QFT in frequency space}. The incoming solution $\phi(\omega,z)$ satisfies the Volterra equation \eqref{eq: Volterra equation for in-coming mode}, which we repeat for convenience:
\begin{equation}
\phi(\omega,z)=\begin{bmatrix}
e^{\i\omega z} \\
0
\end{bmatrix}-\int_z^{+\infty}dz'\begin{bmatrix}
0 & e^{\i\omega(z-z')}\\
e^{-\i\omega(z-z')} & 0
\end{bmatrix}
m\Omega(z')\phi(\omega,z')
\end{equation}
This Volterra equation admits an iterative expansion around the free solution
\begin{equation}
\phi^{(0)}(\omega,z):=\begin{bmatrix}
e^{\i\omega z}\\0
\end{bmatrix}
\end{equation}
Equivalently, one can package the interaction with $m\Omega(z)$ into a (retarded) kernel along the $z$ direction:
\begin{equation}
\phi(\omega,z)=\phi^{(0)}(\omega,z)+\int_z^{+\infty}dz'\begin{bmatrix}
C_1(z,z')\\
C_2(z,z')
\end{bmatrix}e^{\i\omega z'}
\end{equation}
where $C_{1,2}(z,z')$ are supported only for $z<z'$. Plugging this ansatz into the Majorana equation \eqref{eq: eom of QFT in frequency space} gives the kernel equations
\begin{equation}
(\partial_z+\partial_{z'})C_1(z,z')=m\Omega(z)C_2(z,z'),\ \ (\partial_z-\partial_{z'})C_2(z,z')=m\Omega(z)C_1(z,z'), \ \ m\Omega(z)=-2C_2(z,z)
\end{equation}
Setting $z'=z$ in the first equation yields $\partial_zC_1(z,z)=m\Omega(z)C_2(z,z)$. Using the third relation to eliminate $C_2$ gives
\begin{equation}
m^2\Omega(z)^2=-2\partial_zC_1(z,z)
\label{eq: E5}
\end{equation}
To determine $C_1(z,z')$, we now take an inverse Fourier transform:
\begin{equation}
\begin{aligned}
\begin{bmatrix}
C_1(z,z')\\
C_2(z,z')
\end{bmatrix}
&=\int_{-\infty}^{+\infty}\frac{d\omega}{2\pi}e^{-\i\omega z'}\left[\phi(\omega,z)-\phi^{(0)}(\omega,z)\right]\\
&=\int_{-\infty}^{+\infty}\frac{d\omega}{2\pi}e^{\i\omega z'}\left[\phi(-\omega,z)-\phi^{(0)}(-\omega,z)\right]\\
&=\int_{-\infty}^{+\infty}\frac{d\omega}{2\pi}e^{\i\omega z'}\left[\mS(\omega)\sigma_1\phi(\omega,z)+\frac{\sigma_1\tilde\varphi(\omega,z)}{\mathcal J(\omega)}-\phi^{(0)}(-\omega,z)\right]
\end{aligned}
\label{eq: E6}
\end{equation}
where in the first step we relabeled the dummy variable $\omega\to-\omega$, and in the last step we used \eqref{eq: expand physical solution onto in and out mode} and \eqref{eq: C22}.

Using \eqref{eq: unnormalized physical solution} to rewrite $\phi(-\omega,z)$ in terms of $\tilde\varphi(\omega,z)$, it is convenient to split the integrand into four pieces:
\begin{equation}
\begin{aligned}
&\left[\mS(\omega)\sigma_1\phi(\omega,z)+\frac{\sigma_1\tilde\varphi(\omega,z)}{\mathcal J(\omega)}-\phi^{(0)}(-\omega,z)\right]=\\
&\ \ \ \ \ \ \ \ \underbrace{\left[\mS(\omega)-\mS_\infty\right]\sigma_1\left[\phi(\omega,z)-\phi^{(0)}(\omega,z)\right]}_{\text{(I)}}+\underbrace{\left[\mS(\omega)-\mS_\infty\right]\sigma_1\phi^{(0)}(\omega,z)}_{\text{(II)}}\\
&\ \ \ \ \ \ \ \ +\underbrace{\mS_{\infty}\sigma_1\left[\phi(\omega,z)-\phi^{(0)}(\omega,z)\right]}_{\text{(III)}}+\sigma_1\underbrace{\left\{\frac{\tilde\varphi(\omega,z)}{\mJ(\omega)}-\left[-\mS_{\infty}\sigma_1\phi^{(0)}(\omega,z)+\phi^{(0)}(-\omega,z)\right]\right\}}_{\text{(IV)}}
\end{aligned}
\end{equation}

The key simplification is that (III) and (IV) do not contribute once we do the $\omega$ contour integral. For (III), using the bound \eqref{eq: D35} we have $|e^{\i\omega z'}(\phi(\omega,z)-\phi^{(0)}(\omega,z))|\sim e^{-(\Im\omega)(z+z')}$ for $\Im\omega>0$. We can therefore close the contour in the upper half-plane, and analyticity of $\phi(\omega,z)$ there (cf. \eqref{eq: bound on phi mode function}) implies
\begin{equation}
\int_{-\infty}^{+\infty}\frac{d\omega}{2\pi}e^{\i\omega z'}\text{(III)}=0.
\label{eq: E8}
\end{equation}
Similarly, for (IV) the bound \eqref{eq: D39} gives $|e^{\i\omega z'}\text{(IV)}|\sim e^{-(\Im\omega)(z'-z)}$ for $\Im\omega>0$. Since $z'>z$, we again close the contour upstairs, and analyticity (cf. \eqref{eq: D38}) gives
\begin{equation}
\int_{-\infty}^{+\infty}\frac{d\omega}{2\pi}e^{\i\omega z'}\text{(IV)}=0.
\label{eq: E9}
\end{equation}

Thus only (I) and (II) remain. Define:
\begin{equation}
\tilde S(z)=\int_{-\infty}^{+\infty}\frac{d\omega}{2\pi}e^{\i\omega z}[\mS(\omega)-\mS_\infty]
\end{equation}
Combining this definition with \eqref{eq: E6} gives the Marchenko equation for the kernels $C_{1,2}(z,z')$:
\begin{equation}
C_1(z,z')=\int_{z}^{+\infty}dz''\tilde S(z'+z'')C_2(z,z''),\ \ C_2(z,z')=\tilde S(z+z')+\int_{z}^{+\infty}dz''\tilde S(z'+z'')C_1(z,z'')
\label{eq: marchenko equation}
\end{equation}
It is convenient to introduce
\begin{equation}
\tilde h(z_1,z_2):=\tilde S(z_1+z_2),\ \tilde h_z(z_1,z_2):=\tilde S(z_1+z_2)\Theta(z_2-z)
\end{equation}
and to regard $C_1,C_2,\tilde h,\tilde h_z$ as integral operators on $L^2(\mathbb R)$. In operator notation, the Marchenko system becomes
\begin{equation}
C_1=C_2\tilde h_z^\intercal,\ \ C_2=\tilde h+C_1\tilde h_z^\intercal
\end{equation}
Eliminating $C_2$ yields
\begin{equation}
C_1=\tilde h\cdot \tilde h_z^\intercal(1-\tilde h_z^{\intercal2})^{-1}
\end{equation}
By \eqref{eq: E5}, the conformal factor is determined by the diagonal of $C_1$, which we extract using a Fredholm determinant identity:
\begin{equation}
\begin{aligned}
\partial_z&\log\det(\textbf{1}-\tilde h_z^{\intercal2})\\
&=-\Tr[(1-\tilde h_z^{\intercal2})^{-1}(\partial_z\tilde h_z^\intercal\cdot \tilde h_z^\intercal+\tilde h_z^\intercal \cdot\partial_z\tilde h_z^\intercal)]\\
&=-\int dz_1dz_2dz_3(1-\tilde h_z^{\intercal2})^{-1}_{(z_1,z_2)}\left[(\partial_z\tilde h_z)_{(z_3,z_2)}(\tilde h^\intercal_z)_{(z_3,z_1)}+(\tilde h^\intercal_z)_{(z_2,z_3)}(\partial_z\tilde h_z)_{(z_1,z_3)}\right]\\
&=\int dz_1dz_2dz_3(1-\tilde h_z^{\intercal2})^{-1}_{(z_1,z_2)}\left[\tilde S(z_3+z_2)\delta(z-z_2)(\tilde h_z^\intercal)_{(z_3,z_1)}+\tilde S(z_1+z_3)\delta(z-z_3)(\tilde h_z^\intercal)_{(z_2,z_3)}\right]\\
&=\int dz_1dz_3\cdot \tilde S(z+z_3)(\tilde h_z^\intercal)_{(z_3,z_1)}(1-\tilde h_z^{\intercal2})^{-1}_{(z_1,z)}+\int dz_1dz_2\cdot \tilde S(z+z_1)(1-\tilde h_z^{\intercal2})^{-1}_{(z_1,z_2)}(\tilde h_z^\intercal)_{(z_2,z)}\\
&=(\tilde h\tilde h_z^\intercal(1-\tilde h_z^{\intercal2})^{-1})_{(z,z)}+(\tilde h(1-\tilde h_z^{\intercal2})^{-1}\tilde h_z^\intercal)_{(z,z)}\\
&=2(\tilde h\tilde h_z^\intercal(1-\tilde h_z^{\intercal2})^{-1})_{(z,z)}\\
&=2C_1(z,z)
\end{aligned}
\end{equation}
Combine with \eqref{eq: E5}, we obtain:
\begin{equation}
m^2\Omega(z)^2=-\partial_z^2\log\det(\textbf{1}-\tilde h_z^{\intercal2})
\end{equation}
To streamline notation, introduce the projector $P_z=\textbf{1}_{[z,+\infty)}$ so that $\tilde h_z=\tilde hP_z$. Using the expansion of $\log\det$ in traces and the identity $\Tr[(P_z\tilde h)^{2n}]=\Tr[(P_z\tilde hP_z)^{2n}]$, we obtain
\begin{equation}
m^2\Omega(z)^2=-\partial_z^2\log\det[\textbf{1}-(P_z\tilde hP_z)^2]
\label{eq: PhP in appendix}
\end{equation}
Relative to \eqref{eq: mOmega in terms of P_zhP_z} in the main text, the only difference is that here we used $\mS(\omega)-\mS_\infty$ in defining $\tilde S(z)$. Physically, subtracting $\mS_\infty=-1$ removes the hard-wall contribution from the boundary scatterer at $z=0$. Since we are interested in $\Omega(z)$ for $z>0$, this subtraction is inessential: at leading order (cf. \eqref{eq: B8}) it only produces a $\delta(z)$ term, and in the pole expansion the constant piece does not affect the sum.

Accordingly, we may drop the $-\mS_\infty$ subtraction and recover \eqref{eq: mOmega in terms of P_zhP_z} exactly as stated in the main text.

To connect with \eqref{eq: main result1} and \eqref{eq: main result2}, we now use the pole expansion of $S(z)$ on the upper half-plane (cf. \eqref{eq: 2.47} in the main text):
\begin{equation}
S(z)=\sum_i\beta_ie^{-\lambda_i z},\ z>0
\label{eq: E18}
\end{equation}
We only need the projected operator $\mh_z:=P_zhP_z$, which acts on functions supported on $[z,\infty)$. To make the $z$-dependence explicit, shift variables $z_i=z+x_i$ with $x_i>0$ and work on $\mathcal H_{\text{con}}:=L^2(\mathbb R_{>0})$. Then $\mh_z$ acts as
\begin{equation}
\begin{aligned}
(\mh_zf)(x_1)&=\int_0^{+\infty}dx_2\ S(2z+x_1+x_2)f(x_2)\\
&=\sum_{n}\beta_ne^{-\lambda_n2z}e^{-\lambda_nx_1}\int_0^{+\infty}dx_2\ e^{-\lambda_nx_2}f(x_2)
\label{eq: E19}
\end{aligned}
\end{equation}
This factorized form suggests introducing an auxiliary discrete space $\mathcal H_{\text{dis}}:=\operatorname{span}\{|n\rangle\}$ and operators
\begin{equation}
T_1:\mathcal H_\text{dis}\rightarrow\mathcal H_\text{con},\ \langle x|T_1|n\rangle:=\beta_ne^{-\lambda_n2z}e^{-\lambda_nx}
\end{equation}
\begin{equation}
T_2:\mathcal H_\text{con}\rightarrow\mathcal H_\text{dis},\ \langle n|T_2|x\rangle:=e^{-\lambda_nx}
\end{equation}
so that $\mh_z=T_1T_2$. Using Sylvester's determinant theorem, $\det(\textbf{1}-T_1T_2T_1T_2)=\det(\textbf{1}-T_2T_1T_2T_1)$, we obtain
\begin{equation}
m^2\Omega(z)^2=-\partial_z^2\log\det(\textbf{1}-\mh_z^2)=-\partial_z^2\log\det(\textbf{1}_{\mathcal H_\text{con}}- T_1T_2T_1T_2)=-\partial_z^2\log\det(\textbf{1}_{\mathcal H_\text{dis}}- T_2T_1T_2T_1)
\label{eq: E24}
\end{equation}
We now introduce the auxiliary matrix $B(z)$ acting on $\mathcal H_\text{dis}$,
\begin{equation}
\begin{aligned}
B(z):=T_2T_1,
\end{aligned}
\end{equation}
with matrix elements
\begin{equation}
\begin{aligned}
B(z)_{nm}&=\langle n|T_2T_1|m\rangle=\int_0^{+\infty} dx\langle n|T_2|x\rangle\langle x|T_1|m\rangle=\int_0^{+\infty}dx\cdot \beta_me^{-\lambda_m2z}e^{-(\lambda_m+\lambda_n)x}\\
&=\frac{\beta_me^{-\lambda_m2z}}{\lambda_n+\lambda_m}
\end{aligned}
\end{equation}
Substituting into \eqref{eq: E24} yields the compact form of \eqref{eq: main result1} and \eqref{eq: main result2}:
\begin{equation}
m^2\Omega(z)^2=-\partial_z^2\log\det[\textbf{1}-B(z)^2]
\end{equation}

\section{Technical details of near horizon algebra}
\label{app: Technical details of near horizon algebra}
This appendix collects technical details related to the near-horizon algebra discussed in \cref{sec: algebra near horizon}. In \cref{app: sl2R algebra in AdS2 Rindler} we benchmark that the definition of $P^\pm$ and $B$ in \eqref{eq: defintion of three generators} matches the exact sl(2,$\mathbb R$) generators for a massive Majorana fermion in AdS$_2$--Rindler with an asymptotic boundary and standard normalizable boundary conditions. In \cref{app: derive eq 6.13} we derive \eqref{eq: action of [P,P] is B on bifurcate horizon modes}, showing that the commutator $[P^-,P^+]$ acting on bifurcate-horizon modes is proportional to the boost generator.
\subsection{sl(2,$\mathbb R$) algebra in AdS$_2$-Rindler}
\label{app: sl2R algebra in AdS2 Rindler}
Consider a massive Majorana fermion on AdS$_2$-Rindler, with metric:
\begin{equation}
ds^2=\Omega(z)^2(-dt^2+dz^2),\ \Omega(z)=\frac{1}{\sinh z},\ z\geq0
\end{equation}
On any constant $t$ slice, one can write down the three generators of sl(2,$\mathbb R$) represented in terms of fermion bilinear:
\begin{equation}
\begin{aligned}
&H(t):=\frac{\i}{4\pi}\int_0^{+\infty} dz\cdot \Psi_L\partial_z\Psi_L-\Psi_R\partial_z\Psi_R-2m\Omega(z)\Psi_L\Psi_R\\
&L^+(t):=\frac{\i}{4\pi}e^{t}\int_0^{+\infty} dz\cdot e^{z}\Psi_L\partial_z\Psi_L-e^{-z}\Psi_R\partial_z\Psi_R-2m(\coth z)\Psi_L\Psi_R\\
&L^-(t):=\frac{\i}{4\pi}e^{-t}\int_0^{+\infty} dz\cdot e^{-z}\Psi_L\partial_z\Psi_L-e^{z}\Psi_R\partial_z\Psi_R-2m(\coth z)\Psi_L\Psi_R
\end{aligned}
\label{eq: three AdS2 generators}
\end{equation}
The boost generator $H(t)$ is the Hamiltonian, which follows from the action \eqref{eq: action of QFT}. The ladder operators $L^\pm$ can be constructed by writing the corresponding Killing vector field $\xi^t\partial_t+\xi^z\partial_z$, then replacing $\partial_t$ by the Hamiltonian density $\mathcal H(t,z)\propto \Psi_L\partial_z\Psi_L-\Psi_R\partial_z\Psi_R-2m\Omega(z)\Psi_L\Psi_R$ and $\partial_z$ by the momentum density $\mathcal P(t,z)\propto \Psi_L\partial_z\Psi_L+\Psi_R\partial_z\Psi_R$, and finally integrating along a constant-time slice.

A straightforward calculation shows that the sl(2,$\mathbb R$) algebra holds:
\begin{equation}
[H,L^\pm]=\pm \i L^\pm,\  [L^-,L^+]=-\i RH,\ \text{with }R=-2
\label{eq: F3}
\end{equation}
where $R$ is the Ricci scalar. Owing to the explicit factors $e^{\pm t}$ in $L^\pm$, one finds that the ladder operators are in fact time independent:
\begin{equation}
\frac{d}{dt}L^{\pm}(t)=\partial_t L^{\pm}+[\i H,L^\pm]=0.
\end{equation}
Thus they may be pushed to the future/past horizons at $t=\pm\infty$.

First, we consider pushing $L^+(t)$ to past horizon ($t=-\infty$). One first change to lightcone coordinate:
\begin{equation}
L_+(t)=\frac{\i}{4\pi}\int_{t}^{+\infty}dv e^{v}\Psi_L\partial_v\Psi_L-e^{2t-v}\Psi_R\partial_v\Psi_R-e^{t}2m\coth{\frac{v-t}{2}}\Psi_L\Psi_R
\end{equation}
As $t\to-\infty$, the last two terms drop out due to the suppression by $e^{t}$ and $e^{2t}$, and in the first term the fermion $\Psi_L(u,v)|_{u=2t-v}$ becomes the horizon mode $\Psi^p(v)$. Therefore,
\begin{equation}
L^+(-\infty)=\frac{\i}{4\pi}\int_{-\infty}^{+\infty}dv\cdot \Psi^p(v)e^{v}\partial_v\Psi^p(v)\\
=\frac{\i}{4\pi}\int_{-\infty}^0 dV\cdot \Upsilon^p\partial_V\Upsilon^p
\end{equation}
where in the last equality we used \eqref{eq: from Psi to Upsilon}, i.e. $\Psi^p(v)=e^{-v/2}\Upsilon^p(V)$, to pass to Kruskal coordinates. Comparing with \eqref{eq: defintion of three generators}, we find $L^+(-\infty)\equiv P^+$. A similar computation gives $L^-(+\infty)\equiv P^-$. 

Then, we consider pushing $H(t)$ to past horizon ($t=-\infty$). Similarly, one first change to lightcone coordinate:
\begin{equation}
H(t)=\frac{\i}{4\pi}\int_t^{+\infty}dv\cdot\Psi_L\partial_v\Psi_L-\Psi_R\partial_v\Psi_R-2m(\sinh{\frac{v-t}{2}})^{-1}\Psi_L\Psi_R
\end{equation}
As $t\to-\infty$, the third term drops out due to $(\sinh \frac{t}{2})^{-1}$ suppression, and the second term also drops out since $\partial_v\Psi_R\to0$. This is because, in light-cone quantization, the right-mover along the past horizon becomes non-dynamical.  Therefore, we find:
\begin{equation}
H(-\infty)=\frac{\i}{4\pi}\int_{-\infty}^{+\infty}dv\cdot \Psi_L\partial_v\Psi_L=-\frac{\i}{4\pi}\int_{-\infty}^0 dV \cdot \Upsilon^pV\partial_V\Upsilon^p
\end{equation}
Comparing with \eqref{eq: defintion of three generators}, we indeed find $H(-\infty)\equiv B$. A similar calculation shows that  $H(+\infty)\equiv B$.

To summarize, pushing the AdS$_2$ generators defined on a constant-time slice \eqref{eq: three AdS2 generators} to the future/past horizons yields the null-translation and boost generators on the horizon defined in \eqref{eq: defintion of three generators}.

\subsection{Curvature from algebra at bifurcate horizon}
\label{app: Curvature from algebra at bifurcate horizon}
In this appendix, we will explain how does the curvature at horizon shows up from the commutator $[P^-,P^+]$ acting on modes at bifurcate horizon. First, in \cref{app: Preparation: a contour prescription in frequency space} we will describe a useful contour prescription in frequency space. Based on this, in \cref{app: derive eq 6.13}, we will derive \eqref{eq: action of [P,P] is B on bifurcate horizon modes}. In the end, we will benchmark this result on AdS$_2$-Rindler special case in \cref{app: benchmark on AdS2 Rindler case}.
\subsubsection{Preparation: a contour prescription in frequency space}
\label{app: Preparation: a contour prescription in frequency space}
In this section we motivate and develop a contour prescription that will be important in \cref{app: derive eq 6.13}.

Consider the anti-commutator between the bifurcate-horizon mode $\Upsilon^p(0)$ and a boundary operator $\chi(t)$. Since the bifurcate horizon is spacelike separated from any boundary point, one expects $\{\Upsilon^p(0),\chi(t)\}=0$ for all $t$. This can be checked directly:
\begin{equation}
\begin{aligned}
\{\Psi^p(v),\chi(t)\}&=\int dt'\mathcal K^p(t'-v)A(t-t')\\
&=\int d\omega e^{-\i\omega(v-t)}\mathcal K^p(-\omega)\mA(\omega)\\
&=\int d\omega e^{-\i\omega(v-t)}\mathcal R^p(\omega)
\end{aligned}
\end{equation}
In the first line we used the definition of the horizon mode in \eqref{eq: define past and future mode}. In the second and third lines we used $\mathcal K^p(\omega)\mA(\omega)\mathcal K^p(-\omega)=1$ from \eqref{eq: KAK=1, continuous} and $\mathcal R^p(\omega):=\frac{1}{\mathcal K^p(\omega)}$. In \cref{sec: Causal kernel decomposition and horizon modes} we showed that, in the absence of bound states in the upper half-plane, $\mathcal R^p(\omega)$ has the same analyticity properties as $\mK^p(\omega)$. It follows that $\{\Psi^p(v),\chi(t)\}=0$ for $t<v$. Using $\Psi^p(v)=\sqrt{-2\lambda_0V}\Upsilon^p(V)$ and $V=-e^{-2\lambda_0 v}$, we can send $v\to+\infty$ (equivalently $V\to0$) to obtain $\Upsilon^p(0)$ at the bifurcate horizon, and hence $\{\Upsilon^p(0),\chi(t)\}=0$ for all $t$, as expected.

Formally, the above computation corresponds to
\begin{equation}
\lim_{V\rightarrow0}\big[\{\Upsilon^p(V),\chi(t)\}\big]=0.
\end{equation}
It is illuminating to consider the opposite order of limits,
\begin{equation}
\{\lim_{V\rightarrow0}\big[\Upsilon^p(V)\big],\chi(t)\},
\end{equation}
i.e. to first take $V\to0$ in $\Upsilon^p(V)$ and then compute the anti-commutator. Define $\beta_k^p:=\i\operatorname{Res}[\mathcal K^p(\omega\rightarrow i\lambda_k)]$ in the upper half-plane, so that
\begin{equation}
\mathcal K^p(t):=\Theta(-t)\sum_{k}\beta^{p}_ke^{\lambda_k t}.
\end{equation}
Then
\begin{equation}
\Upsilon^p(V)=\frac{1}{\sqrt{2\lambda_0}}e^{\lambda_0v}\int_{-\infty}^vdt'\mathcal K^p(t'-v)\chi(t')=\frac{1}{\sqrt{2\lambda_0}}\int_{-\infty}^vdt'\sum_k\beta^p_ke^{-(\lambda_k-\lambda_0)v}e^{\lambda_kt'}\chi(t').
\end{equation}
As $V\to0$ (i.e. $v\to+\infty$), only the $k=0$ term survives, giving
\begin{equation}
\Upsilon^p(0)=\frac{\beta_0^p}{\sqrt{2\lambda_0}}\int_{-\infty}^{+\infty}dt'e^{\lambda_0t'}\chi(t').
\end{equation}
The corresponding anti-commutator is
\begin{equation}
\{\Upsilon^p(0),\chi(t)\}=\frac{\beta_0^p}{\sqrt{2\lambda_0}}\int_{-\infty}^{+\infty}dt'e^{\lambda_0t'}A(t,t')=\frac{\beta_0^p}{\sqrt{2\lambda_0}}\int_{-\infty}^{+\infty}dt'e^{\lambda_0t'}\sum_{n}c_ne^{-J_n|t-t'|}.
\label{eq: F15}
\end{equation}
The $dt'$ integral diverges as $t'\to+\infty$ whenever there exists any $J_n\leq\lambda_0$. In particular, if $J_0\leq\lambda_0$ then $\{\Upsilon^p(0),\chi(t)\}=\infty$ rather than $0$. This contradiction shows that the two limits do not commute:
\begin{equation}
\lim_{V\rightarrow0}\big[\{\Upsilon^p(V),\chi(t)\}\big]\neq\{\lim_{V\rightarrow0}\big[\Upsilon^p(V)\big],\chi(t)\}.
\end{equation}

The second (naive) computation can nevertheless be repaired by a contour prescription. Rewriting \eqref{eq: F15} in frequency space gives
\begin{equation}
\{\Upsilon^p(0),\chi(t)\}=\frac{\beta_0^p}{\sqrt{2\lambda_0}}\int dt'd\omega e^{\lambda_0t'}e^{\i\omega(t-t')}A(\omega)=\frac{\beta_0^p}{\sqrt{2\lambda_0}}\int dt'd\omega e^{-i(\omega+i\lambda_0)t'}e^{\i\omega t}A(\omega).
\end{equation}
Both $t'$ and $\omega$ are initially integrated along $\mathbb R$. If we shift the $\omega$ contour from $\mathbb R$ to $\mathbb R-\i\lambda_0$, i.e. set $\omega=\omega'-\i\lambda_0$ with $\omega'\in\mathbb R$, then the $dt'$ integral produces $\delta(\omega')$, and we obtain
\begin{equation}
\{\Upsilon^p(0),\chi(t)\}\supset \frac{\beta_0^p}{\sqrt{2\lambda_0}}e^{\lambda_0t}\mA(-\i\lambda_0)=0,
\end{equation}
where we used that $-\i\lambda_0$ is a zero of $A(\omega)$. This reproduces the expected result.

However, in shifting the $\omega$ contour from $\mathbb R$ to $\mathbb R-\i\lambda_0$, one crosses poles of $\mA(\omega)$ at $\{-\i J_n\mid J_n<\lambda_0\}$. These poles generate precisely the divergent real-space contributions discussed below \eqref{eq: F15}. The contour prescription is therefore: \textit{``When shifting the $\omega$ contour from $\mathbb R$ to $\mathbb R-\i\lambda_0$, ignore the poles at $\{-\i J_n\mid J_n<\lambda_0\}$.''}

In the next section we will see that the same prescription is required to derive \eqref{eq: action of [P,P] is B on bifurcate horizon modes}.

\subsubsection{Derive \eqref{eq: action of [P,P] is B on bifurcate horizon modes}}
\label{app: derive eq 6.13}
The physics of the derivation is simple: $[P^-,P^+]$ is written as a bilocal operator on the past horizon, but when it acts on the bifurcate-horizon mode $\Upsilon^p(0)$ only the $V\to0$ region matters, and the bilocal kernel collapses to a local boost action.

Starting from \eqref{eq: kernel of PP commutator} and using \eqref{eq: S-matrix of Upsilon} and \eqref{eq: S(z)}, we can write
\begin{equation}
[P^-,P^+]=-\frac{(2\lambda_0)^{-2}}{8\pi^2}\int_{-\infty}^{+\infty} dv_1dv_2\Psi^p(v_1)\Psi^p(v_2)f(v_1-v_2)
\label{eq: non-local kernel}
\end{equation}
with
\begin{equation}
\text{with: }f(v):=2\sinh(\lambda_0v)\partial_v^2\int_{-\infty}^{+\infty} du e^{-2\lambda_0 u}S(\frac v2-u)S(-\frac v2-u)
\label{eq: definition of f(v)}
\end{equation}
To see how this becomes local, we go to Fourier space,
\begin{equation}
f(v):=\int d\omega e^{\i\omega v}\tilde f(\omega).
\end{equation}
Expanding $\tilde f(\omega)$ around $\omega=0$ packages the bilocal kernel into a local differential operator acting along the horizon coordinate $v$:
\begin{equation}
[P^-,P^+]=-\frac{(2\lambda_0)^{-2}}{4\pi}\int dv \Psi^p(v)\tilde f(-\i\partial_v)\Psi^p(v).
\end{equation}
Now switch to Kruskal $V$ using \eqref{eq: from Psi to Upsilon}. The point is that $\partial_v$ becomes a dilatation operator $2\lambda_0 V\partial_V$, and the extra $\sqrt{-V}$ factors shift the differential operator by $\i\lambda_0$. Concretely,
\begin{equation}
[P^-,P^+]=-\frac{1}{4\pi}(2\lambda_0)^{-2}\int \frac{dV}{-2\lambda_0V}\sqrt{-2\lambda_0V}\Upsilon^p(V)\tilde f(\i2\lambda_0V\partial_V)\left[\sqrt{-2\lambda_0V}\Upsilon^p(V)\right].
\end{equation}
Using $(\i2\lambda_0 V\partial_V)(\sqrt{-2\lambda_0V})=(\sqrt{-2\lambda_0V})(\i2\lambda_0 V\partial_V+\i\lambda_0)$ and iterating, we obtain
\begin{equation}
[P^-,P^+]=-\frac{(2\lambda_0)^{-2}}{4\pi}\int dV\Upsilon^p(V)\tilde f(\i2\lambda_0V\partial_V+\i\lambda_0)\Upsilon^p(V)
\label{eq: F24}
\end{equation}
so that
\begin{equation}
\left[[P^-,P^+],\Upsilon^p(V)\right]=(2\lambda_0)^{-2}\tilde f(2\i\lambda_0V\partial_V+\i\lambda_0)\Upsilon^p(V).
\end{equation}
For generic $V$ this is nonlocal, but at the bifurcate horizon $V=0$ it collapses to
\begin{equation}
\left[[P^-,P^+],\Upsilon^p(0)\right]=(2\lambda_0)^{-2}\tilde f(\i\lambda_0)\Upsilon^p(0).
\end{equation}
Finally, using \eqref{eq: action of boost},
\begin{equation}
[B,\Upsilon^p(0)]=\i\frac12\Upsilon^p(0),
\end{equation}
we conclude:
\begin{equation}
\left[[P^-,P^+],\Upsilon^p(0)\right]=-2\i(2\lambda_0)^{-2}\tilde f(\i\lambda_0)\times [B,\Upsilon^p(0)]
\end{equation}
Comparing with \eqref{eq: action of [P,P] is B on bifurcate horizon modes}, we define the \emph{algebraic curvature}
\begin{equation}
R_\text a:=2\beta_0^{-2}\tilde f(\i\lambda_0).
\end{equation}

Our goal is to show that $R_\text a$ equals the geometric curvature at the horizon, $R_\text h$. Using \eqref{eq: definition of f(v)}, we can express $\tilde f(\i\lambda_0)$ via an inverse Fourier transform:
\begin{equation}
R_\text{a}/m^2=(-2\beta_0^{-2})\int_{-\infty}^{+\infty} dv (e^{2\lambda_0 v}-1)\partial_v^2\int_{-\infty}^{+\infty} du e^{-2\lambda_0 u}S(\frac v2-u)S(-\frac v2-u).
\end{equation}
We split the integrand according to $(e^{2\lambda_0v}-1)$:
\begin{equation}
R_\text a/m^2:=R^{(1)}_\text a/m^2+R^{(2)}_\text a/m^2.
\end{equation}
\begin{equation}
\begin{aligned}
\text{with: }&R^{(1)}_\text a/m^2:=-\frac{\beta_0^{-2}}{2}\int dv_+dv_-e^{-\lambda_0v_+}e^{3\lambda_0 v_-}(\partial_{v_+}-\partial_{v_-})^2S(v_+)S(v_-)\\
&R^{(2)}_\text a/m^2:=2\beta_0^{-2}\int dvdu e^{-2\lambda_0u}\partial_v^2[S(\frac v2-u)S(-\frac v2-u)]
\end{aligned}
\label{eq: Ra(1), Ra(2)}
\end{equation}
where for $R^{(1)}_\text a$ we changed variables to $v_\pm:=\frac v2\pm u$ for later convenience; the integration contours for $v_\pm$ lie along $\mathbb R$.

To analyze potential divergences, we recall the asymptotic behavior of $S(z)$. From \eqref{eq: 2.47}, for $z>0$ we have $S(z)=\sum_k\beta_ke^{-\lambda_kz}$. For $z<0$, the decay is controlled by poles of $S(\omega)$ in the lower half-plane (at $\{-\i J_n\}$), so $S(z)=\sum_n\tilde \beta_ne^{J_nz}$ for $z<0$, with $\tilde\beta_n\propto \operatorname{Res}[\mS(\omega\rightarrow-\i J_n)]$.

Let's first focus on $R^{(1)}_\text a/m^2$. 

For the $v_-$ integral, a potential divergence arises as $v_-\to+\infty$, where the integrand behaves as $e^{(3\lambda_0-\lambda_k)v_-}$. The leading divergence from the $k=0$ term, $\sim e^{2\lambda_0 v_-}$, will be removed below by an appropriate choice of the order of integration. Additional divergences occur whenever there exists $k\geq1$ with $\lambda_k<3\lambda_0$. In particular, if $\lambda_1<3\lambda_0$ then the $v_-$ integral diverges. This divergence is physical and reproduces Case~III in \cref{sec: Curvature near horizon and AdS universality}, where the horizon curvature diverges for $\lambda_1<3\lambda_0$.

\begin{figure}[t]
    \centering
    \includegraphics[width=0.4\linewidth]{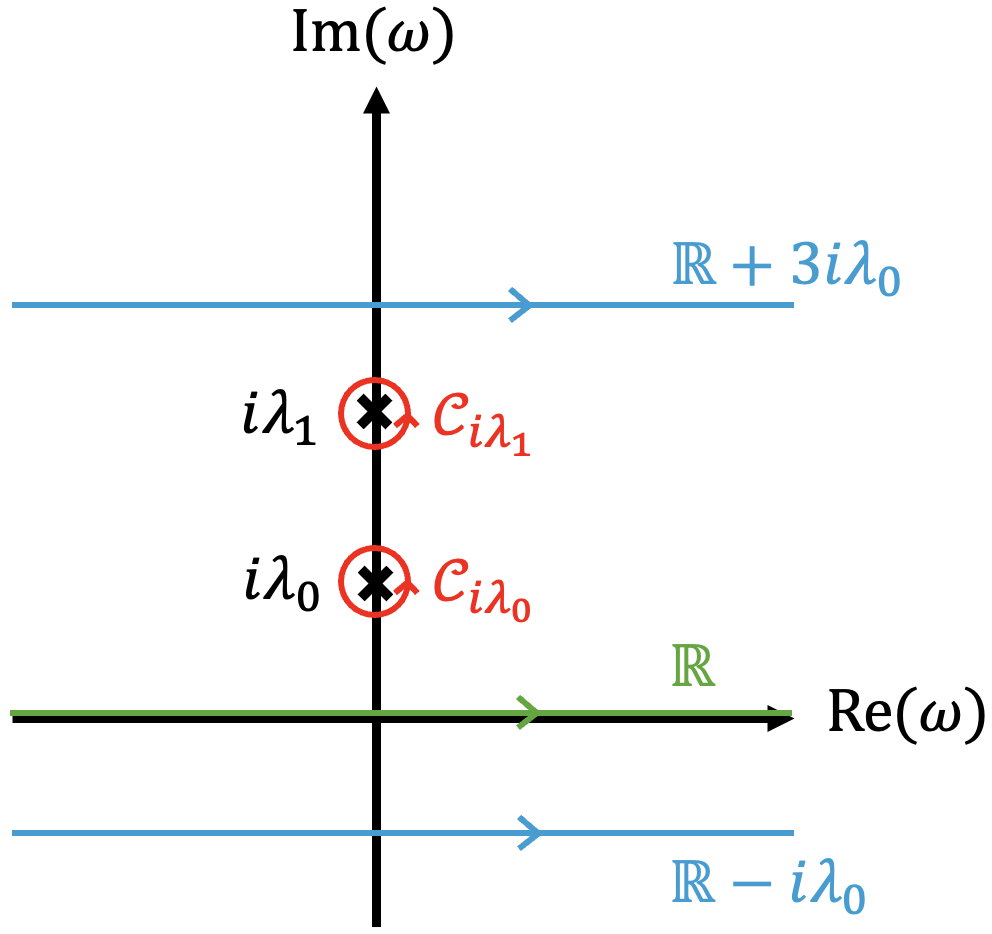}
    \caption{Various contours in the $\omega$ plane, used in \cref{app: derive eq 6.13}.}
    \label{fig: contour of omega}
\end{figure}

For the $v_+$ integral, a potential divergence arises as $v_+\to-\infty$, where the integrand behaves as $e^{-(\lambda_0-J_n)v_+}$. Thus, if there exists any $J_n\leq\lambda_0$ (in particular if $J_0\leq\lambda_0$), the $v_+$ integral diverges. As discussed in \cref{app: Preparation: a contour prescription in frequency space}, this divergence is unphysical and reflects the wrong order of limits in defining $\Upsilon^p(0)$. We remove it by implementing the same contour prescription in Fourier space:
\begin{equation}
R^{(1)}_\text a/m^2:=\frac{\beta_0^{-2}}{2}\int dv_+dv_-\frac{d\omega_+}{2\pi}\frac{d\omega_-}{2\pi}e^{\i(\omega_++i\lambda_0)v_+}e^{\i(\omega_--3i\lambda_0) v_-}(\omega_+-\omega_-)^2\mS(\omega_+)\mS(\omega_-)
\end{equation}
The defining contour of $\omega_\pm$ is along $\mathbb R$. To perform $v_+,v_-$ integral, we will move the $\omega_+,\omega_-$ contour to $\mathbb R-\i\lambda_0$ and $\mathbb R+3\i\lambda_0$ respectively. As explained before, the contour prescription is that, when  moving $\omega_+$ contour, we ignore any possible poles of $\mS(\omega_+)$ at $\{-\i J_n|J_n\leq\lambda_0\}$. When moving $\omega_-$, we will pick up poles of $\mS(\omega_-)$ at $\i\lambda_0$ and possibly at $\i\lambda_k$ if $k\geq1,\lambda_k<3\lambda_0$ (we will use $\lambda_1$ as representative). See \cref{fig: contour of omega}.

Therefore, $R_\text a^{(1)}$ is further separated into three terms denoted as $R_\text a^{(1)}:=R_\text a^{(1,1)}+R_\text a^{(1,2)}+R_\text a^{(1,3)}$:
\begin{equation}
\underbrace{\int_{\mathbb R}d\omega_+\int_{\mathbb R}d\omega_-}_{ R^{(1)}_\text a}\equiv\underbrace{\int_{\mathbb R-i\lambda_0}d\omega_+\int_{\mathbb R+3i\lambda_0}d\omega_-}_{ R^{(1,1)}_\text a}+\underbrace{\int_{\mathbb R}d\omega_+\int_{\mathcal C_{i\lambda_1}}d\omega_-}_{R^{(1,2)}_\text a}+\underbrace{\int_{\mathbb R}d\omega_+\int_{\mathcal C_{i\lambda_0}}d\omega_-}_{R^{(1,3)}_\text a}
\end{equation}

We first evaluate $R^{(1,1)}_\text a$. One first change  $\omega_+:=\tilde\omega_+-\i\lambda_0$, $\omega_-=\tilde\omega_-+3\i\lambda_0$, with $\tilde\omega_{\pm}$ along $\mathbb R$. Then, integration of $v_+,v_-$ along $\mathbb R$ is convergent and impose $\delta(\tilde\omega_+)\delta(\tilde\omega_-)$. We eventually obtain:
\begin{equation}
R^{(1,1)}_\text a/m^2=-2\beta_0^{-2}(2\lambda_0)^2\mS(3\i\lambda_0)\mS(-\i\lambda_0)
\end{equation}
Notice that, $\mS(-\i\lambda_0)=\frac{1}{\mS(\i\lambda_0)}=0$, since $\i\lambda_0$ is a simple pole of $\mS(\omega)$. So, $R^{(1,1)}_\text a$ is zero in general, unless $\mS(3\i\lambda_0)$ hit another pole. This is precisely the critical case when $\lambda_1=3\lambda_0$, where $R^{(1,1)}_\text a$ will become finite. To summarize:
\begin{equation}
R^{(1,1)}_\text a/m^2=\begin{cases}
0, \ \ \ \ \ \ \ \ \ \ \ \ \ \ \ \lambda_1>3\lambda_0\\
(-2)\frac{4\lambda_0^2\beta_1}{\beta_0},\ \  \ \ \lambda_1=3\lambda_0\\
\text{finite value}, \ \ \lambda_1<3\lambda_0
\end{cases}
\label{eq: Ra(1,1)}
\end{equation}

Now, we evaluate $R^{(1,2)}_\text a$. One first integrates over $\omega_-$ along  $\mC_{\i\lambda_1}$, the integrand becomes $e^{\i(\omega_++i\lambda_0)v_+}$ $e^{(3\lambda_0-\lambda_1)v_-}$$(\omega_+-i\lambda_1)^2\mS(\omega_+)\beta_1$. Then, we further change $v_\pm$ back to $u,v$ coordinate, and the two exponential becomes $e^{\i(\omega_+-2\i\lambda_0+\i\lambda_1)\frac v2}e^{\i(\omega_++4\i\lambda_0-\i\lambda_1)u}$. When $\lambda_0<\lambda_1<3\lambda_0$, we have $-\lambda_0<\lambda_1<\lambda_0$. Therefore, we can move the contour of $\omega_+$ from $\mathbb R$ to $\mathbb R+\i(2\lambda_0-\lambda_1)$, to make $v$ integral convergent. By our contour prescription, we ignore the possible poles at $\{-\i J_n|J_n\leq\lambda_0\}$. So, we change variable $\omega_+:=\tilde\omega_++\i(2\lambda_0-\lambda_1)$, $\tilde\omega_+\in\mathbb R$. Integrate over $v$ along $\mathbb R$ impose $\delta(\tilde\omega)$. The remaining $du$ integral along $\mathbb R$ is divergent due to integrand $e^{-(3\lambda_0-\lambda_1)(2u)}(\lambda_0-\lambda_1)^2\mS(2\i\lambda_0-\i\lambda_1)\beta_1$. Therefore, we conclude:
\begin{equation}
R^{(1,2)}_\text a/m^2=\begin{cases}
\infty,  \ \lambda_1<3\lambda_0\\
0,\ \ \ \text{otherwise}
\end{cases}
\label{eq: Ra(1,2)}
\end{equation}

Then, we evaluate $R^{(1,3)}_\text a$. The procedure is similar as  $R^{(1,2)}_\text a$. One first integrates over $\omega_-$ along  $\mC_{\i\lambda_0}$, the integrand becomes $e^{i(\omega_++i\lambda_0)v_+}$$e^{2\lambda_0v_-}$$(\omega_+-\i\lambda_0)^2\mS(\omega_+)\beta_0$. Then, we further change $v_\pm$ back to $u,v$ coordinate, and the two exponential becomes $e^{\i(\omega_+-\i\lambda_0)\frac v2}e^{\i(\omega_++3\i\lambda_0)u}$. Now, to make $v$ integral convergent, we shift contour of $\omega_+$ from $\mathbb R$ to $\mathbb R+\i\lambda_0$. In this process, no pole is crossed. One might worry that $\mS(\omega_+)$ has a simple pole at $\omega_+=\i\lambda_0$, which exactly sit on deformed contour $\mathbb  R+\i\lambda_0$, but the pole is canceled by double zero $(\omega_+-\i\lambda_0)^2$ term in the front. So, we change variable $\omega_+:=\tilde\omega_++\i\lambda_0$, $\tilde\omega_+\in\mathbb R$. Integrate over $v$ along $\mathbb R$ impose $\delta(\tilde\omega_+)$. The integrand of the remaining $du$ integral, $\lim_{\tilde\omega_+\rightarrow0}[e^{\i(4\i\lambda_0+\tilde\omega_+)u}\tilde\omega_+^2\mS(\tilde\omega_++\i\lambda_0)\beta_0]$, is zero, again due to the fact that double zero factor cancels a simple pole and leaving a simple zero. Therefore, if we delay the $du$ integral to the last, we conclude that:
\begin{equation}
R^{(1,3)}_\text a/m^2=0
\label{eq: Ra(1,3)}
\end{equation}

Finally, we calculate $R_\text a^{(2)}$. In Fourier space, \eqref{eq: Ra(1), Ra(2)} becomes:
\begin{equation}
R^{(2)}_\text a/m^2=-\frac{\beta_0^{-2}}{2}\int dudv\frac{d\omega_1}{2\pi}\frac{d\omega_2}{2\pi}e^{\i(\omega_1-\omega_2)\frac v2}e^{\i(-\omega_1-\omega_2+2\i\lambda_0) u}(\omega_1-\omega_2)^2\mS(\omega_1)\mS(\omega_2)
\end{equation}
First imagine moving the first pole slightly above: $\lambda_0\rightarrow\lambda_0+\epsilon$, where one replace $\mS(\omega)$ by $\mS_\epsilon(\omega)$. Then one can safely move the contour of $\omega_{1,2}$ from $\mathbb R$ to $\mathbb R+\i\lambda_0$. So, define $\omega_{1,2}:=\tilde\omega_{1,2}+\i\lambda_0$. In this process no pole is crossed. Then one can integrate $u,v$ along $\mathbb R$ and get:
\begin{equation}
\begin{aligned}
R^{(2)}_\text a/m^2&=-\frac{\beta_0^{-2}}{2}\int d\tilde\omega_1d\tilde\omega_2\delta(\frac{\tilde\omega_1-\tilde\omega_2}{2})\delta(\tilde\omega_1+\tilde\omega_2)(\tilde\omega_1-\tilde\omega_2)^2\mS_\epsilon(\i\lambda_0+\tilde\omega_1)\mS_\epsilon(\i\lambda_0+\tilde\omega_2)\\
&=-2\beta_0^{-2}\int d\tilde\omega_1\delta(\omega_1)\tilde\omega_1^2\mS_\epsilon(\i\lambda_0+\tilde\omega_1)\mS_\epsilon(\i\lambda_0-\tilde\omega_1)\\
&=(-2\beta_0^{-2})\lim_{\tilde\omega_1\rightarrow0}\lim_{\epsilon\rightarrow0}[\tilde\omega_1^2\mS_\epsilon(\i\lambda_0+\tilde\omega_1)\mS_\epsilon(\i\lambda_0-\tilde\omega_1)]\\
&=-2
\end{aligned}
\label{eq: Ra(2)}
\end{equation}
where in the second line, we integrate over $d\tilde\omega_2\delta(\tilde\omega_1+\tilde\omega_2)$, and in the last line we notice that $-\i\beta_0$ is the residual of $\mS(\omega)$ at $\i\lambda_0$.

To summarize, combining $R_\text a^{(1,1)}$, $R_\text a^{(1,2)}$, $R_\text a^{(1,3)}$, and $R_\text a^{(2)}$ from \eqref{eq: Ra(1,1)}, \eqref{eq: Ra(1,2)}, \eqref{eq: Ra(1,3)}, and \eqref{eq: Ra(2)}, we obtain
\begin{equation}
R_\text a/m^2=\begin{cases}
-2, &\lambda_1>3\lambda_0\\
-2(1+\frac{4\lambda_0^2\beta_1}{\beta_0^2}), &\lambda_1=3\lambda_0\\
\infty, &\lambda_1<3\lambda_0
\end{cases}
\end{equation}
This matches the three cases of horizon curvature discussed in \cref{sec: Curvature near horizon and AdS universality}. Therefore,
\begin{equation}
R_\text a=R_\text h.
\end{equation}
This completes the derivation of \eqref{eq: action of [P,P] is B on bifurcate horizon modes}.
\subsubsection{Benchmark on AdS$_2$-Rindler case}
\label{app: benchmark on AdS2 Rindler case}
In the special case of AdS$_2$--Rindler, we expect the nonlocal kernel \eqref{eq: non-local kernel} to become local and $[P^-,P^+]$ to be exactly proportional to $B$. To check this, we first take the inverse Fourier transform of $f(v)$ in \eqref{eq: definition of f(v)} to obtain $\tilde f(\omega)$:
\begin{equation}
\tilde f(\omega)=(\i\lambda_0+\omega)^2\mS(2\i\lambda_0+\omega)\mS(-\omega)-(\i\lambda_0-\omega)^2\mS(2\i\lambda_0-\omega)\mS(\omega)
\end{equation}
For a massive Majorana fermion with non-integer mass and standard normalizable boundary conditions at the asymptotic boundary, the S-matrix is given in \eqref{eq: S matrix of normalizable mode}. Together with $\lambda_0=\frac12$, this yields
\begin{equation}
\tilde f(\omega)=2\i\omega.
\end{equation}
Substituting into \eqref{eq: F24} gives a local operator:
\begin{equation}
[P^-,P^+]=\frac{1}{2\pi}\int dV\Upsilon^p(V)V\partial_V\Upsilon^p(V).
\end{equation}
Comparing with \eqref{eq: defintion of three generators}, we find
\begin{equation}
[P^-,P^+]=2\i B,
\end{equation}
in agreement with the sl(2,$\mathbb R$) algebra \eqref{eq: F3}.

\section{Examples of two-point function that doesn't have a continuous limit}
\label{app: Several examples of two-point function that doesn't have a continuous limit}
Not every boundary two-point function $A(t,t')$ admits a sensible bulk-QFT description in the continuum limit \eqref{eq: definition of continuous limit}. The basic diagnostic is the gate angle: from \eqref{eq: relate gate angle to Omega}, obtaining an $O(1)$ conformal factor requires
$\theta_N=\frac{\pi}{2}-O(\Delta t)$.
If instead $\theta_N$ stays $O(1)$ away from $\pi/2$ (or approaches $0$), the would-be bulk interpretation breaks down. Here we record two explicit examples where this happens.

\paragraph{Example I. }Consider
\begin{equation}
A(t,t')=\frac{1}{\cosh(t-t')}
\end{equation}
which can be obtained by analytically continuing the  large-$q$ SYK two-point function \eqref{eq: syk 2pt} to $q=2$ and infinite temperature. Using the determinant formula \eqref{eq: determinant formula}, we compute the gate angles. The relevant Toeplitz matrix is
\begin{equation}
\small
A_{N+1}=\begin{bmatrix}
1 & \sech(\Delta t) & \sech(2\Delta t) & \cdots &\sech(N\Delta t)\\
\sech(\Delta t) & 1 & \sech(\Delta t) &\cdots & \sech((N-1)\Delta t)\\
\sech(2\Delta t) & \sech(\Delta t) & 1 &\cdots &\sech((N-2)\Delta t)\\
\vdots & \vdots&\vdots & & \vdots\\
\sech(N\Delta t) & \sech((N-1)\Delta t) & \sech((N-2)\Delta t)& \cdots &1
\end{bmatrix}
\end{equation}
Using the block-determinant identity\footnote{We thank Xinyu Sun for explaining the following computation.}
\begin{equation}
\det A_{N+1}=\det\begin{bmatrix}
1 & x_N^\intercal\\
x_N & A_N
\end{bmatrix}=\det(A_N-x_Nx_N^\intercal)
\end{equation}
with $(x_N)_i:=\sech(i\Delta t)$, we find
\begin{equation}
(A_N-x_Nx_N^\intercal)_{ij}=\sech((i-j)\Delta t)-\sech(i\Delta t)\sech(j\Delta t)=(A_N)_{ij}\tanh(i\Delta t)\tanh(j\Delta t)
\end{equation}
and hence, defining $(\Lambda_N)_{ij}=\delta_{ij}\tanh(i\Delta t)$,
\begin{equation}
\det(A_{N+1})=\det(\Lambda_NA_N\Lambda_N)=\det(A_N)\prod_{i=1}^{N}\tanh^2(i\Delta t)
\end{equation}
The determinant formula \eqref{eq: determinant formula} then gives
\begin{equation}
\sin\theta_N=\tanh(N\Delta t)
\end{equation}
so in the continuum limit \eqref{eq: definition of continuous limit},
\begin{equation}
\sin\theta(z)=\tanh(z)
\end{equation}
In particular, $\theta$ remains $O(1)$ rather than approaching $\pi/2$, so this $A(t,t')$ does not admit the desired bulk-QFT limit.

\paragraph{Example II. }Consider
\begin{equation}
A(t,t')=e^{-(t-t')^2}
\end{equation}
A direct computation of the first few gate angles (e.g. in Mathematica; we checked up to $N=12$) gives
\begin{equation}
\sin^2\theta_N=1-e^{-2N(\Delta t)^2}
\end{equation}
In the continuum limit \eqref{eq: definition of continuous limit} this implies $\sin\theta(z)=0$, i.e. the gates approach near-perfect reflection rather than SWAP (cf. \cref{fig: after first layer}(c),(d)). In this case the appropriate scaling limit is instead the Krylov limit \eqref{eq: krylov limit}, which yields an $O(1)$ Krylov coefficient $b_n\equiv\sqrt{2n}$ via \eqref{eq: krylov coeffciient and angle}.

\bibliography{references}

\bibliographystyle{utphys}

\end{document}